\documentclass[]{article}

\usepackage{color}
\usepackage{dsfont,amsmath,amssymb,graphicx,url,bm,mathtools,bbm,dsfont,amsthm}
\usepackage[a4paper, margin=1in]{geometry}
\usepackage[sort,comma]{natbib}
\usepackage[pagebackref=false,colorlinks]{hyperref}
\hypersetup{pdffitwindow=true,linkcolor=blue,citecolor=blue,urlcolor=blue}

\newcommand*{\id}{{\normalfont\hbox{1\kern-0.15em \vrule width .8pt depth-.5pt}}}

\usepackage{siunitx}

\title{Bayesian profile regression for clustering analysis involving a longitudinal response and explanatory variables}

\begin{document}
\begin{center}
{\LARGE\bf Bayesian profile regression for clustering analysis involving a longitudinal response and explanatory variables}
\end{center}
\medskip
\begin{center}
{\large Ana\"is Rouanet, PhD$^{1}$, Rob Johnson, PhD$^{1}$, Magdalena Strauss, PhD$^{1,2}$, Sylvia Richardson, PhD$^{1}$, Brian D Tom, PhD$^{1}$, Simon R White, PhD$^{1,3}$ and Paul D. W. Kirk$^{1,4,\ast}$ \\[15pt]
\emph{$^{1}$MRC Biostatistics Unit, School of Clinical Medicine, University of Cambridge, U.K.}\\
\emph{$^{2}$EMBL-EBI, Wellcome Genome Campus, Hinxton, Cambridgeshire, CB10 1SD, UK}\\
\emph{$^{3}$Department of Psychiatry, University of Cambridge, Cambridge, CB2 3EB, UK}\\
\emph{$^{4}$Cambridge Institute of Therapeutic Immunology \& Infectious Disease (CITIID), Jeffrey Cheah Biomedical Centre, Cambridge Biomedical Campus, University of Cambridge, U.K.}}\\
\smallskip
{$^{\ast}$\url{paul.kirk@mrc-bsu.cam.ac.uk}}\\
\end{center}
\bigskip

\begin{center}
Preprint, \today
\end{center}
\bigskip\bigskip

\begin{abstract}
The identification of sets of co-regulated genes that share a common function is a key question of modern genomics.  Bayesian profile regression is a semi-supervised mixture modelling approach that makes use of a response 
to guide inference toward relevant clusterings.  Previous applications of profile regression have considered univariate continuous, categorical, and count outcomes.  In this work, we extend Bayesian profile regression to cases where the outcome is longitudinal (or multivariate continuous)
and provide PReMiuMlongi, an updated version of PReMiuM, the R package for profile regression.  We consider multivariate normal and Gaussian process regression response models and provide proof of principle applications to four simulation studies. The model is  applied on budding yeast data to identify groups of genes co-regulated during the \textit{Saccharomyces cerevisiae} cell cycle. 
We identify 4 distinct groups of genes associated with specific patterns of gene expression trajectories, along with the bound transcriptional factors, likely involved in their co-regulation process.

\end{abstract}

\section{Introduction}

Understanding gene regulation is a major research challenge in molecular biology \citep{Jacob1961}. To this end, a number of model organisms are studied for their relative simplicity in exploring the underlying regulatory mechanisms, including the budding yeast \textit{Saccharomyces cerevisiae} \citep{tong2004global}. It has been shown that integrating different types of omic data may allow us to further deepen our understanding of the transcription process from DNA to messenger RNA \citep{Savage2010}, including allowing us to identify transcriptional modules \citep{ihmels2002revealing}: sets of functionally related genes that are regulated by the same transcription factor(s). In this paper, we integrate yeast gene expression time course data with chromatin immunoprecipitation microarray data in order to identify such gene sets. This analysis motivates a new adapted clustering approach.

Clustering is typically formulated as an unsupervised method for identifying \textcolor{black}{homogeneous subgroups in a heterogeneous population} \citep{Jain:1999}.  In addition to the perennial problem of how to determine the appropriate number of clusters \citep{Rousseeuw:1987wr,Fraley:2002bg,Sugar:2003fa,Tibshirani:2005if}, a key challenge is how to validate a given clustering structure \citep{,Kerr:2001ht,Brock:2008cm}.  A common approach is to make use of left-out information  to assess if the identified clustering structure provides a relevant stratification of the population \citep{Yeung:2001vf,Handl:2005fo}.  For example, if we had clustered patients on the basis of their blood plasma proteome profiles, then we might check to see if patients allocated to the same cluster also tended to have  responses to treatment which are more similar than for patients in different clusters.  If this were the case, we might conclude that the identified clustering structure was {relevant} for the particular aim of identifying clusters associated with differential treatment responses.  Of course, assessment of relevance is necessarily context- and application-specific.  The left-out information (which we will refer to here as an {outcome} or {response}) is typically very closely related to the true aim of the clustering analysis, and implicitly defines the criterion by which a given clustering structure is assessed to be relevant or irrelevant. 

When clustering \textcolor{black}{datasets with many variables --- some of them possibly not defining} a clustering structure \citep{Law2003,Law2004,Tadesse2005}, or when subsets of variables define a variety of different clustering structures \citep{Cui2007,Niu2010,Guan2010,Li2011,Niu2014,Kirk2018} --- it may be desirable to make use of (potentially highly informative) outcome information directly, in order to guide inference toward the most relevant clustering structures.  That is, we may wish to use the outcome information during the clustering analysis itself, rather than during post-analysis validation.  This is one of the principal motivations for {Bayesian profile regression} \citep{Molitor:2010gm}, a semi-supervised approach for model-based clustering that allows outcome information to be taken into account for the determination of the clustering.  Previous applications of profile regression have considered univariate continuous, categorical, or count outcomes \citep{Liverani:2015jg}. 
In this work, we extend Bayesian profile regression to cases where the outcome is longitudinal (or multivariate continuous) and provide PReMiuMLongi, an updated version of PReMiuM, the R package for profile regression \citep{Liverani:2015jg}.

In the following, we recall the Bayesian profile regression modelling framework in Section 2 and present two extensions to handle a longitudinal outcome in Section 3. In Section 4, we demonstrate through different simulation studies the benefit of integrating a longitudinal outcome in the clustering algorithm and assess the two methods that we propose, in terms of clustering recovery and computation time.  Section 5 features the results of the methods applied to budding yeast data,  uncovering groups of genes co-regulated during the cell cycle,  followed by a discussion in Section 6.

\section[Bayesian profile regression: a semi-supervised clustering model]{\texorpdfstring{Bayesian profile regression:\\ A semi-supervised clustering model}{Bayesian profile regression: a semi-supervised clustering model}}

We suppose that we have data comprising observations on a vector of covariates, ${\bf x}$, and outcomes,~${\bf y}$.  \citet{Molitor:2010gm}, which we follow, permitted their model for ${\bf y}$ to depend upon additional covariates, ${\bf w}$, referred to as fixed effects.  These fixed effects are covariates that may be predictive of the outcome, but do not directly contribute to the clustering, and which come with associated ``global'' (i.e. not cluster-specific) parameters, $\boldsymbol{ \beta}$, for the model linking ${\bf y}$ and~${\bf w}$.  The general model considered in \citet{Molitor:2010gm} for $C$ mixture components is then:
\begin{align}
	p({\bf x}, {\bf y}|\boldsymbol{\phi}, \boldsymbol{\theta}, \boldsymbol{\pi}, \boldsymbol{\beta}, {\bf w}) = \sum_{c=1}^C \pi_c ~f_{\bf y}({\bf y}|\boldsymbol{\theta}_c, {\bf w}, \boldsymbol{ \beta})f_{\bf x}({\bf x}|\boldsymbol{\phi}_c),\label{profRegr}
\end{align}
with  $\pi_c$ the mixture weights,  $\boldsymbol{\phi}_c$ the cluster-specific parameters for the density $f_{\bf x}$,   $\boldsymbol{\theta}_c$ the cluster-specific parameters for the density $f_{\bf y}$ \textcolor{black}{ and $\beta$ the vector of regression parameters in the outcome model}.  The original formulation of this model, which we also adopt here, is in terms of infinite (specifically Dirichlet process) mixture models; however, we note that the model is equally applicable in the case of finite $C$.   A stick-breaking construction \citep{Pitman2002} is adopted for the Dirichlet Process prior	set upon the mixture distribution, \textcolor{black}{also known as a GEM or Griffiths-Engen-McCloskey prior \citep{Pitman2002}, constructed as follows}:
\begin{align}
u_c &\sim \mbox{Beta}(1, \alpha)\notag\\
\pi_1 &= u_1, {\mbox{ and }}
\pi_c = u_c \prod_{r = 1}^{c-1}(1- \pi_r), {\mbox{ for $c \ge 2$}},\label{GEM}
\end{align}    

\noindent \textcolor{black}{where $\alpha$ is a positive parameter, for which we adopt a gamma prior. 
See, for example, \citet{Ishwaran2001} and \citet{Kalli:2009hu} for further background, and \citet{Liverani:2015jg} and \citet{Hastie:2015gc} for details of how inference of the $\pi_c$'s is performed in the Bayesian profile regression model.} \textcolor{black}{Thereafter, we define the component allocation variable $z$ taking values in $1, \cdots, C$. Thus we have $\pi_c=P(z=c)$. }

\textcolor{black}{ Profile regression can accommodate various types of covariates, by specifying the corresponding cluster-specific distributions in Equation (1). In the case of $Q$ categorical covariates, the probability distribution in cluster $c$ for ${\bf x_i}=(x_{i1}, \cdots, x_{i,Q})$ for individual $i$,  $i=1,\cdots,N$ would be written as}
 \begin{equation}
	f_{\bf x}({\bf x_i}|\boldsymbol{\phi}_{z_i})=\prod_{q=1}^Q \prod_{e=1}^{E_q}\phi_{z_i,q,e}^{1_{\{x_{i,q}=e\}}}
	\label{discrete_X}
	\end{equation} 
	
\noindent \textcolor{black}{ 	with $z_i = c$ if the $i$-th individual is allocated to the $c$-th cluster, $\boldsymbol{\phi}_{z_i}=\{\phi_{z_i,q,1},...,\phi_{z_i,q,E_q}\}$  where $\phi_{z_i,q,e}$ is the probability that covariate $q \in \{1,...Q\}$ takes the value $e \in \{1,...,E_q\}$ in cluster $z_i=c$ and $1_{\{x_{i,q}=e\}}=1$ if $x_{i,q}=e$ and 0 otherwise. Conjugate Dirichlet priors are adopted for the parameters $\phi_{z_i,q,e} \sim  ~ \text{Dirichlet}(a_{q,e})$. 
Multivariate Gaussian densities can also be used to handle a set of correlated Gaussian covariates, using a multivariate normal prior for the mean vector and an inverse Wishart prior for the covariance matrix. Finally, a product of both types of probability distributions accommodate mixed mixtures of Gaussian and discrete covariates, assuming independence given the cluster allocations.}

\textcolor{black}{Several distributions are proposed in the current profile regression model for the outcome, such as Bernoulli, Poisson, Binomial, etc. In the binary case for example, the probability distribution $f_Y$ for individual $i$ is given by}:
	
	\begin{equation*}
		P({\bf y_i}=1|\theta_{z_i}, \beta, {\bf w_i})= \text{expit}(\theta_{z_i}+ \beta^{\top} {\bf w_i})
		\end{equation*}
\textcolor{black}{where $\text{expit}(x)=\frac{\exp(x)}{1+\exp(x)}$, $\theta_{z_i}$  as well as the regression parameters  $\beta$ follow $t$ location-scale distributions.} 

The profile regression model described in Equation (\ref{profRegr}) implicitly assumes that the covariates,~${\bf x}$, and outcomes,~${\bf y}$, are linked via a shared clustering structure, but are otherwise conditionally independent. 
The inclusion of fixed effects ${\bf w}$ in Equation \eqref{profRegr} allows for the possibility that there are additional nuisance covariates that might be predictive of the observed ${\bf y}$, but which we do not wish to include in the clustering analysis.  For example, in an analysis of data from a genome wide association study of lung cancer considered in \citet{Papathomas2012}, ${\bf x}$ comprises genetic covariates (SNPs), while ${\bf w}$ comprises potentially confounding covariates including age,  sex, and smoking status. 
Adopting Bair's terminology \citep{Bair:2004fo}, we refer to this as a {semi-supervised} mixture model, since clustering is guided by an outcome variable.  

\textcolor{black}{A variable selection procedure selects the covariates ${\bf x}$ that contain clustering information. The parameters in Equation (\ref{profRegr}) are replaced by composite ones defined as: 
	\begin{align}
	\phi^*_{z_i,q}= &~{ {\gamma_{z_i,q}}}~\phi_{z_i,q} + (1-{ {\gamma_{z_i,q}}})~\phi_{0,q}\nonumber \\
	\gamma_{z_i,q} \sim&~ \text{Bernoulli}(\rho_q) \label{rho}
	\end{align}
	where $\phi_{0,q}$ is the empirical proportion of genes with $x_q=1$ in the whole population  and $\gamma_{z_i,q}$ is a relevance indicator for covariate $q$ in cluster $z_i$,  with a sparsity inducing hyperprior $\rho_q \sim 1_{\{\omega_q=0\}} \delta_0(\rho_q) + 1_{\{\omega_q=1\}}~ \text{Beta}(0.5,0.5)$ and $\omega_q\sim \text{Bernoulli}(0.5)$.  Values of the hyperprior $\rho_q$ close to 1 indicate a difference in the cluster-specific distributions of variable $q$  compared to its distribution in the whole sample. See \citet{Liverani:2015jg} for more details about the alternative ``soft variable selection'' procedure, also implemented in PReMiuM and PReMiuMlongi.
}

The profile regression model is fitted via Markov Chain Monte Carlo (MCMC) using a blocked Gibbs sampler. A representative clustering structure is identified using  postprocessing techniques of the MCMC output, based 
on the posterior similarity matrix which represents the estimated 
co-clustering probabilities for each pair of individuals, 
and also quantifies the uncertainty associated with the cluster allocations. 
We refer the reader to \citet{Liverani:2015jg} for further details.

\section{Clustering guided by a longitudinal outcome}\label{longOut}

In previous applications of Bayesian profile regression, the outcome ${\bf y}$ has been assumed to be univariate \citep{Molitor:2010gm,Papathomas2012}.  Here we extend the original model to consider cases in which ${\bf y}$ comprises longitudinal continuous data.  \textcolor{black}{We present the multivariate normal (MVN) and Gaussian process (GP) regression response models in Sections \ref{multivar} and \ref{funData}, respectively. In Section~\ref{multivar}, we restrict attention to situations in which we have a longitudinal outcome measured on individuals at a common set of time points.  
In Section \ref{funData}, individuals may have different outcome measurements with specific time points. }

\subsection{Multivariate normal outcome model}\label{multivar}

We propose to model the outcome as multivariate normal, refering to this as the MVN response model, when all individuals/units have response measurements at a common set of time points.  Such a  model may also be appropriate when each individual's longitudinal response is summarised through a collection of statistics, as in \citet{Hathaway1993}.  For flexibility, we make no assumptions about the structure of the covariance matrix in our model, and defer until Section \ref{funData} the discussion of a model in which we explicitly try to capture the time-ordering of the data. \textcolor{black}{In this section, the vector of $M$ time points, ${\bf t}=\{t_1, \ldots, t_M \}$, is assumed to be the same for all individuals, the outcome for the $i$-th individual is then of the form ${\bf y}_i = [y_{i,1}, \ldots, y_{i,M}]^\top \in \mathbb{R}^M$, where $y_{is}$ denotes the value for the response measured on the $i$-th individual at $s$-th time point $t_s$ and it is assumed that there are no missing outcome values.}

For the MVN response model, the component-specific parameters $\boldsymbol{\theta}_c$ in Equation~\eqref{profRegr} comprise mean vector, $\boldsymbol{\mu}_c$, and covariance matrix, $\boldsymbol{\Sigma_c}$.  In the simple case where there are no confounding covariates ${\bf w}$, the density $f_{\bf y}({\bf y} | \boldsymbol{\theta}_c)$ is simply a multivariate normal density with mean $\boldsymbol{\mu}_c$ and covariance matrix $\boldsymbol{\Sigma_c}$.  When we have confounding covariates, we assume that they act on the response via a constant shift of the mean.  In particular, if ${\bf w}_i \in \mathbb{R}^R$, then $\boldsymbol{\beta}\in \mathbb{R}^R$, and the MVN response model is as follows:

\begin{align}
	f_{\bf y}({\bf y}_i|&\boldsymbol{\theta}_c, \boldsymbol{ \beta}, {\bf w}_i, {\bf t})= f_{\bf y}({\bf y}_i|\boldsymbol{\mu}_c, \Sigma_c, \boldsymbol{\beta}, {\bf w}_i, {\bf t})\notag\\
	&= \frac{1}{\sqrt{(2\pi)^M|\Sigma_c|}}\exp\left(-\frac{1}{2}({\bf y}_i-(\boldsymbol{\mu}_c + (\boldsymbol{\beta}^\top {\bf w}_i){\bf 1}_M))\Sigma_c^{-1}({\bf y}_i-(\boldsymbol{\mu}_c + (\boldsymbol{\beta}^\top{\bf w}_i){\bf 1}_M))^{\top} \right),\label{MVN}
\end{align}
where ${\bf 1}_M$ denotes a vector of ones of length $M$ and we note that $\boldsymbol{\beta}^\top{\bf w}_i$ is a scalar.  To perform inference, we adopt conjugate normal-inverse Wishart \textcolor{black}{\citep{Gelman2013}} priors  $\boldsymbol{\mu_c}|\boldsymbol{\Sigma_c}\sim\mathcal{N}(\boldsymbol{\mu_0},\boldsymbol{\Sigma_c}/\kappa_0)$, and  $\boldsymbol{\Sigma_c}\sim\mathcal{W}^{-1}(\boldsymbol{R_0},\nu_0)$, where $\boldsymbol{\mu_0} = \frac{1}{N}\sum_{i=1}^{N}{\bf y_i}$, where $N$ is the number of participants, $\kappa_0 = 0.01$, $\nu_0 = M$ and 
\begin{equation}
\boldsymbol{R_0} = \left(
\frac{\nu_0}{N}\sum_{i=1}^{\textcolor{black}{N}}({\bf y_i}-\boldsymbol{\mu_0})({\bf y_i}-\boldsymbol{\mu_0})^T\right)^{-1}.\label{R0}
\end{equation}

\subsection{Gaussian process outcome model: a Bayesian functional data approach}\label{funData}
In this section, we consider a more general setting in which the number of observations and the time points vary across individuals. In this case, we write ${\bf y}_i = y_i({\bf t}_i)=[y_i(t_{i,1}), \ldots, y_i(t_{i,M_i})]^\top$ with ${\bf t}_i$ 
the vector of the $M_i$ time points associated with the $i$-th individual, and $y_i(t)$ refers to the value of the response for the $i$-th individual at time~$t$.    In the outcome model, we assume that the complete longitudinal trajectory of the $i$-th individual's response is expressed in the following form: 
\begin{equation} 
{\bf y_i}(t) = g_i(t) +  \boldsymbol{\beta}^\top{\bf w}_i + \epsilon_{i}(t),\label{responseModel}
\end{equation}
where 
$g_i$ is a continuous function of time; $\epsilon_{i}(t)=\epsilon_{i,t} \sim \mathcal{N}(0, \sigma_{i}^2)$ is assumed to be i.i.d. Gaussian noise; and $ \boldsymbol{\beta}$ and ${\bf w}_i$ are the vectors of regression parameters and of the individual covariates, respectively, defined as in the MVN specification.   In practice, of course, we only observe $y_i(t)$ at a discrete set of time points, $\{t_{i,1}, \ldots,  t_{i,M_i}\}$; however, this functional approach proves useful when dealing with individuals who have observations at different time points.

If individuals $i$ and $j$ are both in cluster $c$, then we assume that $g_i(t) = g_j(t): = g^{(c)}(t)$, where $g^{(c)}$ denotes the function associated with the $c$-th cluster.  Moreover, $\sigma_{i}^2 = \sigma_{j}^2 := \sigma_{c}^2$, where $\sigma_{c}^2$ denotes the noise variance associated with the $c$-th cluster.   The conditional distribution of $y_i$ at time $t$, given that the $i$-th individual is allocated to the $c$-th component, is then:
\begin{align}
	{\bf y_i}(t)|z_i = c \quad &\sim\quad \mathcal{N}(g^{(c)}(t) + \boldsymbol{\beta}^\top{\bf w}_i , \sigma_{c}^2).\label{likeli1}
\end{align}

In order to proceed, we could specify a parametric form for the functions~$g^{(c)}$ (e.g. we might specify that $g^{(c)}$ is a polynomial of degree $d$), whose component-specific parameters we would need to infer in order to fit the model.  However, as we now describe, here we adopt a Bayesian non-parametric approach, and take a Gaussian process prior for the unknown function $g^{(c)}$.

\subsubsection{Gaussian process priors for unknown functions}

A Gaussian process is a collection of random variables, any finite number of which have a joint Gaussian distribution \citep{Rasmussen2006}.   This definition simply means that a (potentially infinite) collection $v_1, v_2, \ldots $ of random variables defines a Gaussian process if and only if any finite subcollection of the variables is jointly distributed according to a Gaussian distribution.   

In GP regression, a GP prior is assumed for the outputs of an unknown function~$g^{(c)}$\textcolor{black}{\citep{Rasmussen2006}}.  This means that we assume a priori that $\{g^{(c)}(t_1), g^{(c)}(t_2), \ldots, g^{(c)}(t_T)\}$ are jointly distributed according to a $T$-variate Gaussian distribution for any vector of times $\{t_1, t_2, \ldots, t_T\}$ of finite length $T$.  To fully specify a GP prior, we require a mean function, $m^{(c)}$, and a covariance function, $k^{(c)}$, which define the mean vectors and covariance matrices of the Gaussian distributions associated with each finite subcollection of the variables.  We write $g^{(c)} \sim \mathcal{GP}(m^{(c)}, k^{(c)})$ to indicate that we have assumed a Gaussian process prior with mean function $m^{(c)}$ and covariance function $k^{(c)}$  for the function $g^{(c)}$, so that:
\begin{quote}
	$g^{(c)} \sim \mathcal{GP}(m^{(c)}, k^{(c)})$ if and only if, for any finite collection $\{t_1, t_2, \ldots, t_T\}$ of times, we have $[g^{(c)}(t_1), \ldots, g^{(c)}(t_T)]^\top \sim \mathcal{N}_T({\bf m}^{(c)}, {\bf K^{(c)}}),$ where ${\bf m}^{(c)}_j = m^{(c)}(t_j)$, ${\bf K}^{(c)}_{j, j'} = k^{(c)}(t_j, t_{j'})$, and $\mathcal{N}_T$ denotes a $T$-variate normal distribution.
\end{quote}

There are many possibilities for $m^{(c)}$ and $k^{(c)}$;  see, for example, \citet{Rasmussen2006}.  In practice, we take the standard default choice of setting $m^{(c)}$ to be the zero function, so that $m^{(c)}(t_j) = 0$ for all $t_j$. \textcolor{black}{Note that if one prefers to use a non-zero mean function, the specification with a null mean function can then be applied to the difference between observations and the specified non-zero mean function.} We also take $k^{(c)}$ to be the widely-used {squared exponential} covariance function  or Gaussian kernel \textcolor{black}{ which is infinitely differentiable (i.e. very smooth), characterized by only two parameters, and has previously provided robust performances in the context of clustering \citep{Strauss2020}}. Thus we have:
\begin{align}
k^{(c)}(t_j,t_{j'})& = a_c\exp\left(-\frac{(t_j-t_{j'})^2}{2l_{c}}\right),\label{kernel}
\end{align}

\noindent where here the hyperparameters $a_c$ and $l_{c}$ are respectively the {signal variance} and {length-scale} for the squared exponential covariance function $k^{(c)}$.  Likewise, we define the corresponding function on time vectors: $\mathcal{K}^{(c)}({\bf t}, {\bf s})$ whose output is a matrix with $(i,j)^{\text{th}}$ element equal to $k^{(c)}(t_i,s_{j})$. All hyperparameters ($a$, $l$ and the variance of the measurement error $\sigma^2$) must be positive.  Following \citet{Neal1999}, we deal throughout with the logarithms of these quantities, eliminating the positivity constraint (and thereby making sampling of these quantities more straightforward).  For convenience, we adopt independent, standard normal priors for each of $\log(a)$, $\log(l)$ and $\log(\sigma^2)$ as in \citet{Kirk2012} and \citet{mcdowell2018}.\\

\textcolor{black}{
In non-parametric Bayesian clustering, the tuning of the variance prior is of particular importance. Depending on the application at hand, standard lognormal priors may be too vague and the initialization may lead to the algorithm getting stuck into states with too few or too many clusters, leading to spurious results. \citet{Ross2013} proposed to set the hyperparameters using empirical Bayes strategies. The lognormal hyperpriors adopted for the covariance functions $\mathcal{K}_c$ can be tuned with an empirical variogram of the data \citep{Diggle1990}, which quantifies the correlation between observations as a function of the time interval and provides empirical estimates of the signal variance and length-scale parameters.  
In addition, and based on our experience, it may be necessary to constrain the variance of the measurement error, as large values may lead to a small number of clusters and conversely, small  values may lead to a high  number. This can be diagnosed by unstructured posterior similarity matrices (i.e. with unclear separations). We thus propose a specification where the measurement error variance is proportional to the signal variance, in each cluster: $ a_{c} = r*\sigma_{c}^2$, where the $r$ ratio is fixed by the user. Sensitivity analyses are then necessary to validate this variance ratio value.}

\subsubsection{Likelihood}
We adopt a GP prior, with component specific hyperparameters $a_c$ and $l_c$, for each unknown function~$g^{(c)}$.  Suppose there are $n_c$ individuals associated with component $c$.  For notational convenience, we assume that these are individuals $1, \ldots, n_c$, i.e. we assume that $z_i = c$ for $i = 1, \ldots, n_c$ and $z_i \ne c$ for all other $i$.  Recalling that 
${\bf y}_{i} = [y_i(t_{{i},1}), \ldots,  y_i(t_{{i},M_{i}})]^\top$
and that ${\bf t}_i=[t_{{i},1}, \ldots,  t_{{i},M_{i}}]$ denotes the vector of the $M_{i}$ time points associated with the ${i}$-th individual, we define the ``corrected'' vector of observations for the $i$-th individual to be ${\bf y'}_i = {\bf y}_{i} - (\boldsymbol{\beta}^\top {\bf w}_i){\bf 1}_{M_i}$.  It then follows from our model for the response (Equations \eqref{responseModel} and \eqref{likeli1}) and the GP prior, that:  
\medskip
\begin{align}
	\label{GPlik}
	[{\bf y'}_1, \ldots, {\bf y'}_{n_c} ]^\top | g^{(c)}, \sigma_c^2 \quad &\sim\quad \mathcal{N}([{\bf g}^{(c)}_1, \ldots, {\bf g}^{(c)}_{n_c} ]^\top , \sigma_{c}^2 ~I_{\mathcal{M}_c})  \\\medskip 
	\label{GPprior}
	[{\bf g}^{(c)}_1, \ldots, {\bf g}^{(c)}_{n_c} ]^\top | a_c, l_c &\sim \mathcal{N}({\bm 0}, {\bf K}^{(c)}) 
\end{align}
\medskip
where ${\bf g}^{(c)}_i = [g^{(c)}(t_{{i},1}), \ldots, g^{(c)}(t_{{i},M_{i}})]^\top$, 
$\mathcal{M}_c = \sum_{i = 1}^{n_c} M_i$ and

\begin{equation}
{\bf K}^{(c)} = \left( \begin{array}{ccc}
{\bf K}^{(c)}_{{\bf t}_1,{\bf t}_1} & \ldots & {\bf K}^{(c)}_{{\bf t}_1,{\bf t}_{n_c}} \\
\vdots & \ddots & \vdots \\
{\bf K}^{(c)}_{{\bf t}_{n_c},{\bf t}_1} & \ldots & {\bf K}^{(c)}_{{\bf t}_{n_c},{\bf t}_{n_c}}\end{array}\right)
\text{ and }
{\bf K}^{(c)}_{{\bf t}_i,{\bf t}_j} = \left( \begin{array}{ccc}
k^{(c)}(t_{i,1}, t_{j,1}) & \ldots & k^{(c)}(t_{i,1}, t_{j,M_j}) \\
\vdots & \ddots & \vdots \\
k^{(c)}(t_{i,M_i}, t_{j,1}) & \ldots & k^{(c)}(t_{i,M_i}, t_{j,M_j})\end{array}\right)\label{18}
\end{equation}
with ${\bf K}^{(c)}$ and ${\bf K}^{(c)}_{{\bf t}_i,{\bf t}_j}$  matrices with dimensions $\mathcal{M}_c \times \mathcal{M}_c$  and $M_i \times M_j$, respectively, and individuals $i$ and $j$ in cluster $c$.

This defines the conditional likelihood given the function ${\bf g}$:
\begin{equation}
[{\bf y}_1, \ldots, {\bf y}_{n_c} ]^\top | {\bf g}^{(c)}, \sigma_c^2 \sim \mathcal{N}([(\boldsymbol{\beta}^\top {\bf w}_1){\bf 1}_{M_1}, \ldots, (\boldsymbol{\beta}^\top {\bf w}_{n_c}){\bf 1}_{M_{n_c}}]^\top + {\bf g}^{(c)\top},  \sigma_c^2 I_{\mathcal{M}_c}) \label{lik:cond}
\end{equation}
with ${\bf g}^{(c)}=[{\bf g}^{(c)}_1, \ldots, {\bf g}^{(c)}_{n_c} ]$. 
Alternatively, marginalising $g^{(c)}$, we see from Equations \eqref{GPlik} and \eqref{GPprior} that
\begin{equation}
[{\bf y'}_1, \ldots, {\bf y'}_{n_c} ]^\top | a_c, l_c, \sigma_c^2 \sim \mathcal{N}({\bf 0}, {\bf K}^{(c)} + \sigma_c^2 I_{\mathcal{M}_c}).
\end{equation}

We then can express the marginal likelihood: 
\begin{equation}
[{\bf y}_1, \ldots, {\bf y}_{n_c} ]^\top | a_c, l_c, \sigma_c^2 \sim \mathcal{N}([(\boldsymbol{\beta}^\top {\bf w}_1){\bf 1}_{M_1}, \ldots, (\boldsymbol{\beta}^\top {\bf w}_{n_c}){\bf 1}_{M_{n_c}}]^\top, {\bf K}^{(c)} + \sigma_c^2~ I_{\mathcal{M}_c}). \label{lik:marg}
\end{equation}

\subsection{Model inference}\label{prior}
The model is estimated via MCMC. The DP parameter $\alpha$ is updated using a Metropolis-Hastings step and cluster-specific covariate parameters $\phi_c$ are updated using Gibbs steps.  For the MVN specification of the outcome model, the mean vectors $\boldsymbol{\mu}_c$, covariance matrices $\boldsymbol{\Sigma}_c$ are updated using Gibbs sampler. Regarding the inference of the model with the GP specification, the logarithms of the hyperparameters of the squared exponential covariance function are updated in a Metropolis-within-Gibbs step. 
	Details on the complete Gibbs samplers are given in Section A in the  Supplementary Materials.\\

For the GP specification, we implemented two sampling algorithms of the hyperparameters $\boldsymbol{\theta}_c=(\log(a_c), \log(l_c), \log(\sigma_c^2))$, either marginalising or  sampling the $g^{(c)}$ function. 

\paragraph{\textcolor{black}{Marginal algorithm}}
In the first case, the hyperparameters are sampled from the following posterior distribution:
\begin{equation*}
	p(\boldsymbol{\theta}_c|\textbf{z},\boldsymbol{\beta},{\bf y}^{(c)},{\bf t}^{(c)},{\bf w}^{(c)}) \propto p(\boldsymbol{\theta}_c)\cdot p({\bf y}^{(c)}|\textbf{z},\boldsymbol{\theta}_c,\boldsymbol{\beta},{\bf t}^{(c)},{\bf w}^{(c)})
\end{equation*}
with ${\bf y}^{(c)}$ and ${\bf t}^{(c)}$ the vectors of observations and time points, respectively, of all the individuals in cluster $c$, and ${\bf w}^{(c)}$ the corresponding covariate matrix. The likelihood is defined in Equation (\ref{lik:marg}) and involves inverting a  covariance $\mathcal{M}_c\times \mathcal{M}_c$-matrix at each update of either $\textbf{z}$, $\beta$ or $\boldsymbol{\theta}_c$. To reduce computation time and, to a greater extent, handle large datasets,  we used the Woodbury matrix identity \citep{woodbury1950} to update the allocation variable $\textbf{z}$ in the corresponding Gibbs sampling step \citep{Strauss2019} (see Sections B.1 and B.2 of the  Supplementary Materials). We also implemented an approximation to inverse cluster-specific covariance matrices $K^{(c)}$ using an intermediate grid of regular time points \citep{snelson2006} (Section B.3 of the Supplementary Materials). This approximation is used if either the grid of time points or its size is pre-specified by the user. In the second case, the grid step is then defined from the range of observed time points.

\paragraph{\textcolor{black}{Conditional algorithm}}
\textcolor{black}{In the case where this approximation is not sufficient to reduce computation times (either the size of the time grid or the number of time points for all individuals in a same cluster is high)}, 
we implemented a second algorithm in which the function $\textsl{\bf g}^{(c)}$ is sampled and the conditional posterior distributions of the hyperparameters $\boldsymbol{\theta}^{1,2}_c=(\log(a_c), \log(l_c))$ and $\boldsymbol{\theta}^{3}_c=\log(\sigma_c
^2)$ are given by:
\begin{align*}
	p(\boldsymbol{\theta}^{1,2}_c|\textbf{z},\textsl{\bf g}^{(c)},{\bf t}^{(c)}) &\propto  p(\boldsymbol{\theta}^{1,2}_c)\cdot  p(\textsl{\bf g}^{(c)}|\textbf{z},\boldsymbol{\theta}^{1,2}_c,{\bf t}^{(c)})\\
	p(\boldsymbol{\theta}^{3}_c|\textbf{z},\textsl{\bf g}^{(c)},\boldsymbol{\beta},{\bf y}^{(c)},t^{(c)},{\bf w}^{(c)}) &\propto  p(\boldsymbol{\theta}^3_c)\cdot  p({\bf y}^{(c)}|\textbf{z},\boldsymbol{\theta}^3_c,\textsl{\bf g}^{(c)},\boldsymbol{\beta},{\bf t}^{(c)},{\bf w}^{(c)})
\end{align*}
where the likelihood corresponds to a $\mathcal{M}_c$-variate normal density with null mean and diagonal covariance matrix, with diagonal elements equal to $\sigma_c^2$. The posterior distribution of $\textsl{\bf g}^{(c)}(t^*)$, with $t^*$ any input time vector, is a multivariate Gaussian density with mean and covariance matrices:

\begin{align}
	{\bf m}^{*(c)} &=  \mathcal{K}^{(c)}(t^*,t^{(c)})~\times ~ [\mathcal{K}^{(c)}(t^{(c)},t^{(c)}) + \sigma_c^2~ I_{\mathcal{M}_c}]^{-1} ~\times~ {\bf y}^{(c)} \label{postGP}\\
	{\bf K}^{*(c)}&=\mathcal{K}^{(c)}(t^*,t^*) -\mathcal{K}^{(c)}(t^*,t^{(c)})~\times~  [\mathcal{K}^{(c)}(t^{(c)},t^{(c)}) + \sigma_c^2~ I_{\mathcal{M}_c}]^{-1}~\times~  \mathcal{K}^{(c)}(t^{(c)},t^*)\label{var_postGP}
\end{align}

\textcolor{black}{
	When the cluster-specific covariance matrices have low dimensions (equal to the sum of the time points of all individuals allocated to the given cluster), the marginalized algorithm will be faster, as hyperparameters are not sampled. However, if these dimensions are high, solving the necessary matrix inversions will become prohibitively time consuming or even encounter numerical difficulties. The computational complexity of matrix inversion being at least $O(n^{2.37})$ according to state-of-the-art blockwise inversion algorithm \citep{alman2021}, as a rule of thumb, we recommend using the conditional algorithm in the case where the mean number of time points per individual multiplied by the number of individuals is relatively high with respect to the computing power (e.g. higher than 1000 on a laptop with a 2.2 GHz dual core processor)}. When, in addition to this, time points are irregular across individuals, we recommend using the approximation mentioned above to compute the inverse of the covariance matrices.

\textcolor{black}{We recommend using at least as many iterations as data points for the sampling phase and 10\% iterations for the burn-in phase. In general, one should keep increasing the number of iterations while convergence of the MCMC chain is not reached. This convergence (or lack thereof) can be checked in such mixture models via the trace of the global parameters ($\alpha$, $\beta$ and the number of non-empty clusters) and the marginal partition posterior defined as $P(Z|D)$ by \citet{Hastie:2015gc}.}

\textcolor{black}{Finally, the MVN specification does not handle missing outcome values and all individual should have the same number of observations, considered to be collected at the same observation times. However, the GP specification handles missing outcome values and loss to follow-up under the assumption that data are missing either completely at random or at random.}
The R page for PReMiuMlongi, \textcolor{black}{\url{https://github.com/premium-profile-regression/PReMiuMlongi}}, contains both the package and detailed documentation.

\subsection{Estimated cluster trajectories and individual predictions}

A representative partition is obtained using post-processing methods: for each number of clusters \textcolor{black}{from 2 }to a specified threshold, a Partitioning Around Medoid (PAM) algorithm is performed on the posterior dissimilarity matrix $1-S$, where $S$ is a $N\times N$-matrix estimating the posterior probability of co-clustering for each pair of individuals. Then the optimal partition is chosen by maximising the average silhouette width across the best PAM partitions, as in \citet{Liverani:2015jg}. \\

The uncertainty on cluster allocations is then integrated into the estimation of the cluster-specific parameters, following \citet{Molitor:2010gm}. The GP hyperparameters of the $c^{\text{th}}$ cluster of the optimal partition are estimated by considering the $N_c$ individuals allocated to it, and averaging over the sampled hyperparameters corresponding to the clusters these individuals were allocated to, at each MCMC iteration:
\begin{equation}
	\hat{\boldsymbol{\theta}}_{c}= \frac{1}{N_c}\frac{1}{H}\sum_{h=1}^H  \sum_{i|z_i^*=c}\boldsymbol{\theta}^{(h)}_{z_i^{(h)}} 
	\label{profile_estim_stat}
\end{equation}
with $z_i^*$ the cluster of individual $i$ in the optimal partition and $\boldsymbol{\theta}^{(h)}_{z_i^{(h)}}$ the hyperparameters corresponding to the cluster $z_i^{(h)}$ individual $i$ was allocated to at iteration $h$. 
Following the same procedure, we can compute individual longitudinal predictions  (or predictions for a given profile of individuals) \textcolor{black}{while accounting for the uncertainty on cluster allocations: at each iteration, the given individual is assigned to a cluster based on his/her posterior membership probabilities given the data; then, the final longitudinal predictions $\hat{Y}_{i}$ are computed by averaging over his/her cluster-specific predictions at each MCMC iteration}. In the MVN specification, the predicted trajectory is obtained by
\begin{align*}
	\hat{\bf Y}_{i}&= \frac{1}{H} \sum_{h=1}^H  \hat{\boldsymbol{ \mu}}^{(h)}_{z^{(h)}_i}+\left(\hat{\boldsymbol{\beta}}^{(h)\top} {\bf  w_i}\right){\bf 1}_{M} \\
\end{align*}
with 
 $\hat{\boldsymbol{\beta}}^{(h)}$ 
the vector of estimated fixed effects 
at iteration $h$ 
and \textcolor{black}{$\boldsymbol{\hat{\mu}}^{(h)}_{z_i^{(h)}}$ the estimated mean} corresponding to the cluster $z_i^{(h)}$ the individual was allocated to, at iteration $h$.
In the GP specification, the predicted trajectory is obtained by the \textcolor{black}{point-wise} median of the posterior cluster-specific mean functions computed on a specified time grid $\textbf{t}$ of length $n_t$, over the $H$ MCMC iterations:

\begin{equation*}
	\hat{\bf Y}_{i}(\textbf{t})= \text{median}_h \{ {\bf m}^{*(h)}_{z^{(h)}_i}(\textbf{t})+ (\boldsymbol{\beta}^{(h)\top} {\bf w_i}){\bf 1}_{n_t}\} 
\end{equation*}
with ${\bf m}^{*(h)}_{z^{(h)}_i}$ the posterior mean function (Equation \ref{postGP}) in cluster $z^{(h)}_i$ the individual was allocated to at iteration $h$. The 90\% credible intervals were computed by considering the 5$^{\text{th}}$ and 95$^{\text{th}}$ percentiles. Section F of the Supplementary Materials  describes how these models have been incorporated into PReMiuMlongi, and the outputs that are generated.

\section{Simulation studies}\label{sim}

In this section, we present four simulation studies. The first two demonstrate our motivation in using outcomes in profile regression to guide clustering: first, where the clustering structure in the covariates is weak and, second, where there are multiple clustering structures in the covariates, of which only one has clinical relevance. The third demonstrates the MVN and GP model formulations, comparing them with each other in terms of duration and efficacy in recovering the generated clustering structure. The methods are also assessed as the number of observation times is varied, and as the clustering structures overlap.
The fourth study assesses the GP model on unequal time points.

Efficacy in recovering the target clustering structure is assessed via Rand indices \citep{Rand1971}. The Rand index is a metric ranging from 0 to 1 that reflects the similarity between two clustering structures. The index is 1 for identical structures. 
The adjusted Rand index corrects for the extent of agreement that would be expected by chance, given the size of the clusterings \citep{Hubert1985}. Hence, it is possible for the adjusted Rand index to take a negative value. In this section, we use the posterior expectation adjusted Rand (PEAR) criterion \citep{Fritsch:2009kd} that \textcolor{black}{corresponds to the adjusted Rand index between the optimal partition obtained from postprocessing the fitted model and the generated partition}. This index is computed using the R package \texttt{mcclust} \citep{Fritsch:2009kd}.


\subsection{Inference}
\textcolor{black}{In the following simulation studies, we adopt a GEM prior for the mixture distribution (Equation \ref{GEM}), with $\alpha\sim \text{Gamma}(2,1)$. The covariate model is defined by Equation (\ref{discrete_X}) with $\phi_{c,q,e} \sim  ~ \text{Dirichlet}(1)$ for each cluster $c$, covariate $q$ and covariate level $e$. In Simulations 1, 2 and 3, we use the MVN specification for the outcome model, defined in Equation (\ref{MVN}) with  $\kappa_0 = 0.01$ and $\nu_0$ equal to the number of time points $M$. In Simulations 3 and 4, the outcome model with the GP specification is defined by Equations \ref{GPlik} and \ref{GPprior}, adopting a null mean function for the GP prior and a squared exponential function for the variance covariance, with standard log normal priors for the hyperparameters $[\text{log}(a_c), \text{log}(l_c), \text{log}(\sigma^2_c)]\sim\mathcal{N}({\bf 0},\mathcal{I}_3)$. The initial number of clusters is set to 20. 
}

\subsection{Simulation study 1: Semi-supervised vs. unsupervised clustering}\label{sim1}

Here we demonstrate the importance of using outcome data in identifying latent clustering structures. We define two clusters, each with 50 individuals, and simulate covariate data and 0 to 4 outcome variables.  We use PReMiuMlongi with an MVN response model and we compare their estimated clustering structures with the generated one. 

Our focus is on the difference between inference using no outcome data ($M=0$), and inference using $M\geq 1$ observation times. Inference using no outcome data corresponds to studies in which, first, clustering is performed on covariate data and, second, cluster memberships are leveraged to explain some outcome. The rationale of profile regression is that, when the purpose of clustering is to group individuals according to their outcomes, using outcome data leads to more meaningful cluster labels.

\subsubsection{Design}

The covariates are discrete variables generated from a categorical distribution. We set the number of covariates to 10, denoted $q=1,...,10$. Each covariate $q$ has $R_q=R=3$ categories, with multinomial $R-$parameter $\boldsymbol{\phi}_{c,q}$ generated as follows:
$\boldsymbol{\phi}_{c,q} = v\textcolor{black}{\boldsymbol{\phi}^{(0)}_{c,q}}+\frac{1}{R}(1-v)\textbf{1}_R$ with 
$\textcolor{black}{\boldsymbol{\phi}^{(0)}_{c,q}} \sim \text{Dir}(\alpha_0\textbf{1}_{R})$
where $\textcolor{black}{\boldsymbol{\phi}^{(0)}_{c,q}}$ and $\textbf{1}_R$ are vectors of length $R$, and $\alpha_0=0.01$. 


\textcolor{black}{The $v$ parameter influences the cluster separability in terms of covariate profiles. A value of $v$ close to $0$ leads to similar profiles, as the probability $\phi_{c,q}$ for each covariate $q$ tends towards $1/R$ for all clusters $c$. A value of $v$ close to $1$ leads to distinct profiles, as the small value of $\alpha_0$ leads to most people in cluster $c$ having the same value for covariate $q$, which is unlikely to be the same between the clusters for all ten covariates.
We choose $v=0.4$ as an intermediate value to demonstrate the added benefit of integrating repeated outcome measures for recovering the clustering}.

The outcome is generated from a multivariate normal distribution with means $\textbf{1}_M$ and $4\cdot \textbf{1}_M$ for the first and second clusters, respectively, and covariance matrices $0.5 \mathcal{I}_M$, where $\mathcal{I}_M$ is the identity matrix of dimension $M$, the number of observation times.

\subsubsection{Results}

\begin{figure}[ht!] 
	\begin{center}
		\includegraphics[width=0.5\textwidth]{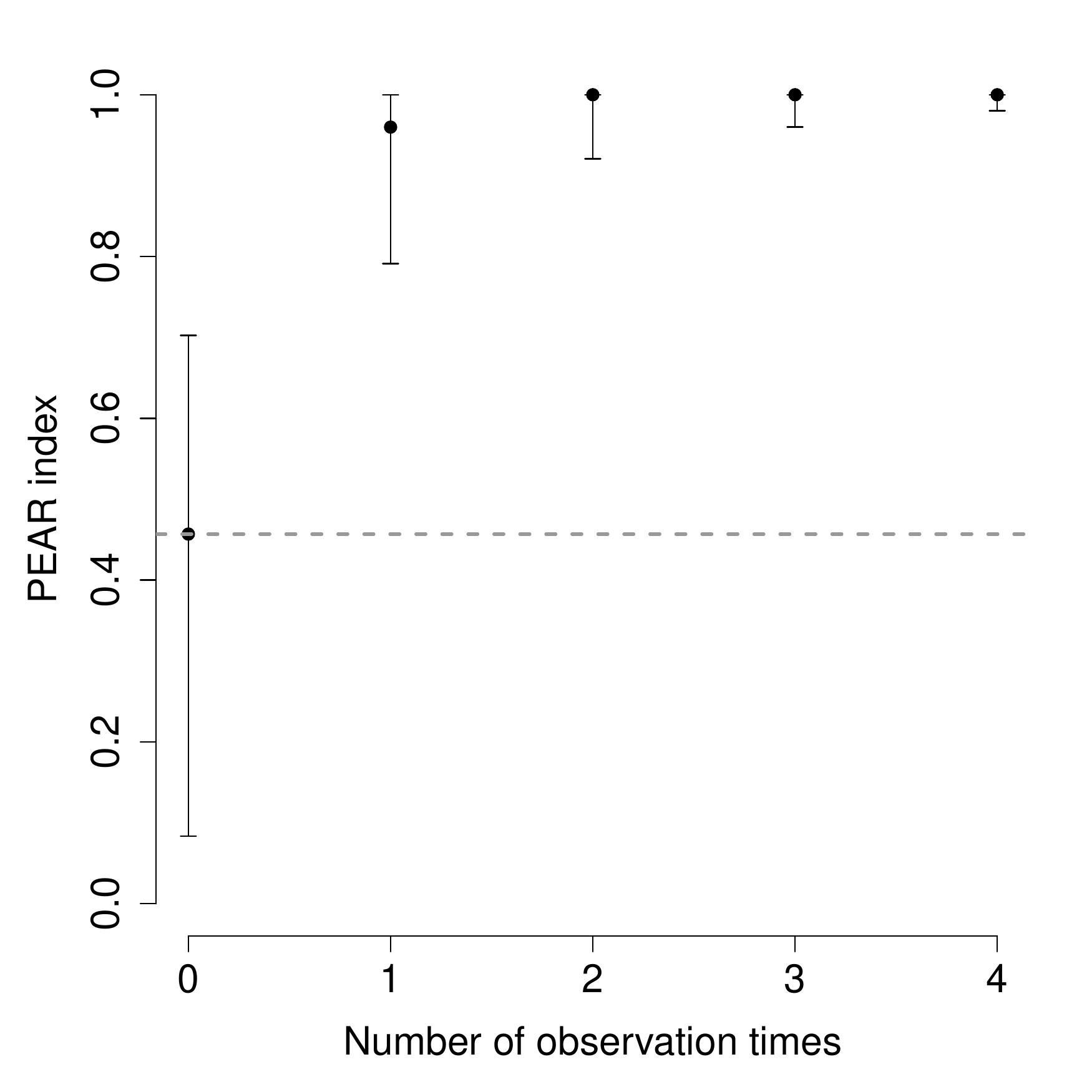} 
		\caption{The effect of number of observation times on PEAR index. Points are median PEAR indices for 1000 simulations with error bars showing 0.05 and 0.95 quantiles. In each simulation, data are generated for 100 individuals belonging to one of two clusters. Each individual has 10 covariates and 0 to 4 time points (or MVN outcomes). }
		\label{nOutcomes}
	\end{center}
\end{figure}

We ran PReMiuMlongi with the MVN response model and $\max(2000,2000M)$ iterations each for burn in and sampling. The effect of the number of observation times on the resulting PEAR indices is shown in Figure \ref{nOutcomes}. The x axis represents 5 simulation settings with all individuals having 0 to 4 time points. Each setting is simulated 1000 times. \textcolor{black}{When the outcome is unaccounted for, the PEAR indices are low as the separability parameter between the covariate profiles is relatively low ($\nu$ = 0.4)}. The major PEAR improvement is from zero to one observation time. This is due to the way the longitudinal pattern is simulated: the clustering structure is weak in the covariates and stronger in the outcomes. 
It is possible to recover, to some extent, the clustering structure in the covariates without use of outcome data. However, inclusion of outcome data improves cluster-structure (Figure \ref{nOutcomes}). 

\subsection{Simulation study 2: Semi-supervised clustering and variable selection}\label{sim2}

\textcolor{black}{Here we adapt the simulation study of Section \ref{sim1}  to a case in which there are two sets of covariates, that correspond to two different clustering structures of the population. The outcome data shares the clustering structure of one set of covariates and is independent from the other one (which we refer to as  “alternative clustering”). Moreover, we use the variable selection feature of PReMiuMlongi, in order to select the covariates that guide the final clustering. In the absence of outcome data, there is no reason for one set of covariates to be selected over the other one. Outcome data are therefore required for identifying meaningful clustering structures when more than one structure is present in the data according to different covariate subsets.}

\subsubsection{Design}

\textcolor{black}{Covariate data consist of four variables, two of which correspond to a first clustering structure (which we describe as the ``alternative clustering'', and denote $\check{c}$) and two of which correspond to a second clustering structure, which retains the label $c$. We simulate covariates as in Simulation 1, while choosing $\nu=1$ for all covariates, so that each clustering structure has a strong signature when those covariates are viewed alone (see Supplementary Figure 10). Extending the notation from Section \ref{sim1}, where each individual has a cluster assignment $\check{z}_i$ for $\check{c}$ and a cluster assignment $z_i$ for $c$, covariate $x_{q,i}|\check{z}_i=\check{c}$ has parameter
${\phi}_{\check{c},q}$ for $q=1,2$, and $x_{q,i}|z_i=c$ has parameter
${\phi}_{c,q} $ for $q=3,4$.}

\textcolor{black}{Response data are simulated as in Section \ref{sim1}, according to clustering structure $c$: }
$$y_i|z_i=c \sim \left\{
\begin{array}{lr}
{MVN}(\textbf{1}_M,0.5 I_M)  &  c=1 \\
{MVN}(4\cdot \textbf{1}_M,0.5 I_M)  &  c=2
\end{array}
\right\}.$$
\textcolor{black}{The response variable is aligned with clustering structure $c$, and is independent of the  alternative clustering structure $\check{c}$. The two covariates with clustering structure $c$ therefore align with the response variable. Again we vary the number of observation times from 0 to 4 in five simulation scenarios. We use this model to illustrate how inclusion of outcome data influences variable selection, and how these together affect PEAR indices.}

\subsubsection{Results}

\begin{figure}[ht!] 
	\begin{center}
		\includegraphics[width=0.5\textwidth]{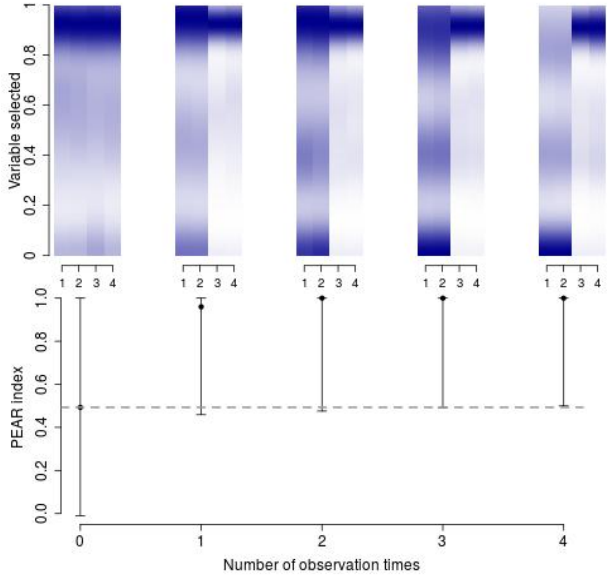} 
		\caption{The effect of number of observation times on PEAR indices when there are two clustering structures in the covariate data. Top: density plots showing selection for the four covariates. Bars show the weights given to the covariate in each of 1000 simulations, where 0 means unselected and 1 selected. Darker blue indicates higher density. Covariates 3 and 4 align with the outcome data. Bottom: points are median PEAR indices for 1000 simulations with error bars showing 0.05 and 0.95 quantiles. 
		}
		\label{var-select}
	\end{center}
\end{figure}

Inclusion of response variables enables recovery of the \textcolor{black}{target clustering structure (or generated clustering structure) }(Figure \ref{var-select}). The top row of Figure \ref{var-select} shows density plots of variable selection for each number of observation times. The bottom row represents  the corresponding 5$^{\text{th}}$, 50$^{\text{th}}$ and 95$^{\text{th}}$ percentiles of PEAR indices for the 1000 simulations of each scenario, when comparing the estimated clustering structure with the one that aligns with the outcome. When the response is excluded (number of observation times is zero), covariates 1--4 are not differentially selected, and correspondence between the inferred and generated clustering structures is low (median PEAR index around 0.5). When four time points (or MVN outcomes) are included, covariates one and two are deselected in favour of covariates three and four, resulting in identification of the clustering structure that aligns with the response.

Here we have shown that, for a dataset in which there is more than one clustering structure, use of covariate variable selection with outcome data in profile regression improves recovery of the \textcolor{black}{target clustering} structure. 

\subsection{Simulation study 3: Comparison of MVN and longitudinal specifications}\label{mvnvslong}

To draw out the differences between the two response models, we use each in turn to simulate data, and use both to infer clusters and profiles. For the sake of comparison, we simulate data that are longitudinal, with all individuals having observations at the same time points, so that both the MVN and GP model formulations may be applied to the data generated. 
We compare the two methods in terms of PEAR indices and the time taken by the algorithm, and then assess the effects of a) the data-generating model, b) the number of observation times, and c) the extent to which the generated clusters are \textcolor{black}{identifiable}.

\subsubsection{Design}

In this section, we generate two clusters of 30 individuals. Covariate data consist of two discrete variables. \textcolor{black}{In order to better highlight the difference between MVN and GP methods and avoid help from the covariates for cluster identification, this simulation study features no cluster structure in the covariate data.}

Outcome data for the two clusters are simulated from a mean vector of $10\cdot\textbf{1}_M$ and $\textcolor{black}{10(1-\xi \cdot j)}$, with $j=0, \cdots 10$, respectively, with $M$ the number of equally spaced time points and  $\xi$ a gradient (or separability parameter) ranging from -1 to 0. 
However, the covariance matrices are the same for both clusters. When generating MVN data, we use a matrix of dimension $M$ with diagonal values of 1 and lag-1 values of 0.5. When generating from a GP response model, we use the covariance function defined in Equation \ref{kernel} with parameters $\{\log(a), \log(l), \log(\sigma^2)\}=\{-0.5,-0.1,-0.5\}$.

\begin{figure}[ht!]
	\begin{center}
		\includegraphics[width=0.6\textwidth]{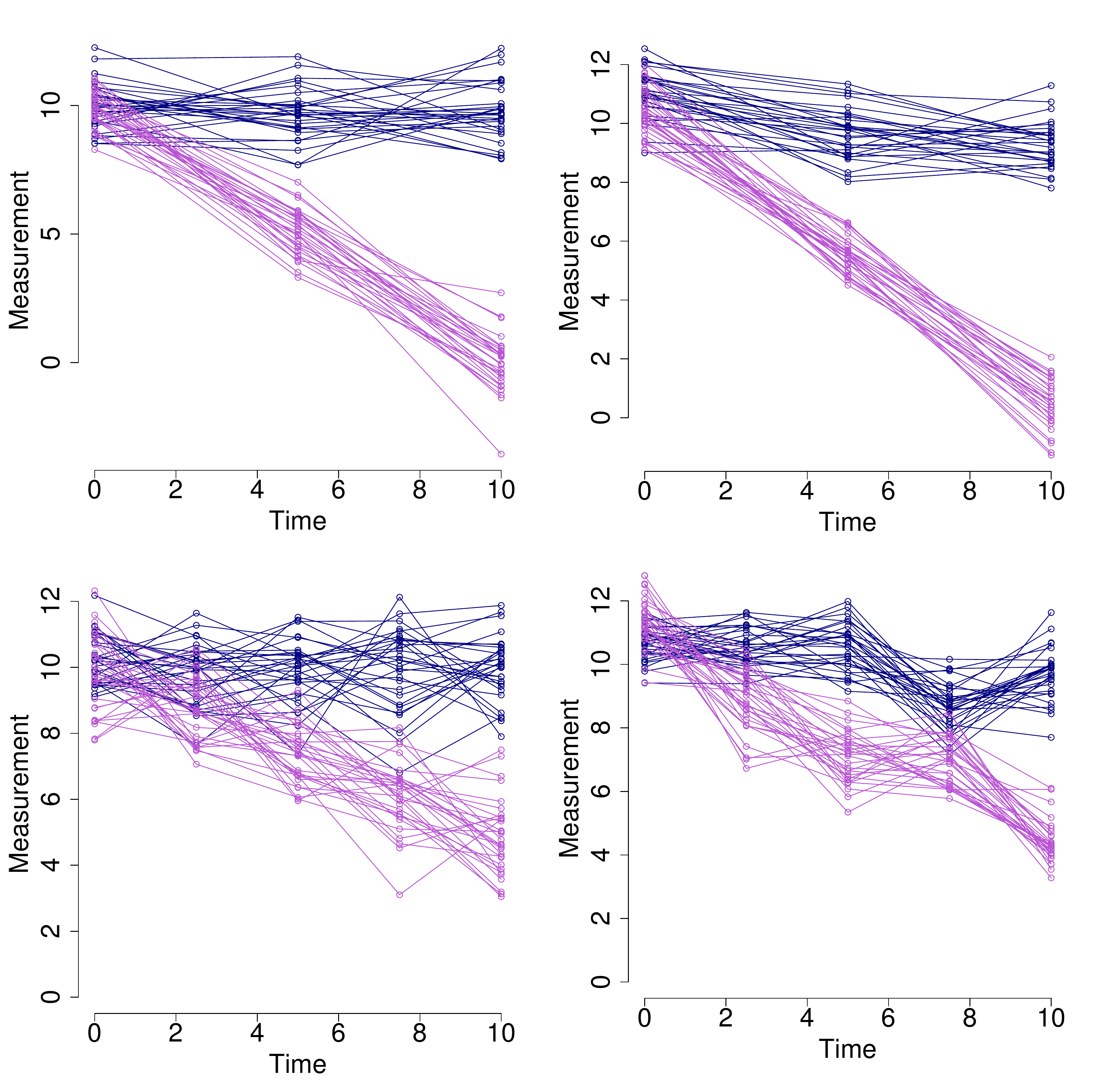}  
		\caption{Simulated data. Examples of data generated with the  MVN model (left) and the GP model (right). Top line: three data points across the time period 0--10. The mean gradient is -1 for the second (purple) cluster. Bottom line: five data points across the time period 0--10. The mean gradient is -0.5 for the second (purple) cluster }
		\label{simdata}
	\end{center}
\end{figure}
We consider several scenarios by varying the data-generating model (MVN or GP), the number of observation times $M$ (3 to 6),  the gradient $\xi$ (-1 to 0). Some examples of simulated data are given in Figure \ref{simdata}. Both response models are estimated on each scenario, with $5000(M-2)$ iterations each in the burn-in phase and in the sampling phase for the MVN response model and 5000 iterations for each phase for the GP one. Each scenario is repeated 1000 times.

\subsubsection{Results}

\begin{figure}[ht!] 
	\begin{center}
		\includegraphics[width=0.9\textwidth]{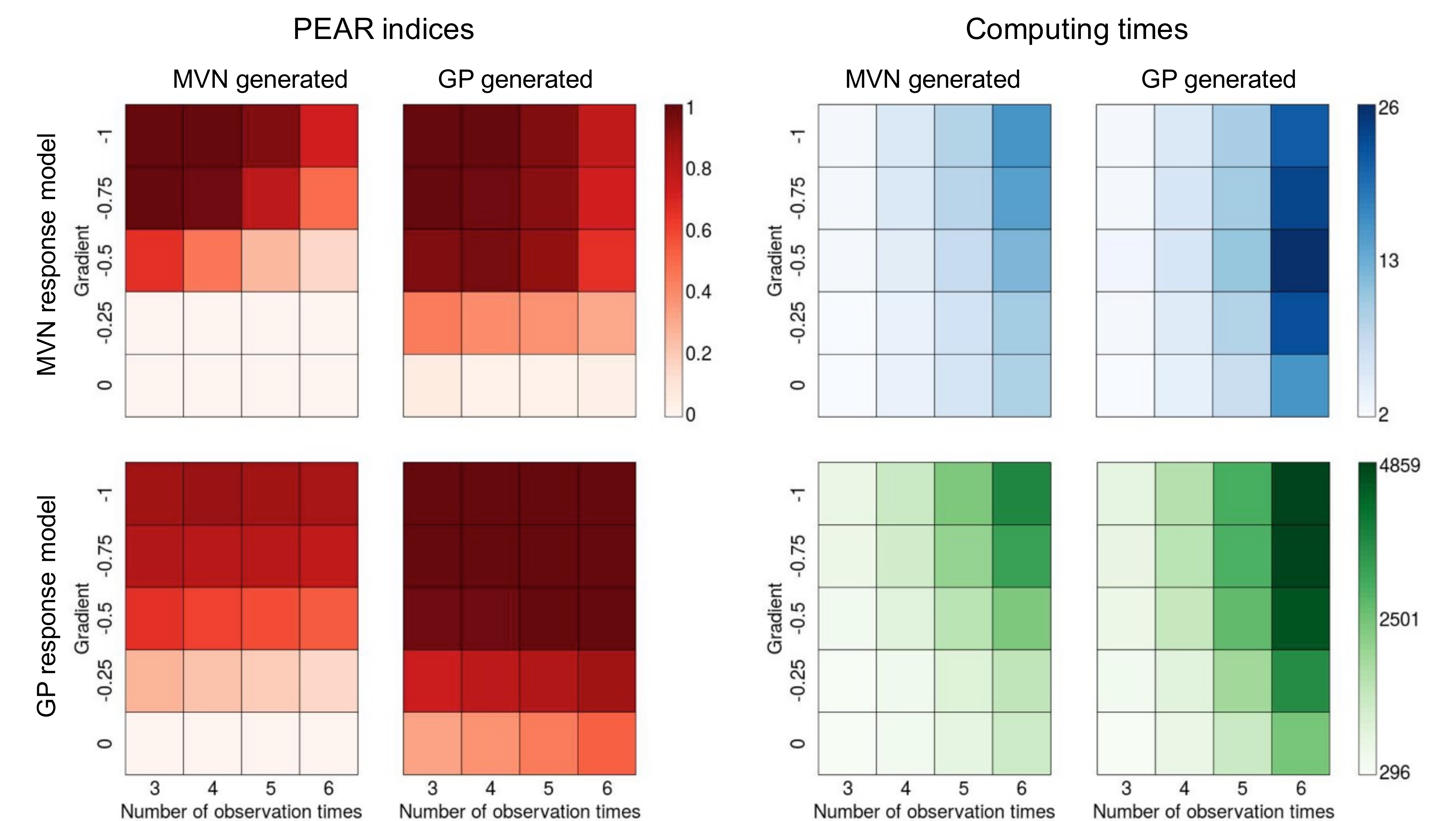}
		\caption{Simulation study results and timings. The first two columns of red heatmaps show PEAR indices following inference with the MVN response model (top row) and the GP response model (bottom row). Data were generated with a MVN model (first column) and a GP model (second column). The last two columns of heatmaps show the corresponding duration, in seconds, of inference. Each square within each heatmap represents a single experimental set-up: the gradient (which indicates how separable the clusters are) varies on the $y$ axis, and the number of observation times varies on the $x$ axis. The colour in the first two columns of heatmaps corresponds to the average PEAR index of 1000 repetitions, where darker red indicates a higher index. In the last two columns, the colour corresponds to the average duration of 1000 repetitions, where darker blue/green indicates a longer time. PReMiuMlongi was run with 5000 iterations each for burn in and sampling for the GP response model. For the MVN response model, the number of iterations for the burn in and sampling were each $5000(M-2)$, where $M$ is the number of observation times} 
		\label{simpear}
	\end{center}
\end{figure}

Figure \ref{simpear} summarises the effects of the three variables of interest on the two response models. 
A small gradient ($\xi$<-0.5) between the two generated clusters seems to have a strong negative impact on the MVN response model performances. \textcolor{black}{Also, we observe that the target clustering structure is better recovered by the MVN response model when there are fewer observation times, for a budgeted computation of 10000(M-2) iterations. Since the model dimension increases with more observation times, more iterations of the sampler are required  to reach convergence, as illustrated in Supplementary Figure 11.}

For the GP response model, the generated clustering structure is well recovered for the GP-generated data, 
as these data have an inherent structure that the response model can uncover. This response model has only three parameters to infer per cluster regardless of the number of observation times, in comparison with $\frac{1}{2}M(M-1)$ for the MVN response model, so fewer iterations are required for the GP response model. Both models perform better when the clusters are more separable, as we would expect.

Figure \ref{simpear} summarises the time taken for PReMiuMlongi to run. The MVN response model is much quicker than the GP model, by two orders of magnitude, despite a larger number of iterations in the burn-in phase and the sampler for datasets with more than three observation times.

\subsection{Simulation study 4: Handling irregularly spaced time points}
\subsubsection{Design}
\begin{figure}[ht!]
	\centering
	\includegraphics[width=0.5\linewidth]{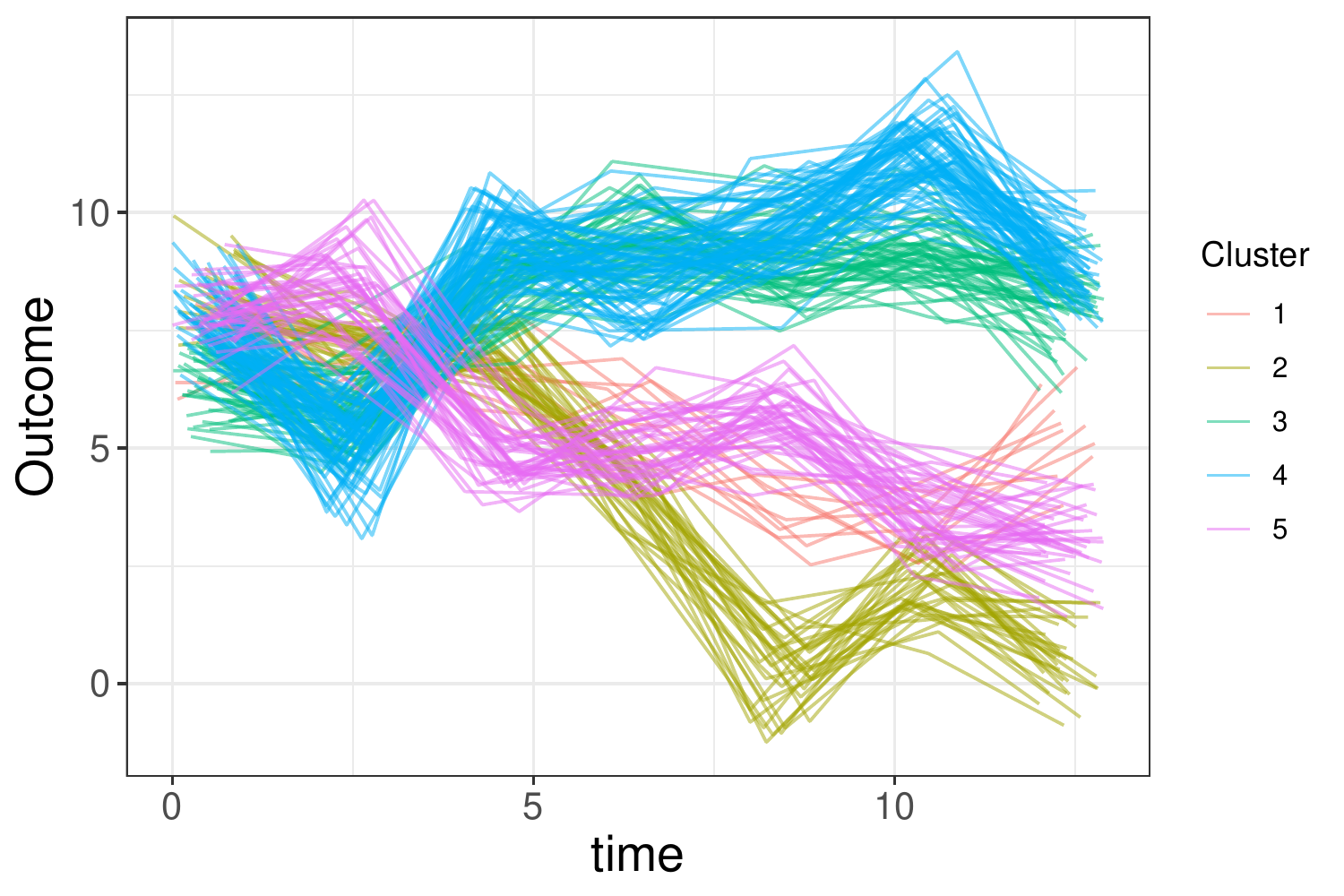}
	\caption{Generated longitudinal trajectories for 200 individuals, coloured by clusters.}
	\label{spaghettiplot}
\end{figure}

We generated one dataset of 200 individuals, allocated to 5 clusters of sizes 10, 30, 50, 70 and 40 individuals, respectively. Conditionally on the cluster allocations, a Gaussian outcome was generated from a Gaussian Process $Y_{i,j} = \textcolor{black}{g_c}(t_{ij}) + \epsilon_{i,j}$ where $ \epsilon_{i,j} \sim \mathcal{N}(0,\sigma^2_c)$ and $\textcolor{black}{g_c} \sim GP(m_c, \mathcal{K}_c)$ with a null prior mean function $m_c \equiv 0$ and a squared exponential covariance function $ \mathcal{K}_c(s,t)=a_c ~ {\rm exp}\big(\frac{-(s-t)^2}{l_c}\big)$. Outcome observations were generated for each individual at 7 visits (${\bf v}=[0, 2, 4, 6, 8, 10, 12]$ years) with individual-specific time points ($t_{ij} \sim \mathcal{U}([v_j, v_j+0.9]), i=1, \cdots 200, j=1, \cdots 7$). Figure \ref{spaghettiplot} displays the individual outcome trajectories for the 200 individuals, coloured according to the cluster allocations.  A set of 5 categorical covariates, with 3 categories each, were generated for each individual given the cluster allocations, with the cluster-specific parameters shown in Table \ref{hyper}.

\smallskip

\begin{table}[ht!]
	\begin{tabular}{c| c c c}
		Cluster & 	${\bf m}_c$& $\boldsymbol{\theta}_c$ &$\boldsymbol{\phi}_c$ \\\hline
		1&[9,8.5,8,6,5,4,3]&[0.5,0.1,-0.7]& 	[0.8,0.1,0.1] [0.8,0.1,0.1] [0.8,0.1,0.1] [0.8,0.1,0.1] [0.8,0.1,0.1]\\
		2&[9,7,6,4,2,1,0]&[0.6,0.2,-0.3] & [0.1,0.8,0.1] [0.1,0.8,0.1] [0.1,0.8,0.1] [0.1,0.8,0.1] [0.1,0.1,0.8]\\
		3&	[7,5,8,9,10,11,9]& [0.1,0.3,-0.7] &[0.4,0.5,0.1] [0.5,0.4,0.1] [0.1,0.4,0.5] [0.5,0.1,0.4] [0.5,0.3,0.2] \\
		4&[8,5,8,9,10,11,9] &[0.3,0.4,-0.5] &[0.2,0.7,0.1] [0.7,0.2,0.1] [0.3,0.6,0.1] [0.2,0.1,0.7] [0.6,0.3,0.1] \\
		5&[7,7.5,6,5,5,3,2] & [0.1,0.5,-0.7]&[0.6,0.3,0.1] [0.2,0.1,0.7] [0.3,0.6,0.1] [0.7,0.2,0.1] [0.2,0.7,0.1] \\
	\end{tabular}
	\caption{Hyperparameters for data generation (with $\boldsymbol{\theta}_c = \log((a_c, l_c, \sigma^2_c)^{\top})$)} 
	\label{hyper}
\end{table}

The model was then estimated using cluster-specific Gaussian Processes with standard normal hyper priors for the variance logparameters: $\log((a_c, l_c, \sigma^2_c)^{\top})\sim \mathcal{N}({\bf 0},\mathcal{I}_3)$. 
We adopted a Gamma prior for the concentration parameter of the DP, $\alpha \sim  ~ Gamma(2,1)$, and Dirichlet priors for the cluster-specific parameters of the covariate models $\phi_{c,q,e} \sim  ~ \text{Dirichlet}(1)$ for cluster $c$, covariate $q$ and covariate level $e$. The initial number of clusters was set to 20.

\subsubsection{Results}

We ran 7 chains on the generated dataset, with 10000 iterations and compared the traces of $\alpha$, number of clusters (nClus), number of non empty clusters (Figure \ref{diag}) and posterior similarity matrices in Supplementary Figure 12. 

\begin{figure}[ht!]
		\includegraphics[width=\linewidth]{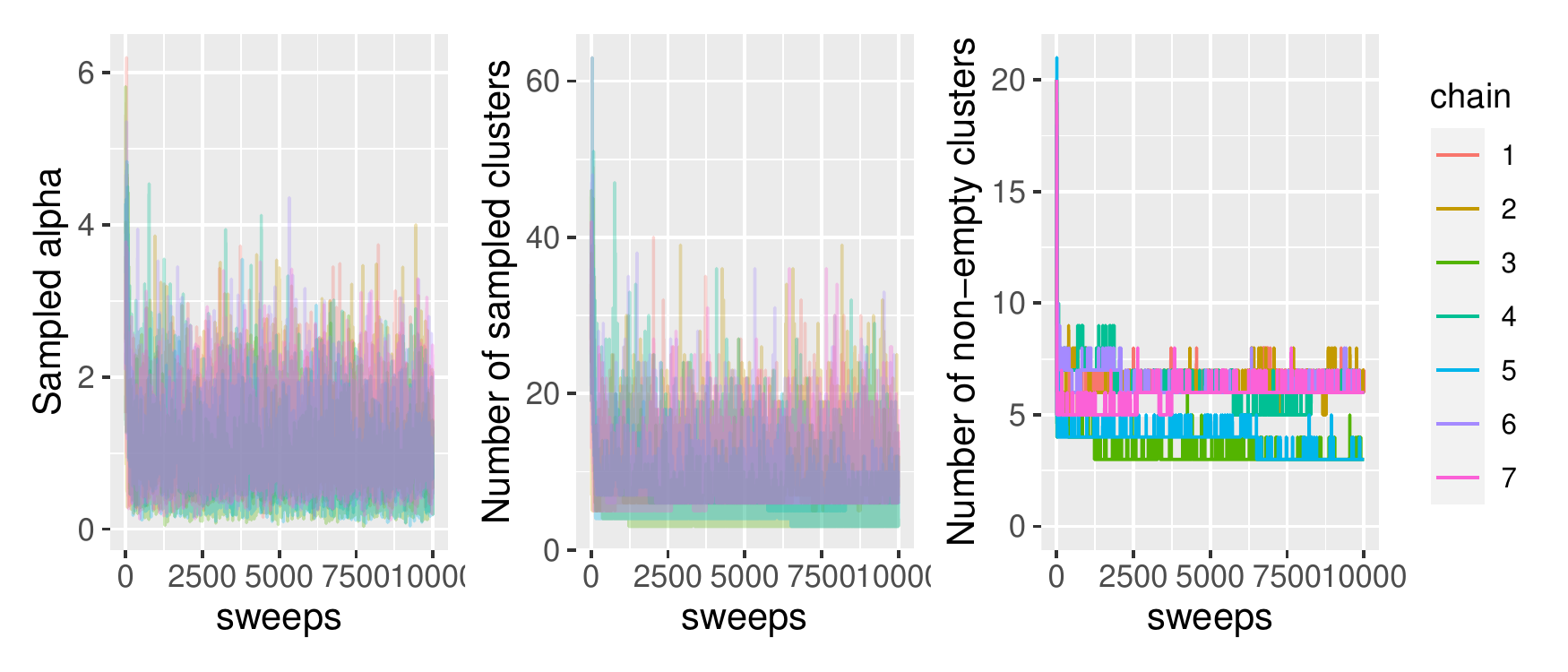}
	\caption{From left to right: Traces of the $\alpha$ DP parameter, of the number of clusters and number of non-empty clusters for the 7 chains.}
	\label{diag}
\end{figure}

The traces of $\alpha$ overlap and show convergence of the 7 chains. The traces of the number of clusters across chains also overlap but have different minimal values, bounded by the number of non-empty clusters, shown on the third plot. Even though the traces of non-empty clusters vary between 3 and 7 across the iterations, the similarity matrices (Supplementary Figure 12) show a good recovering of the generated clustering, represented in the first column of each plot. Posterior similarity matrices are highly contrasted because of the overall clear separability of the clusters. However, the third and fifth chains merge clusters 1 and 5, whose longitudinal trajectories are actually overlapping (Figure \ref{spaghettiplot}). This may imply that the chains are stuck in local minima, which is a common challenge in mixture models,  
and we could improve the algorithm by altering their proposal distribution. Split and merge moves could be added as well in the PReMiuMlongi sampler as in Jain and Neal\citep{Jain2007}  but this is beyond the scope of the current work. 

\textcolor{black}{We present a similar simulation study with 500 individuals in Section D in the Supplementary Materials, to mimic the size of the data application.}

\section{Application}\label{app}
We applied our methodology on budding-yeast data. Yeast is widely studied in genetic research as, like humans, they are eukaryotic organisms, with DNA information enclosed in cell nuclei. Additionally, they are unicellular, and therefore easier to study.  The genome of the yeast species \textit{Saccharomyces cerevisiae} has been entirely sequenced and is publicly available. During the cell cycle,  gene expression is regulated by transcription factors, proteins inducing or repressing  the transcription of DNA into mRNA information by binding to specific DNA sequences.  Similar expression patterns across genes suggest a co-regulation process, indicating that these genes may be regulated by a common set of transcription factors.  The protein-DNA interactions can be investigated using chromatin immunoprecipitation (ChIP) microarrays, which identifies the DNA binding sites.  The objective was to analyze jointly ChIP and gene expression data, to uncover groups of genes co-regulated during the  budding yeast cell cycle and identify the transcription factors involved in this process, in order to better understand the complex relationships between genes.

\subsection{Data}
\begin{figure}[ht!]
	\centering
	\includegraphics[width=0.6\linewidth]{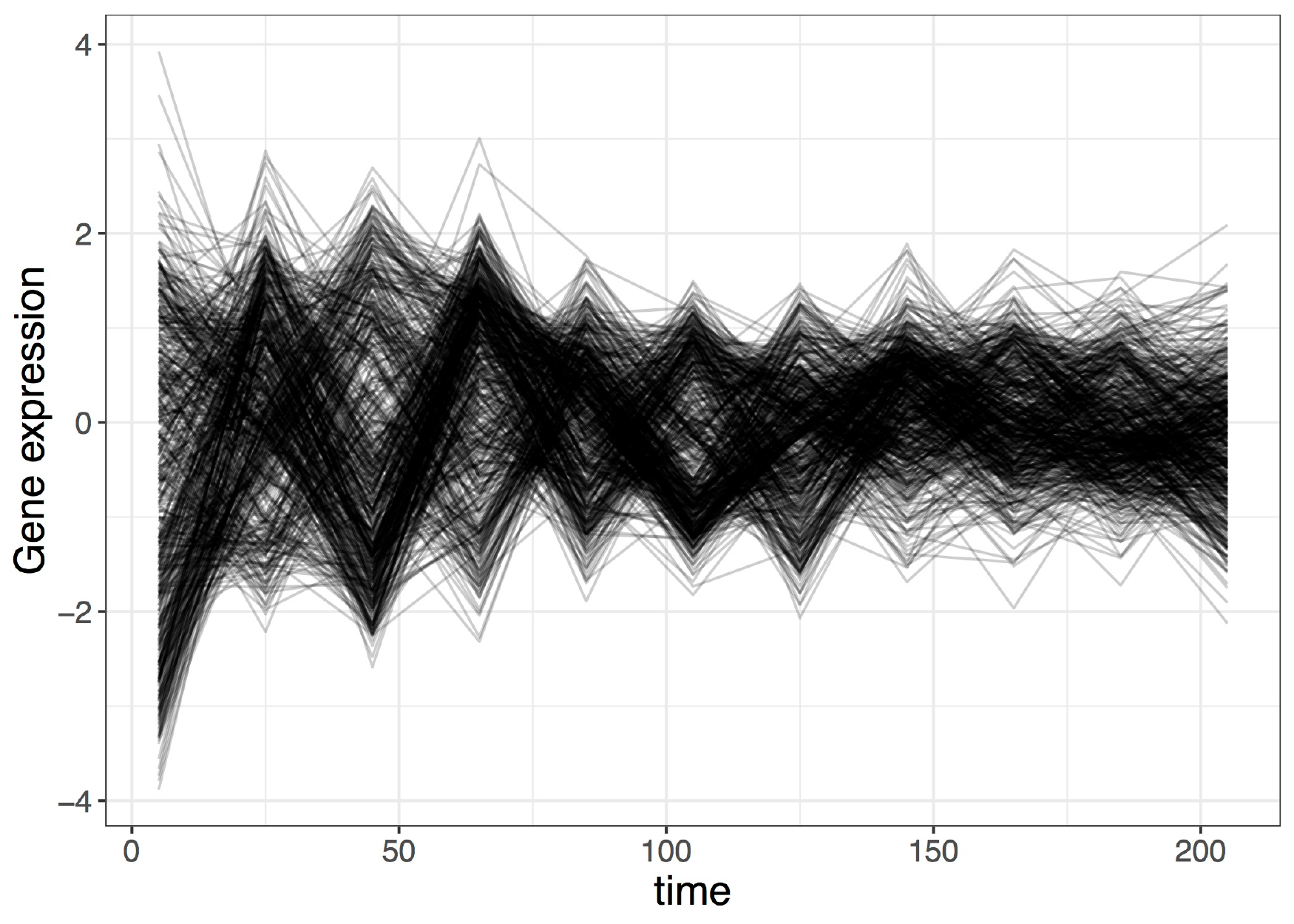}
	\caption{Standardised gene expression for 551 genes of budding yeast observed over time, in minutes}
	\label{spaghetti_plots}
\end{figure}

We analyzed gene expression time course data collected by \citet{Granovskaia2010} and ChIP data provided by \citet{Harbison2004}, also analyzed in \citet{Kirk2012} and  \citet{Zurauskiene2016}. Gene expression was quantified over time by collecting high-resolution, strand-specific tiling microarray profiles of RNA expression every 5 minutes over three Saccharomyces cerevisiae cell cycles, for a total of 41 measures. The gene measurements were then standardised at the gene level, so that the mean and variance of the expression measures of each gene were 0 and 1, respectively.
The binary ChIP data contained binding information for 117 transcription factors, \textcolor{black}{informing which transcription factors bound to which genes, during the cell cycle. } 
We selected 551 genes with periodic expression patterns identified in \citet{Granovskaia2010}  for which ChIP data were available and considered the transcription factors which bound to at least one of these genes, totalling 80. To alleviate computational burden, we reduced the set of repeated gene expression measurements to a set of 11 regularly spaced time points. The corresponding trajectories for the 551 genes are depicted in Figure \ref{spaghetti_plots}.

\subsection{Model}
We analysed the gene expression time course $Y$ and the transcription factors $X_q$, $q=1,..., 80$, specifying the following model:
\begin{align*}
	f(X,Y;\Theta)=&\sum_{c=1}^{\infty} \pi_c ~  f(Y|\theta_c)\prod_{q=1}^{80} f(X_q|\phi_c)\\
	\pi \sim &~GEM(\alpha) \\
	X_{i,q} \sim & ~Bin(\phi_{c,q})\\
	Y_{i}=& ~g_c(t_{i}) + \epsilon_i(t_{i}) \text{ with } \epsilon_i(t_{i})\sim \mathcal{N}(0,\sigma^2_c ~\mathcal{I}_{n_i})\\
\end{align*}
with $Y_i$, $t_i$ the vectors of individual repeated measures and time points, respectively, $X_{i,q}$ the binary variable indicating whether transcription factor $q$ binds to gene $i$,  $\alpha$ the concentration parameter of the Dirichlet Process prior influencing the number of clusters, $\Theta$ the vector of all parameters, $\phi_{c,q}$ the probability that transcription factor $q$ equals 1 in cluster $c$. 
Hyperpriors were defined as:
\begin{align*}
	\alpha \sim & ~ Gamma(2,1)\\
	\phi_{c,q} \sim & ~ Dirichlet(1) \\
	g_c\sim &~ \mathcal{GP}(m_c, \mathcal{K}_c) \text{ with }   \log(a_{c})\sim  ~\mathcal{N}(1.48,0.5)\text{ and }
	\log(l_{c}) \sim  ~\mathcal{N}(5,0.1)
\end{align*}
with $m_c$ defined as the null function and $\mathcal{K}_c$ a squared exponential function (defined in Equation (\ref{kernel})).
The lognormal hyperpriors adopted for the covariance functions $\mathcal{K}_c$ were chosen based on an empirical variogram of the data \citep{Diggle1990}. \textcolor{black}{We constrained the variance of the measurement error, defining it as proportional to the cluster variance: $ a_{c} = r*\sigma_{c}^2$ with $r=4$. }
Finally, a variable selection approach was used to identify the transcription factors likely involved in the co-regulation process of the genes in each group.

\subsection{Results}

\begin{figure}[ht!]
	\centering
	\includegraphics[width=0.7\linewidth]{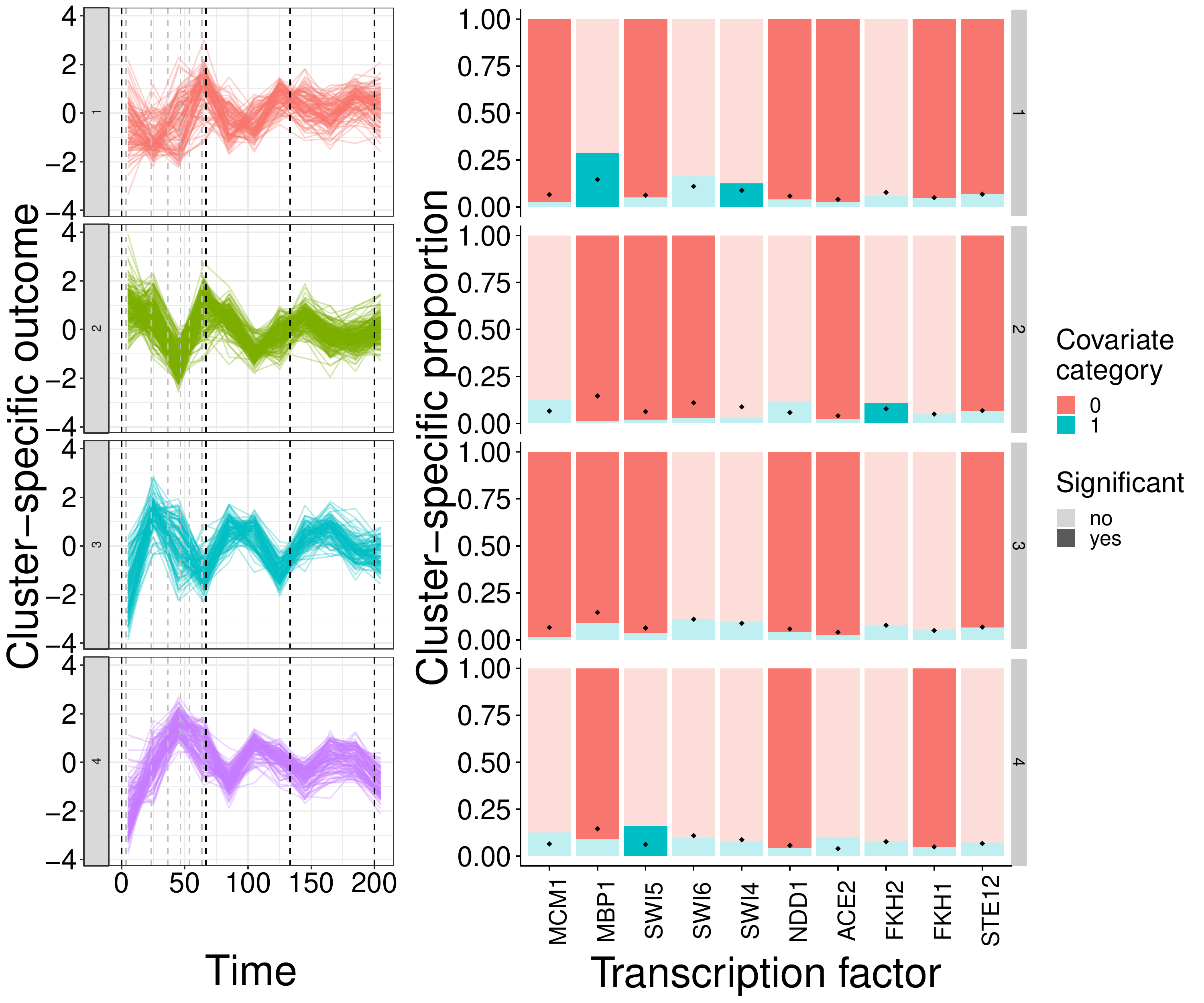}
	\caption{Trajectories of gene expression and covariate profiles for the 4 clusters of the final partition. On the left panels, trajectories of the genes in the color associated with the cluster they are allocated to. Dashed vertical black lines delimit the three cell cycles and dashed vertical gray ones delimit the different phases, in the first cycle: M/G1, G1, S, G2, G2/M, M and M/G1 again.  On the right panels: cluster-specific profiles with regard to the 10 selected transcription factors: MCM1, MBP1, SWI5, SWI6, SWI4, NDD1, ACE2, FKH2, FKH1 and STE12.  In each cell, associated with cluster $k=1,...,4$ (represented by rows) and transcription factor $q=1,...,10$ (represented by columns), \textcolor{black}{the black pip represents the empirical proportion of genes with $x_q=1$ in the whole sample. The turquoise filled square and the red one represent the estimated cluster-specific proportions $\hat{P}(x_{c,q}=1)$ and $\hat{P}(x_{c,q}=0)$, respectively. Squares are dark filled if the 95\% credibility interval of $\hat{P}(x_{c,q}=1)$  does not contain the empirical proportion, meaning that the considered category is significantly different in the cluster compared to the whole sample. Red squares are dark filled if the 5th percentile is above the empirical mean, and blue ones are dark filled if the 95th is below.}}
	\label{GP_results}
\end{figure} 

\begin{figure}[ht!]
	\centering
	\includegraphics[width=0.4\linewidth]{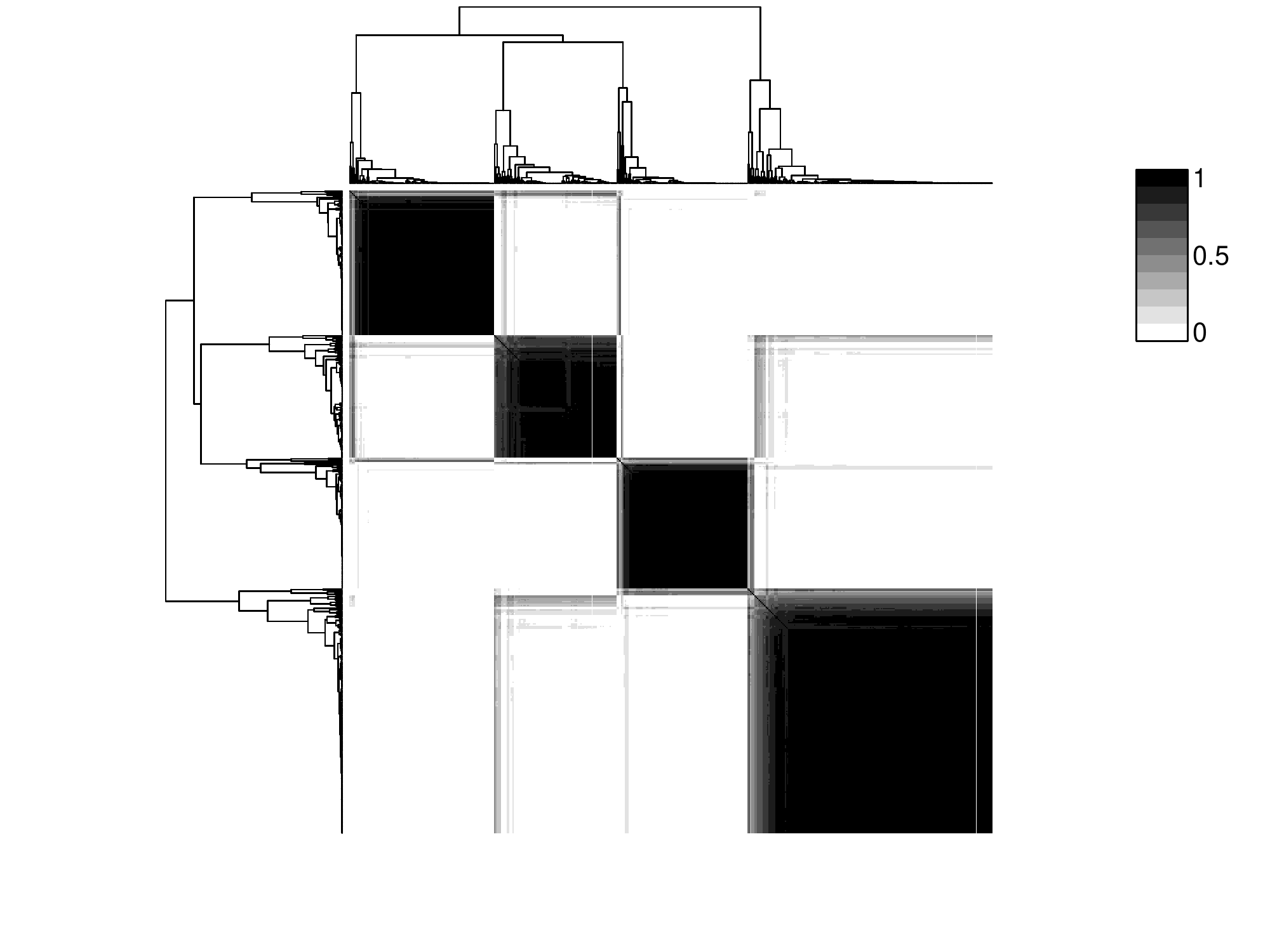}
	\caption{Posterior similarity matrix (PSM) representing the probability of pairwise co-clustering for the 551 genes. Each element $(i,j)$ of this $551\times 551$-matrix is proportional to the number of times genes $i$ and $j$ were allocated to the same cluster, over the 10000 MCMC iterations.}
	\label{PDM}
\end{figure}

The model was estimated by MCMC with 10000 iterations. 
Based on the posterior similarity matrix (Figure \ref{PDM}), 
we identified a final partition of 4 clusters of 109, 206, 113 and 123 genes respectively. Based on a 10\% lower threshold for the variable-specific relevance indicator $\rho_q$ (Equation \ref{rho}), the variable selection identified 10 transcription factors which had different distributions across the clusters and were thus likely involved in the co-regulation process: 
MCM1, MBP1, SWI5, SWI6, SWI4, NDD1, ACE2, FKH2, FKH1 and STE12. 
The gene expression trajectories in these 4 clusters are depicted in Figure \ref{GP_results}, exhibiting distinct patterns with different peak times, occurring in different phases of the cycle. The cell cycle is divided into 4 phases: Gap 1 (G1) during which the cell grows and prepares for duplication, synthesis (S) where DNA replication occurs, Gap 2 (G2) during which the cell keeps growing to prepare for division, and mitosis (M). The 4 clusters seem to be involved in the M/G1, G1, G1 and G2 phases, respectively.  The description of the same clusters regarding all the transcription factors is depicted in Supplementary Figure 17. The right panels of figures \ref{GP_results} and 13 were produced with the R package premiumPlots (\url{https://github.com/simisc/premiumPlots}).

\textcolor{black}{We assessed the relevance of the clustering solution using the Gene Ontology Term Overlap (GOTO) scores \citep{Mistry2008}. GOTO scores can be interpreted as quantifying the homogeneity of Gene Ontology annotations within clusters of genes: they represent the average number of shared GO annotations for a pair of genes from the same cluster. Having similar values for our clustering compared to previous results on this dataset (see Supplementary Table 2) thus validates biological function homogeneity for our identified clusters.}

\textcolor{black}{Finally, we analyzed the same dataset using the MVN specification in Section E.2 of the Supplementary Materials. The results show that the GP specification over the MVN specification allows the apparent underlying clustering structure of the gene expression data to be recovered, notably thanks to its flexibility and reduced number of parameters to estimate, compared to the MVN outcome model with an unstructured covariance matrix.}


\subsection{Convergence diagnosis and sensitivity analysis}

We ran two additional chains to assess the convergence of the model, and compared the trace of the concentration parameter $\alpha$ of the Dirichlet processes, as well as the number of clusters across the MCMC iterations, shown in Supplementary Figure 18. The cluster-specific parameters cannot be traced as the number of clusters varies from one iteration to another. Supplementary Figure 19 presents the two posterior similarity matrices which are very similar to the one in Figure \ref{PDM}, which also confirms the convergence of the model.

We estimated the same model with different values for the  ratio between the signal variance parameter and the measurement error variance $r= \frac{a_{c}}{\sigma_{c}^2}$ in a sensitivity analysis. All the models were specified in the same way, as described above. We compared in terms of PEAR indices the partitions obtained with different values of $r$ (Supplementary Table 3). 
We observe very stable results when $r$ varies between 2 and 6, showing little sensitivity to the setting of this parameter.

\section{Discussion}
We extended the semi-supervised profile regression approach to uncover the intrinsic clustering structure in a heterogeneous population from  a longitudinal outcome and a set of correlated covariates. Outcome patterns and covariate profiles, defined as combinations of covariate values, are linked through a non-parametric clustering structure. A Dirichlet Process prior is used on the mixing distribution to handle an unconstrained number of clusters, also allowing us to quantify the uncertainty in cluster allocations. We implemented two modelling approaches to describe the repeated outcome, based on either a multivariate Gaussian distribution to handle regular time points for all individuals or Gaussian Process regression, accommodating either irregular time points and/or a large number of time points. 
The cluster-specific parameters are estimated while incorporating the uncertainty in the cluster allocations, and individual marginal predictions can be computed, accounting for the population heterogeneity. 
The simulation study showed that the MVN modelling recovered better the clustering structure and was robust with a low number of time points, while the GP modelling had better clustering performances with more than 5 or 6 time points, even though it required a longer computation time regardless of the number of time points and separability of the clusters. Applying the model to  budding yeast data, we integrated gene expression time course measurements and transcription factor binding data to identify clusters of co-regulated genes and associated transcriptional modules.

\textcolor{black}{In the current version of the algorithm, we adopted an Inverse Wishart prior distribution for the variance covariance matrix in the MVN specification, which is advantageous for its conjugacy. However, with a high number of time points, a Hierarchical Inverse Wishart prior may provide more flexibility to the model \citep{alvarez2014}}.  We also implemented a squared exponential kernel for the GP prior covariance function. This is a common choice for such priors, but can lead to some instability when
	fitting longitudinal data with highly overlapping trajectories. Indeed, the
	variance hyperprior has an important impact on the final number of non-empty
	clusters, with high signal variances leading to a small number, and vice versa.
	In our application, we fixed the variance ratio to ensure that the signal
	variance is larger (or at least equal) to the measurement variance for each
	cluster, in conjunction with weakly informative hyperpriors for the signal
	variance. Additionally, we performed a sensitivity analysis showing the
	stability of our results with different values for this variance ratio.

\citet{Duvenaud2013} considered other base kernels such as linear, quadratic, rational quadratic or periodic ones, and proposed composite kernels defined as combinations of the base ones. Such extension in profile regression could potentially improve the recovering of known structures in the dataset. Besides, conditionally on the cluster allocations, the variability of the outcome is entirely modelled by this covariance function. Cluster-specific fixed effects could refine the cluster identification and help disentangle the effect of a covariate across the different clusters. Finally, hierarchical Gaussian Process regression \citep{Hensman2013} would allow the handling of multiple longitudinal markers, all correlated through an underlying cluster-specific process. However, the estimation of such models would require a large amount of data to estimate all the levels of the hierarchical GP modelling.

The dataset analysed in this paper contained no missing values, but our model used with the GP specification could also be applied to incomplete data, such as epidemiological cohorts. However, it relies on the assumption that missing outcome data are missing at random. \textcolor{black}{The MVN specification does not handle missing outcome data, but multiple imputation methods \citep{Rubin1987,MICE} could be considered for a future extension}. For handling not-at-random missing data, we intend to extend the algorithm to model jointly a longitudinal outcome and a time-to-event, using a cluster-specific proportional hazards model. The missingness process would then be linked to the outcome through the clusters, which would capture entirely the correlation between the two quantities. Multiple endpoints, such as time to disease onset and disease progression, could also be accommodated by combining a multi-state model, associating specific risks of disease with the different clusters, as in \citet{Rouanet2016}.  

Finally, profile regression assumes a common clustering structure to all data types integrated in the analysis. More flexible approaches consider different structures for the longitudinal outcome and the covariates, with possible common clusters \citep{Savage2010, Kirk:2012hj, Kirk2018}. This enables the identification of the meaningful clusters with regard to all the biomarkers, while discarding the irrelevant stratifications in the population, possibly linked to the inherent structure of a given set of variables. This would also be an interesting area for development.

\section*{Acknowledgements}
This work was supported by the National Institute for Health Research [Cambridge Biomedical Research Centre at the Cambridge University Hospitals NHS Foundation Trust]. The views expressed are those of the authors and not necessarily those of the NHS, the NIHR or the Department of Health and Social Care.

\section*{Funding}
AR, SR and BDT: MRC-funded Dementias Platform UK (RG74590). \\ SR: MRC programme MC\_UU\_00002/10. BDT: MRC programme MC\_UU\_00002/2.\\ RJ: MRC programme MR/R015600/1.
\\SRW: MRC programme MC\_UU 00002/2 and NIHR BRC-1215-20014.
\\PDWK: MRC programme MC\_UU 00002/13.

%
%

\bibliographystyle{apalike}

\bibliography{wileyNJD-AMA}

\section*{Appendix: GOTO analysis for biological assessment of clustering}

We assessed the relevance of the clustering solution using the Gene Ontology Term Overlap (GOTO) scores \citep{Mistry2008}. \textcolor{black}{As explained in the Web Supplementary of Kirk et al. (2012), the GOTO score for a pair of genes is calculated as the number of annotations associated to both genes. These annotations are relative to three different ontologies: biological process (bp), molecular function (mf) and cellular component (cc). The GOTO score for a cluster c is the average of the GOTO scores for the Nc(Nc-1)/2 pairs of genes allocated to cluster c, with Nc the cluster size. Finally, the GOTO scores presented in Table 2 are the average over the 4 clusters identified in the application of the cluster-specific GOTO scores, weighted by the cluster sizes.
GOTO scores can be interpreted as quantifying the homogeneity of Gene Ontology annotations within clusters of genes. }

Table \ref{GOTO} compares the GOTO scores obtained with PReMiuMlongi and 
iCluster \citep{Shen2009}, 
a clustering method allowing integration of multiple datasets (gene expression time course data collected by \citet{Granovskaia2010}  and ChIP data provided by \citet{Harbison2004})  using  a joint latent variable model. \citet{Shen2009} recommend choosing the number of clusters that minimizes the proportion of deviation score, indicating stronger cluster separability. On this dataset, the proportion of deviation score was minimized with 2 clusters (as shown in Supplementary Materials in \citet{Kirk:2012hj}). The comparison shows very similar results, the advantage of PReMiuMlongi being that it automatically infers the number of clusters instead of using heuristic approaches \textit{a posteriori}. In this case, we found 4 clusters as best describing the population heterogeneity.

\begin{table}[h!]
	\begin{center}
		\begin{tabular}{c c c c c}
			Method & GOTO	&GOTO	&GOTO& Number\\
			&(bp) & (mf) & (cc) & of genes\\\hline
			PReMiuMlongi&5.77&	0.88	&8.14& 551\\
			iCluster (k=2)&5.90& 0.89 &8.18& 551\\
		\end{tabular}
	\end{center}
	\caption{GOTO scores associated with the biological process (bp), molecular function (mf), and cellular component (cc) ontologies, for PReMiuMlongi and iCluster methods.}
	\label{GOTO}
\end{table}

\textcolor{black}{Finally, Table \ref{iclustercomp} presents the comparison of the clustering structures obtained by iCluster and PReMiuMlongi with the GP specification. The correspondence between the clusters is not clear as clusters 1-4 of the latter method are scattered in both clusters 1 and 2 of the former method. The individual evolutions by clusters obtained by iCluster (Supplementary Figure 20) are much more heterogeneous, which demonstrates the improved fit of the PReMiuMlongi clustering with the observed trajectories.}

\begin{table}[h!]
	\begin{center}
\begin{tabular}{ccc}
     &1 &  2\\\hline
1  &77  &32\\
2   &4 &202\\
3  &74 & 39\\
4 &123   &0\\
\end{tabular}
	\end{center}
\caption{Cross-table of the clustering structures obtained by PReMiuMlongi with the GP specification (rows) and iCluster (columns).}
\label{iclustercomp}
\end{table}

\section*{Supplementary Materials for Bayesian profile regression for clustering analysis involving a longitudinal response and explanatory variables}
 \appendix

 \section{Conditionals for Gibbs sampling}\label{conditionals}
 
 The full joint posterior distribution can be written 
 \begin{equation}
 p(\boldsymbol{\phi},\textbf{z},\alpha,\boldsymbol{\theta},\boldsymbol{\beta}|\textbf{x},\textbf{y},t,w)  \propto  p(\alpha)\cdot p(\textbf{z}|\alpha)\cdot p(\boldsymbol{\phi})\cdot  p(\textbf{x}|\textbf{z},\boldsymbol{\phi})\cdot p(\boldsymbol{\beta})\cdot p(\boldsymbol{\theta})\cdot  p(\textbf{y}|\textbf{z},\boldsymbol{\theta},\boldsymbol{\beta},t,w)\label{postdistr}
 \end{equation}
 
\textcolor{black}{Of note, the prior distributions parameters are omitted in the remaining of the text, for the sake of brevity.} Recall that for the MVN response, $\boldsymbol{\theta}=\{\boldsymbol{\mu}_{c},\boldsymbol{\Sigma_c}\}_{c\in 1\cdots C}$; for the GP, $\boldsymbol{\theta}=\{\text{log}(a_c), \text{log}(l_c), \text{log}(\sigma_c^2)\}_{c\in 1\cdots C}$ for a marginalised function $\textbf{\textsl{g}}$, and $\boldsymbol{\theta}=\{\textbf{\textsl{g}}_c,\text{log}(a_c), \text{log}(l_c), \text{log}(\sigma_c^2)\}_{c\in 1\cdots C}$ for the case that $\textbf{\textsl{g}}$ is sampled, with $C$ the number of clusters.

 \noindent The joint posterior distribution is sampled through sequential sampling of the following conditional distributions: 
 \begin{align}
 	p(&\boldsymbol{\phi}|\textbf{z},\textbf{x}) \propto p(\phi)\cdot p(\textbf{x}|\textbf{z}, \boldsymbol{\phi})\nonumber\\
 	p(&\textbf{z}|\boldsymbol{\phi},\alpha,\boldsymbol{\theta},\boldsymbol{\beta},\textbf{x},\textbf{y},\textbf{t},\textbf{w}) \propto  p(\textbf{z}|\alpha)\cdot p(\textbf{x}|\textbf{z},\boldsymbol{\phi})  \cdot p(\textbf{y}|\textbf{z},\boldsymbol{\theta},\boldsymbol{\beta},\textbf{t},\textbf{w})\nonumber\\
 	p(&\alpha|\textbf{z})  \propto p(\alpha)\cdot p(\textbf{z}|\alpha)\nonumber\\
 	p(&\boldsymbol{\theta}|\textbf{z},\boldsymbol{\beta},\textbf{y},\textbf{t},\textbf{w}) \propto p(\boldsymbol{\theta})\cdot p(\textbf{y}|\textbf{z},\boldsymbol{\theta},\boldsymbol{\beta},\textbf{t},\textbf{w})\label{theta_conditional}\\
 	p(&\boldsymbol{\beta}|\textbf{z},\boldsymbol{\theta},\textbf{y},\textbf{t},\textbf{w})  \propto p(\boldsymbol{\beta})\cdot p(\textbf{y}|\textbf{z},\boldsymbol{\theta},\boldsymbol{\beta},\textbf{t},\textbf{w})\nonumber
 \end{align}
 
 \subsection{Conditionals for the multivariate normal response model}
 
 For the MVN response model, we define $\boldsymbol{\theta}=\{\boldsymbol{\mu},\boldsymbol{\Sigma}\}$ with $\boldsymbol{\mu}=\{\boldsymbol{\mu}_{c}\}_{c\in 1\cdots C}$ and $\boldsymbol{\Sigma}=\{\boldsymbol{\Sigma_c}\}_{c\in 1\cdots C}$. These have priors $\boldsymbol{\Sigma_c}\sim\mathcal{W}^{-1}(\boldsymbol{R_0},\nu_0)$, $\boldsymbol{\mu_c}|\boldsymbol{\Sigma_c}\sim\mathcal{N}(\boldsymbol{\mu_0},\boldsymbol{\Sigma_c}/\kappa_0)$. Equation \ref{theta_conditional} is written 
 \begin{align}
 	p(&\boldsymbol{\mu}|\phi,\alpha,\textbf{z},\boldsymbol{\Sigma},\textbf{x},\textbf{y}) \propto  p(\boldsymbol{\mu}|\boldsymbol{\Sigma})\cdot p(\textbf{y}|\textbf{z},\boldsymbol{\mu},\boldsymbol{\Sigma}, \boldsymbol{\beta},\textbf{t},\textbf{w})\nonumber\\
 	p(&\boldsymbol{\Sigma}|\boldsymbol{\phi},\alpha,\textbf{z},\boldsymbol{\mu},\textbf{x},\textbf{y}) \propto  p(\boldsymbol{\Sigma})\cdot p(\boldsymbol{\mu}|\boldsymbol{\Sigma}) \cdot \textcolor{black}{p(\textbf{y}|\textbf{z},\boldsymbol{\mu},\boldsymbol{\Sigma}, \boldsymbol{\beta},\textbf{t},\textbf{w})}\nonumber
 \end{align}
 
 The posterior distributions from which we sample are \citep{Murphy:2007ww}
 \begin{equation}\label{iwish}
{\boldsymbol{\Sigma_c}} | \textbf{y} \sim \mathcal{W}^{-1}\left(\boldsymbol{R_c}, \nu_0+n_c\right), \quad\quad\boldsymbol{\mu_c} |\boldsymbol{\Sigma_c},\textbf{y} \sim \mathcal{N}_M\left( \frac{\kappa_0\boldsymbol{\mu_0}+n_c{\bar{\bf y}^{(c)}}}{\kappa_0+n_c} , \frac{\boldsymbol{\Sigma_c}}{\kappa_0+n_c} \right).
 \end{equation}
 
 Here, 
 \begin{equation}
 n_c=\sum_{i=1}^N\id_{[z_i=c]}\nonumber
 \end{equation}
 is the number of individuals in cluster $c$,
 \begin{equation}
 \bar{\bf y}^{(c)} = \frac{1}{n_c}\sum_{i=1}^N\left(\left(\bar{\bf y}_i-\boldsymbol{\beta}^T \textbf{w}_i \right)\cdot\id_{[z_i=c]}\right)\nonumber
 \end{equation}
 is the mean vector of outcomes for individuals in cluster $c$,
 \begin{equation}
 {\bf S_c} = \sum_{i=1}^N(\bar{\bf y}_i-\bar{\bf y}^{(c)})(\bar{\bf y}_i-\bar{\bf y}^{(c)})^T\id_{[z_i=c]}\nonumber
 \end{equation}
 is the scatter matrix for cluster $c$, and
 \begin{equation}
 \boldsymbol{R_c} = \boldsymbol{R_0}+ {\bf S_c}+\frac{\kappa_0n_c}{\kappa_0+n_c}(\bar{\bf y}^{(c)}-\boldsymbol{\mu_0})(\bar{\bf y}^{(c)}-\boldsymbol{\mu_0})^T\nonumber
 \end{equation}
 is the updated scale matrix for cluster $c$.
 
 The marginal likelihood is
 \begin{equation}
 p({\bf y}|z=c, \boldsymbol{\mu_c}) = \frac{1}{\pi^{n_cM/2}}\frac{\Gamma_M((\nu_0+n_c)/2)}{\Gamma_M(\nu_0/2)}\frac{|\boldsymbol{R_0}|^{\nu_0/2}}{|\boldsymbol{R_c}|^{(\nu_0+n_c)/2}}\left(\frac{\kappa_0}{\kappa_0+n_c}\right)^{M/2}.\nonumber
 \end{equation}

 \subsection{Conditionals for the Gaussian process response model}
  
  \subsubsection{Marginal algorithm}\label{margAlg}
  We infer the cluster-specific hyperparameters that define the mean and covariance functions. For their prior distributions, we use three independent log-normal distributions \citep{Kirk:2012hj}:
 \begin{align}
 \theta_c = [\log(a_c), \log(l_c), \log(\sigma_c^2)] &\sim \mathcal{N}_3(\boldsymbol{\mu_{{\theta}_c}},[\boldsymbol{s_{{a}_c}^2},\boldsymbol{s_{{l}_c}^2}, \boldsymbol{s_{\sigma^2_c}^2}]*I_3)\nonumber
 \end{align}

  For the GP response model, when marginalising the function $g^{(c)}$, the Metropolis-Hastings steps of each hyperparameter ${{\theta}}_{c,j}$ $({j=1,\cdots,3})$ 
   are done as follows: \\
  
 \begin{enumerate}
 	\item[] \textbf{for} j=1:3
 	 \begin{enumerate}
 	 	\item Evaluate $T_1=p\left(y^{(c)}\left|\boldsymbol{\theta}_{c,-j},\boldsymbol{\beta}\right.\right)\cdot p({\boldsymbol{\theta}}_{c,j})$
 	\item Sample ${\boldsymbol{\theta}}_{c,j}^*\sim \mathcal{N}({\boldsymbol{\theta}}_{c,j},s^2_{\boldsymbol{\theta}_{j}})$ 
 	\item Evaluate $T^*=p\left(y^{(c)}\left|\boldsymbol{\theta}_{c,-j},\boldsymbol{\beta}\right.\right)\cdot p({\boldsymbol{\theta}}_{c,j}^*)$
 	\item Calculate $A = \text{min}\left\{1,T^*/T_1\right\}$
 	\item Update ${{\theta}}_{c,j}  = \left\{\begin{array}{ll}{{\theta}}_{c,j}^* \quad& \text{with probability } A \\{{\theta}}_{c,j}  & \text{with probability } 1-A\end{array} \right.$
 \end{enumerate} 
\end{enumerate}
 
 This corresponds to the fourth line of Equation (\ref{theta_conditional}).  When marginalising $g^{(c)}$, the individual outcome variables are not independent from one another given the parameters $[\boldsymbol{\theta}, z, \boldsymbol{\beta}]$. Hence, the individual likelihood in cluster $c$ is computed by conditioning on all the data points in the cluster: 
\begin{equation*}\label{gplikelihood}
f_{\textbf{y}}(y_i|y^{(c)}, \boldsymbol{\theta}_c,z_i=c,\boldsymbol{\beta},w )= p(y_i|y^{(c)}, \boldsymbol{\theta}_c,\boldsymbol{\beta},w)=\frac{ p\left(y^{(c)}\left|\boldsymbol{\theta}_c,\boldsymbol{\beta},w\right.\right)}
{ p\left(y_{-i}^{(c)}\left|\boldsymbol{\theta}_c,\boldsymbol{\beta},w\right.\right)}
\end{equation*}
where
\begin{equation*}
 p\left(y^{(c)}\left|\boldsymbol{\theta}_c,\boldsymbol{\beta},w\right.\right)
= \frac{1}{\sqrt{(2\pi)^{\mathcal{M}}|{K^{(c)}}|}}\exp\left\{-\frac{1}{2}(y^{(c)}-\lambda^{(c)}){K^{(c)}}^{-1}(y^{(c)}-\lambda^{(c)})^{T} \right\}
\end{equation*}
 and
 \begin{equation*}
 \lambda^{(c)} = m_0^{(c)} + \boldsymbol{\beta}^{T}\textbf{w}^{(c)}.
 \end{equation*}
with the dimension of $K^{(c)}$ being equal to the number of time points for all individuals allocated to cluster $c$. The step sizes $s_{\boldsymbol{\theta}_{j}}$  
are updated in the same way as the fixed-effect coefficients in PReMiuMlongi. All step sizes are initially set to 1 and we aim for an acceptance rate of 0.44 \citep{Liverani:2015jg}.

  \subsubsection{Conditional algorithm}
 
 \noindent For the GP specification with sampling of $g$, the joint posterior distribution Eq(\ref{postdistr}) is changed as follows:
 \begin{equation*}
 p(\boldsymbol{\phi},\textbf{z},\alpha, {\bf a},{\bf l}, \boldsymbol{ \sigma^2},\textbf{\textsl{g}},\boldsymbol{\beta}|\textbf{x},\textbf{y},t,w) \propto  p(\alpha)\cdot p(\textbf{z}|\alpha)\cdot p(\boldsymbol{\phi})\cdot  p(\textbf{x}|\textbf{z},\boldsymbol{\phi})\cdot p(\boldsymbol{\beta})\cdot p({\bf a})\cdot p({\bf l})\cdot p(\boldsymbol{ \sigma^2})\cdot  p(\textbf{\textsl{g}}|{\bf a}, {\bf l})\cdot p(\textbf{y}|\textbf{z},\textbf{\textsl{g}},\boldsymbol{\beta}, \boldsymbol{ \sigma^2},t,w)\\
 \end{equation*}

 \noindent Also, $\textbf{z}, {\bf a}, {\bf l}, \boldsymbol{ \sigma^2}$ , $\boldsymbol{\beta}$ and $\textbf{\textsl{g}}$ are sampled following the conditionals:
 \begin{align}
 p(&\textbf{z}|\boldsymbol{\phi},\alpha,\textbf{\textsl{g}},\boldsymbol{\beta},\textbf{x},\textbf{y},t,{w}) \propto  p(\textbf{z}|\alpha)\cdot p(\textbf{x}|\textbf{z},\boldsymbol{\phi})  \cdot p(\textbf{y}|\textbf{z},\textbf{\textsl{g}},\boldsymbol{\beta},\boldsymbol{\sigma^2},t,w)\nonumber\\
 p(&{\bf a}|\textbf{z},\textbf{\textsl{g}}) \propto p({\bf a})\cdot p(\textbf{\textsl{g}}|{\bf a})\nonumber\\
 p(&{\bf l}|\textbf{z},\textbf{\textsl{g}}) \propto p({\bf l})\cdot p(\textbf{\textsl{g}}|{\bf l})\nonumber\\
 p(&\boldsymbol{\sigma^2}|\textbf{z},\textbf{\textsl{g}},\boldsymbol{\beta},\textbf{y},t,w) \propto p(\boldsymbol{\sigma^2})\cdot p(\textbf{y}|\textbf{z},\textbf{\textsl{g}},\boldsymbol{\beta},\boldsymbol{\sigma^2},t,w)\nonumber\\
 p(&\boldsymbol{\beta}|\textbf{z},\textbf{\textsl{g}},\textbf{y},t,w)  \propto p(\boldsymbol{\beta})\cdot p(\textbf{y}|\textbf{z},\textbf{\textsl{g}},\boldsymbol{\beta},\boldsymbol{\sigma^2},t,w)\nonumber\\
 \textbf{\textsl{g}}^{(c)}&|_{\textbf{z},\textbf{\textsl{g}},{\bf a}, {\bf l},t,w} \sim \mathcal{N}\left(m^{*(c)}, K^{*(c)}\right)\nonumber
 \end{align}
 \noindent with $m^{*(c)}$ and $K^{*(c)}$ defined as in Eq(15) and (16).

 The hyperparameters are resampled via Metropolis-within-Gibbs steps:
 
 \begin{enumerate}
 	\item[] \textbf{for} j=1,2
 	\begin{enumerate}
 		\item[a'.] Evaluate $T_1=p\left(g^{(c)}\left|\boldsymbol{\theta}_{c,-j}\right.\right)\cdot p({\boldsymbol{\theta}}_{c,j})$
 		\item[b'.] Sample ${\boldsymbol{\theta}}_{c,j}^*\sim \mathcal{N}({\boldsymbol{\theta}}_{c,j},s^2_{\boldsymbol{\theta}_{j}})$ 
 		\item[c'.] Evaluate $T^*=p\left(g^{(c)}\left|\boldsymbol{\theta}_{c,-j}\right.\right)\cdot p({\boldsymbol{\theta}}_{c,j}^*)$
 		\item[d'.] Calculate $A = \text{min}\left\{1,T^*/T_1\right\}$
 		\item[e'.] Update ${{\theta}}_{c,j}  = \left\{\begin{array}{ll}{{\theta}}_{c,j}^* \quad& \text{with probability } A \\{{\theta}}_{c,j}  & \text{with probability } 1-A\end{array} \right.$
 	\end{enumerate} 
 	\item[] \textbf{for} j=3
 	\begin{enumerate}
 		\item[] Steps a to e (in Section \ref{margAlg})
 	\end{enumerate}
 	\item[] sample $g^{(c)}$
 \end{enumerate}

The step sizes are updated in the same way as the fixed-effect coefficients in PReMiuMlongi. All step sizes are initially set to 1 and we aim for an acceptance rate of 0.44 \citep{Liverani:2015jg}.

 \section{Matrix inversion using Woodbury matrix identity}\label{Woodbury}
 In this section, we explicit an efficient inversion technique for updating the cluster-specific variance covariance matrices (Equation 9) needed in the likelihood computation (Equation 10)
 of the Gibbs sampling step for the allocation variable ${\bf z}$. In the following, $M_0= \mathcal{K}(\tau,\tau)$ denotes the current variance covariance matrix of cluster $c$, with $\tau$ the vector of all the observation time points of the individuals in cluster $c$, in ascending order, and $M_{new}$ the updated cluster-specific variance covariance matrix, either after adding individual $i$ (sub-section \ref{WM1}) or removing it (sub-section \ref{WM2}). The vector of the $n_i$ ordered observation time points of individual $i$ is denoted by $\tau_{i}$. In this section, we remove all the subscripts/indices on $c$ for notational convenience. \\
 
 The Woodbury matrix identity states that, given a square invertible $n\times n$ matrix $M$, an $n\times p$ matrix $U$ and an $p\times n$ matrix $V$, provided that $(I_p + VM^{-1}U)$ is invertible, we get:
 \begin{equation}
 (M+UV)^{-1} = M^{-1} -M^{-1} U (I+VM^{-1}U)^{-1}VM^{-1} \label{WM}    
 \end{equation}
 If $M^{-1}$ is known, this equation only requires  inverting a matrix of dimensions $p\times p$, with $p<n$.
 
 \subsection{Adding a new individual to a cluster}\label{WM1}
 
 We aim at inverting the updated variance-covariance matrix $M_{{new}}$, once individual $i$ has been allocated to cluster $c$ ($z_i=c$). We permute $M_{new}$ such that it can be written as:\\
 
 \begin{tabular}{l l l}
 	$ M_{new}= 
 	\begin{pmatrix}
 	M_0 & K^{\top}_{i,0}\\
 	K_{i,0} & K_{i}\\
 	\end{pmatrix}$&
 	with&
 	$\begin{matrix}
 	K_{i,0}=(\tau_{i}, \tau)  \\
 	K_{i}=(\tau_{i},\tau_{i})
 	\end{matrix}$
 \end{tabular}\\

 \noindent By applying Equation (\ref{WM}) and defining $B=M_0^{-1} K^{\top}_{i,0}$ and $A=K_{i} - K_{i,0}B$, we obtain:
 
 \begin{equation*}
 	M_{{new}}^{-1}= 
 	\begin{pmatrix}
 		M_0^{-1} + BA^{-1} K_{i,0}M_0^{-1}& -BA^{-1}\\
 		-A^{-1}B^{\top} & A^{-1}\medskip
 	\end{pmatrix}
 \end{equation*}
 Also,  $det(M_{{new}})= det(M_0) \times det(A)$.
 
 \subsection{Removing an individual from a cluster}\label{WM2}
 
 In this case, we aim at inverting $M_{new}$ after removing individual $i$ from $M_0$. We permute $M_0$ so that:
 \begin{equation*}
 	M_0=
 	\begin{pmatrix}
 		M_{new} & K^{\top}_{i,0} \\ 
 		K_{i,0} &K_i
 	\end{pmatrix}
 \end{equation*}
 We obtain $M_{new}^{-1}$ using Eq(\ref{WM}) twice, based on the following equations:
 \begin{align*}
 	\begin{pmatrix}
 		M_{new} & 0 \\ 
 		0 &K_i
 	\end{pmatrix}^{-1}
 	&= \Big[
 	\begin{pmatrix}
 		M_{new} & K^{\top}_{i,0} \\ 
 		0 &K_i
 	\end{pmatrix}
 	+
 	\begin{pmatrix}
 		K^{\top}_{i,0} \\ 
 		0 
 	\end{pmatrix}
 	\begin{pmatrix}
 		0_{n_i} &
 		-I_{n_i} 
 	\end{pmatrix}
 	\Big]^{-1}\\
 	\begin{pmatrix}
 		M_{new} & K^{\top}_{i,0} \\ 
 		0 &K_{i}
 	\end{pmatrix}^{-1}
 	&= \Big[
 	M_{0} 
 	+
 	\begin{pmatrix}
 		0_{n_i,0}  \\ 
 		I_{n_i} 
 	\end{pmatrix}
 	\begin{pmatrix}
 		-K_{i,0} &
 		0_{n_i} 
 	\end{pmatrix}
 	\Big]^{-1}
 \end{align*}
 
 \noindent Also, we obtain: $det(M_{{new}})= \frac{det(M_0)}{det(A)}$.

 \subsection{Sparse Gaussian Process}\label{WM3}
 When the number of time points is large or when the time points are too irregular across the individuals, we propose to approximate the inverses and determinants of GP covariance matrices using a time grid \citep{snelson2006} in order to reduce computation time. We denote by ${\bf u}$ a vector of $n_u$ regularly spaced time points. The variance covariance matrix of the vector $\tau$ of time points in ascending order can then be approximated by:\\
 
 \begin{minipage}[c]{0.2\linewidth}
 	\begin{align*}
 		K_{\tau,\tau} + \sigma^2 I &\approx Q_{\tau,\tau} + diag(K_{\tau,\tau}-Q_{\tau,\tau})+ \sigma^2 I \\
 		&\approx Q_{\tau,\tau} + \Lambda
 	\end{align*}
 \end{minipage}
 \hspace{0.5cm} with
 \begin{minipage}[c]{0.1\linewidth}
 	\begin{align*}
 		Q_{\tau,\tau}&=K_{\tau,u}K_{u,u}^{-1}K_{u,\tau}  \\
 		\Lambda&=diag(K_{\tau,\tau}-Q_{\tau,\tau})+ \sigma^2 I
 	\end{align*}
 \end{minipage}\\\bigskip
 
 \noindent Based on Equation (\ref{WM}) and Sylvester's theorem \citep{Akritas1996} and defining $A_{u\tau}= (\text{chol}(K_{u,u})^{\top})^{-1}K_{u,\tau}$ so that $A^{\top}_{u\tau}A_{u\tau}=Q_{\tau,\tau}$, we obtain:
 
 \begin{align*}
 	(Q_{\tau,\tau}+\Lambda)^{-1}&= \Lambda^{-1} - \Lambda^{-1}A^{\top}_{u\tau} (I_{n_u}+A_{u\tau}\Lambda^{-1}A^{\top}_{u,\tau})^{-1}A_{u\tau}\Lambda^{-1}\\
 	det(Q_{\tau,\tau}+\Lambda)&= det(I_{n_u}+A^{\top}_{u\tau} \Lambda^{-1}A_{u\tau})det(\Lambda)
 \end{align*}

\textcolor{black}{An example of this approximation can be found in Reid and Wernisch(2016)\citep{reid2016}.}
 
  \section{Supplementary figures}

 \begin{figure}[ht] 
	\begin{center}

		\includegraphics[width=0.5\textwidth]{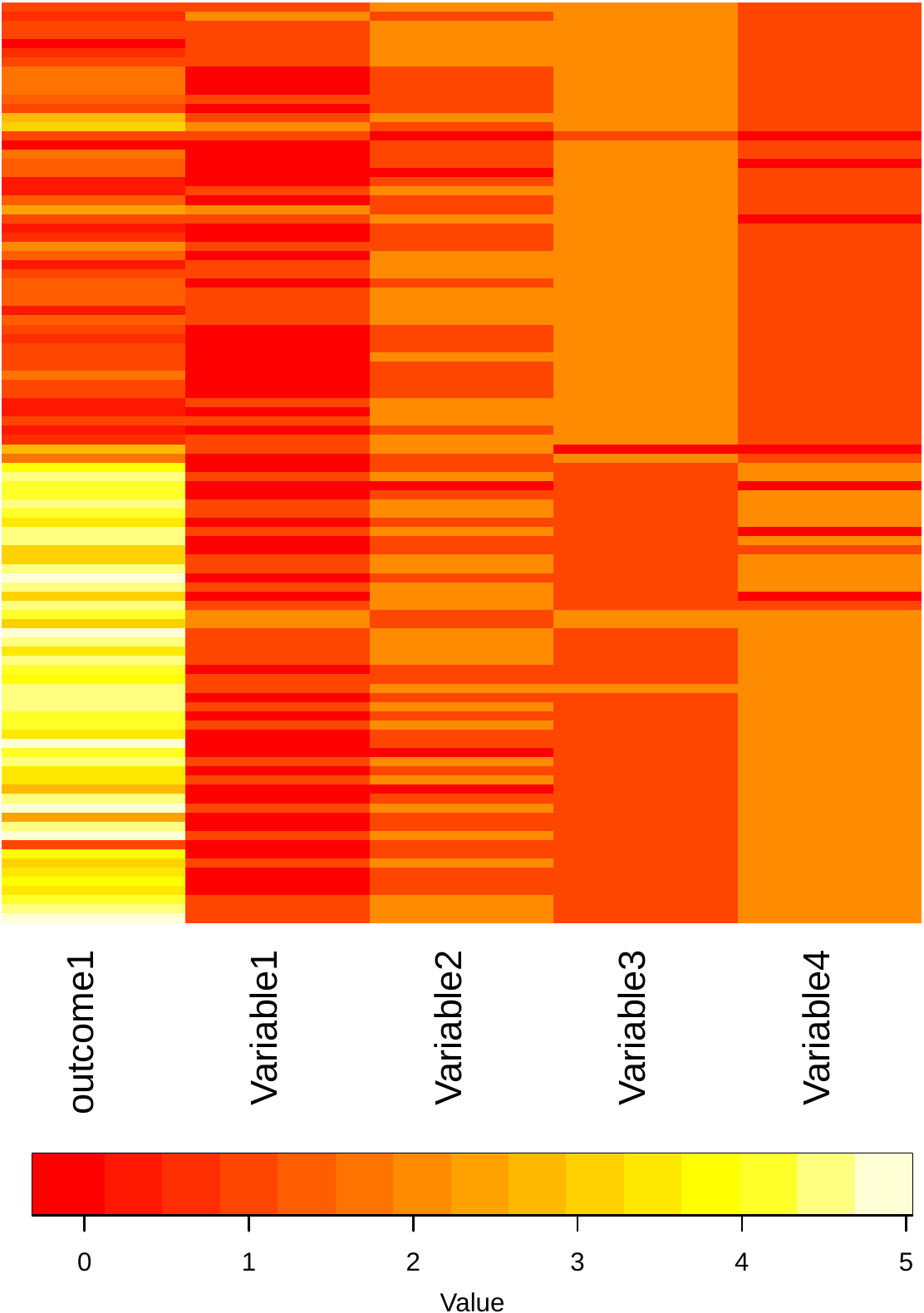}
		\caption{Simulated outcome and covariate data for the second simulation study (Section 4.3). 
			Each row represents five measurements of one person. Of the four variables, the first two do not align with the outcome (columns 2 \& 3). The last two variables align with the outcome (columns 4 \& 5).}
		\label{varselectdata}
	\end{center}
\end{figure}
 
 \begin{figure}[ht] 
 	\begin{center}
 		\includegraphics[width=0.5\textwidth]{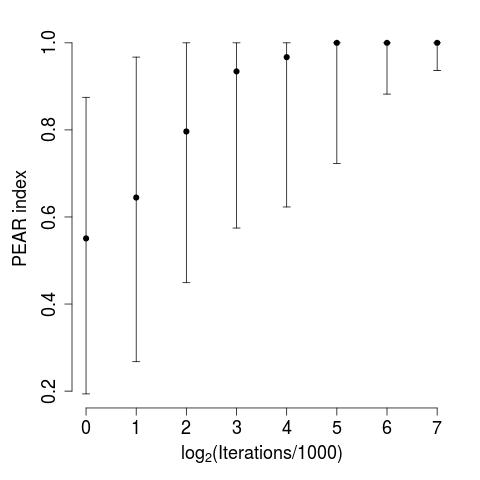}
 		\caption{Efficacy of MVN model in recovering a true clustering structure as the number of iterations in PReMiuMlongi changes. This plot corresponds to a single square in Figure 4:  
 			the top-right square of the top-left sub-figure. Each point represents the median of 1000 instances of inference of the clustering structure given data generated as described in Section 4.3, 
 		and error bars show 0.05 and 0.95 quantiles. The data-generating model is MVN, the response model is MVN, the number of time points is six, and the gradient of the second cluster's response mean is -1 (i.e. the responses of the two clusters are most clearly separable). Inference is trialled with 1000, 2000, 4000, 8000, 16000, 32000, 64000 and 128000 iterations in each of the burn-in and sampling phases. As the number of iterations increases, the true underlying clustering structure is better recovered, as evidenced by the increase in PEAR index.}
 		\label{convergence}
 	\end{center}
 \end{figure}
 
 \begin{figure}
 	\begin{minipage}{\linewidth}
 		\vspace{-0.5cm}
 		\centering
 		\includegraphics[width=0.45\linewidth]{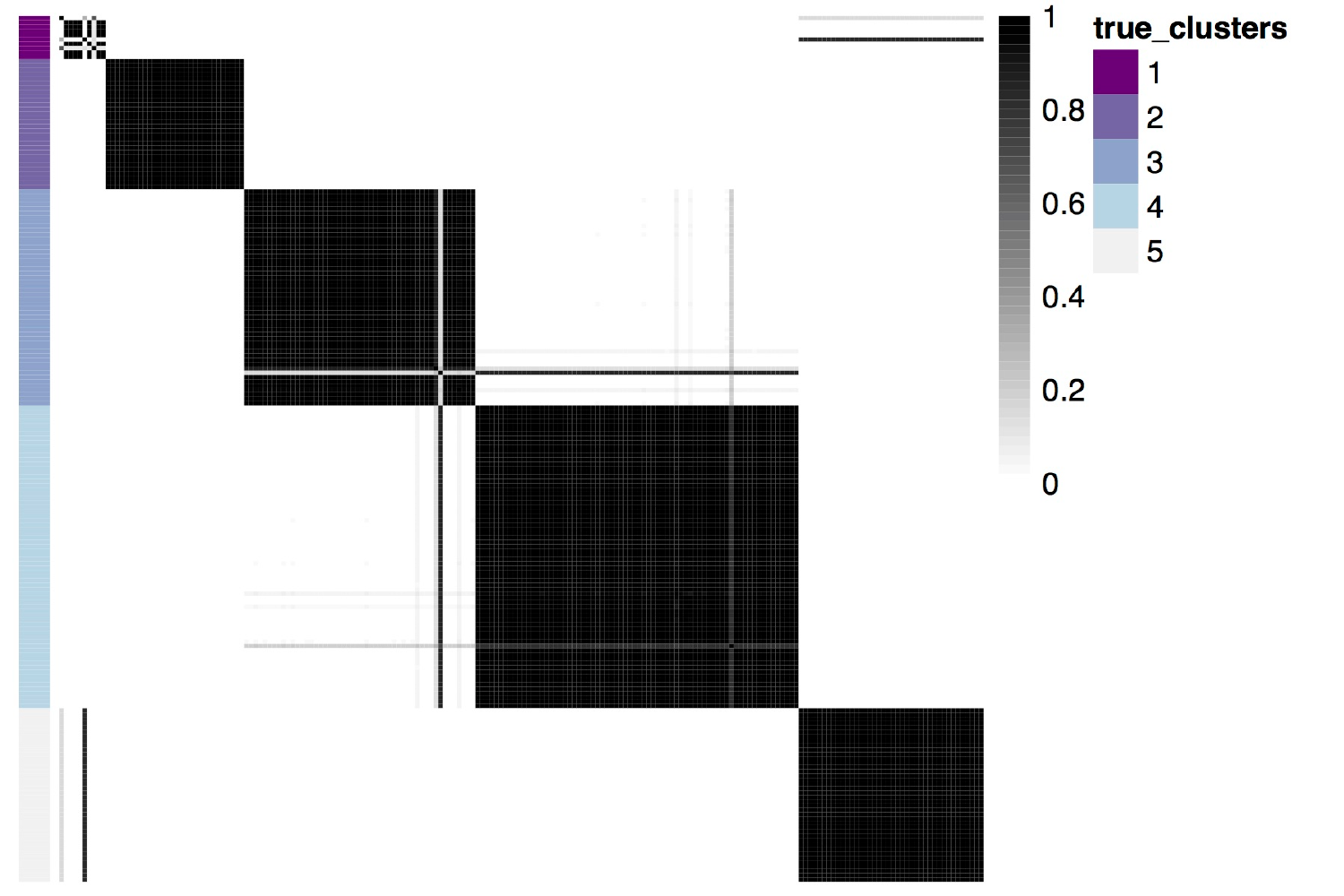}
 		\includegraphics[width=0.45\linewidth]{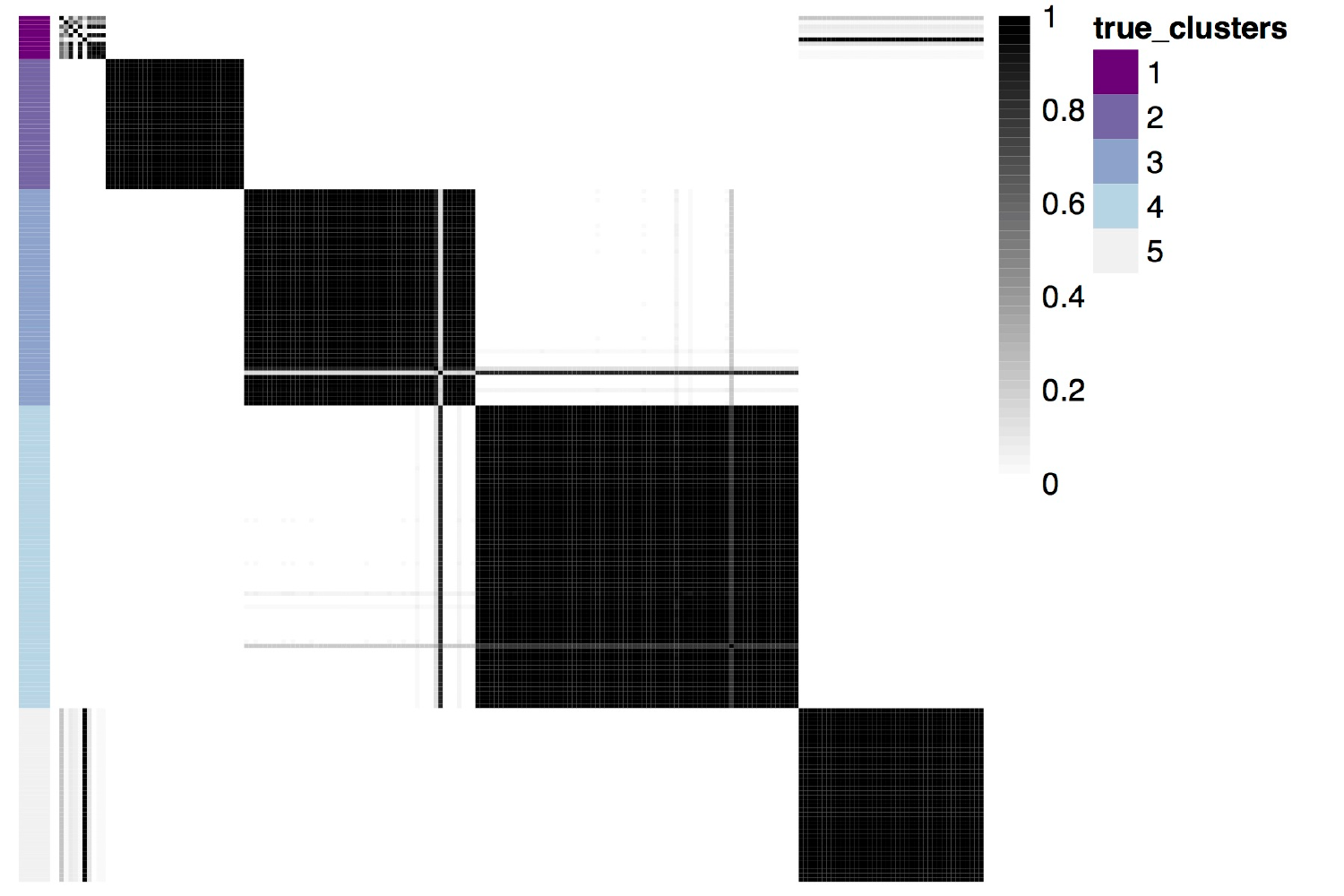}\\ \vspace{1cm}
 		\includegraphics[width=0.45\linewidth]{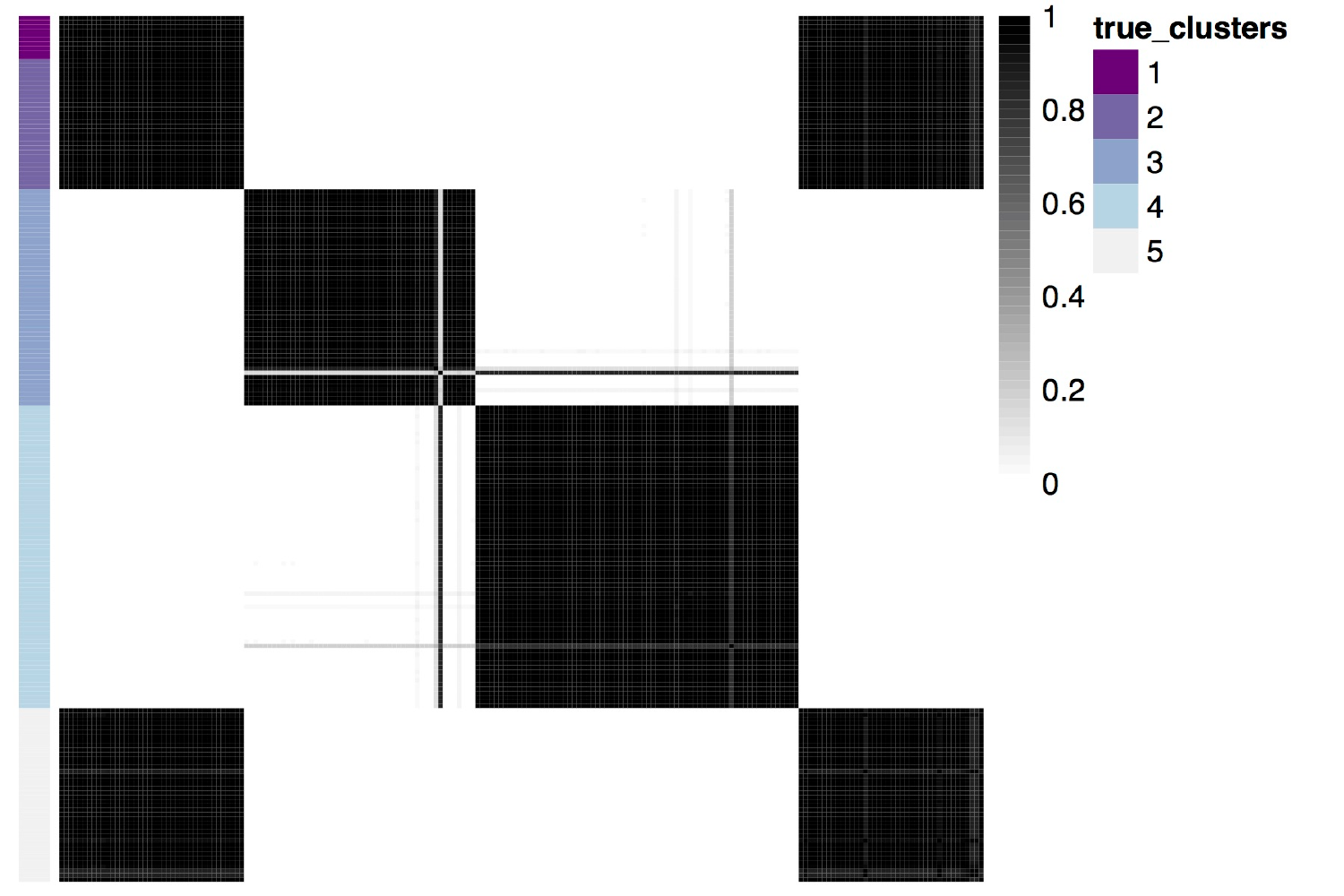}
 		\includegraphics[width=0.45\linewidth]{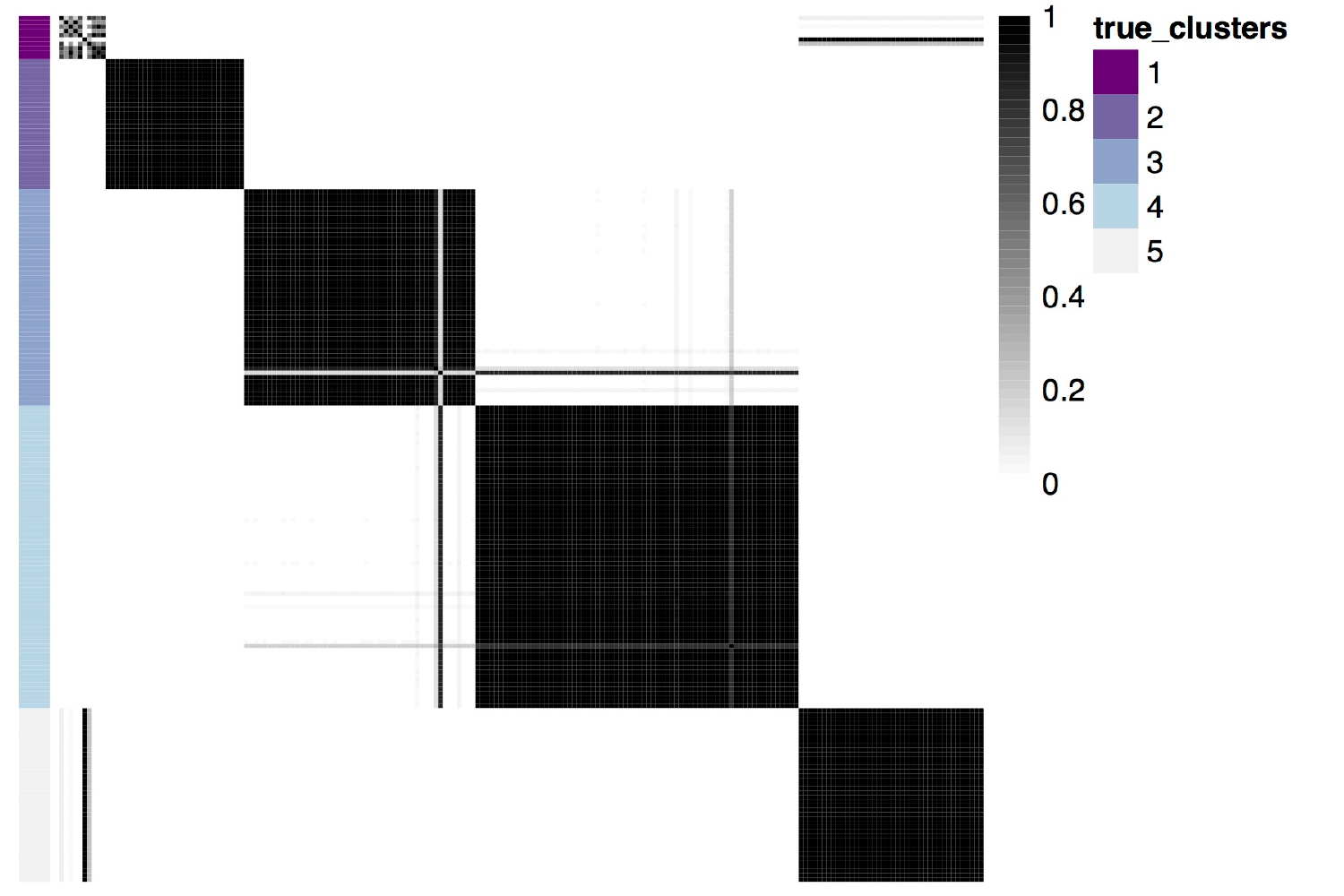}\\ \vspace{1cm}
 		\includegraphics[width=0.45\linewidth]{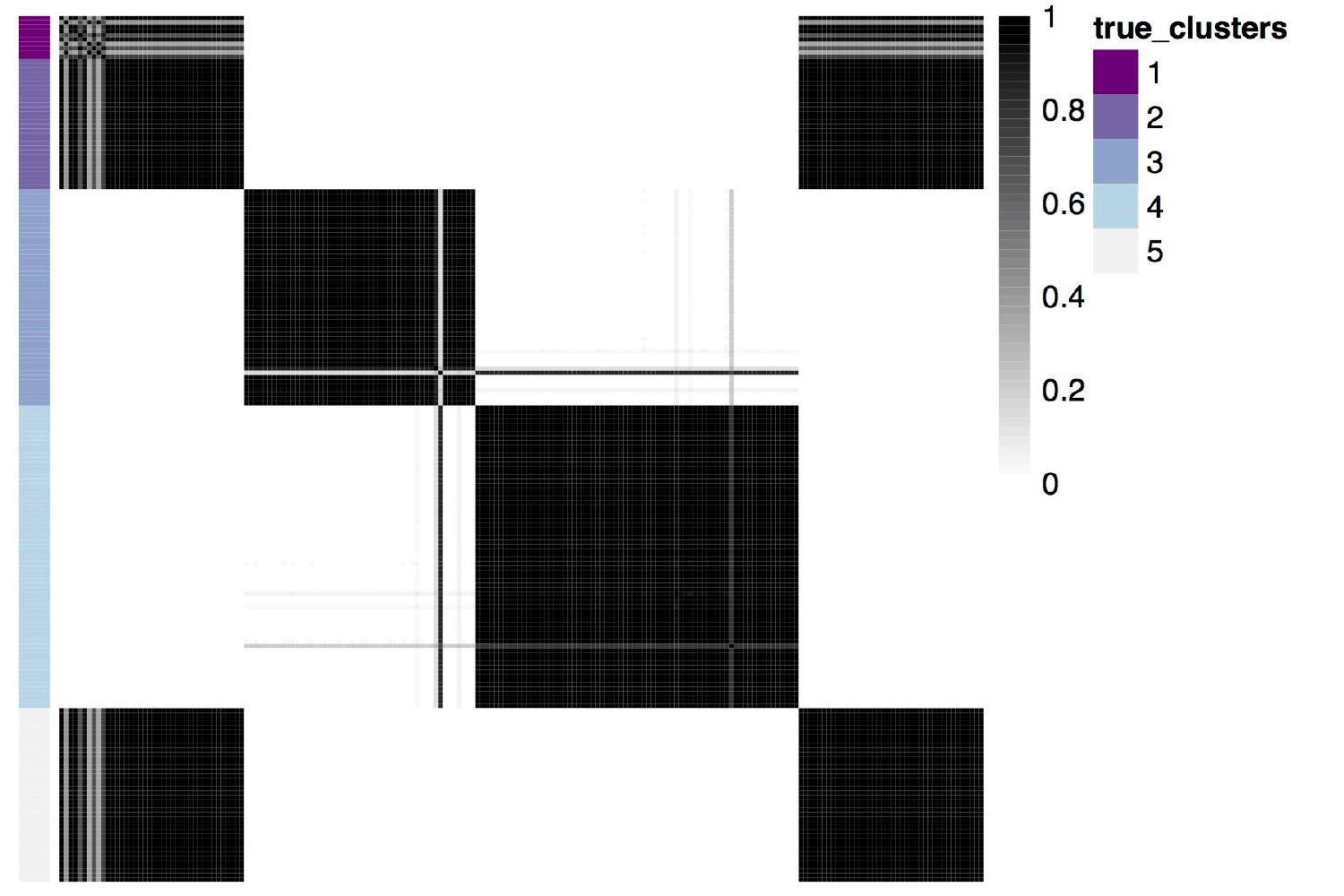}
 		\includegraphics[width=0.45\linewidth]{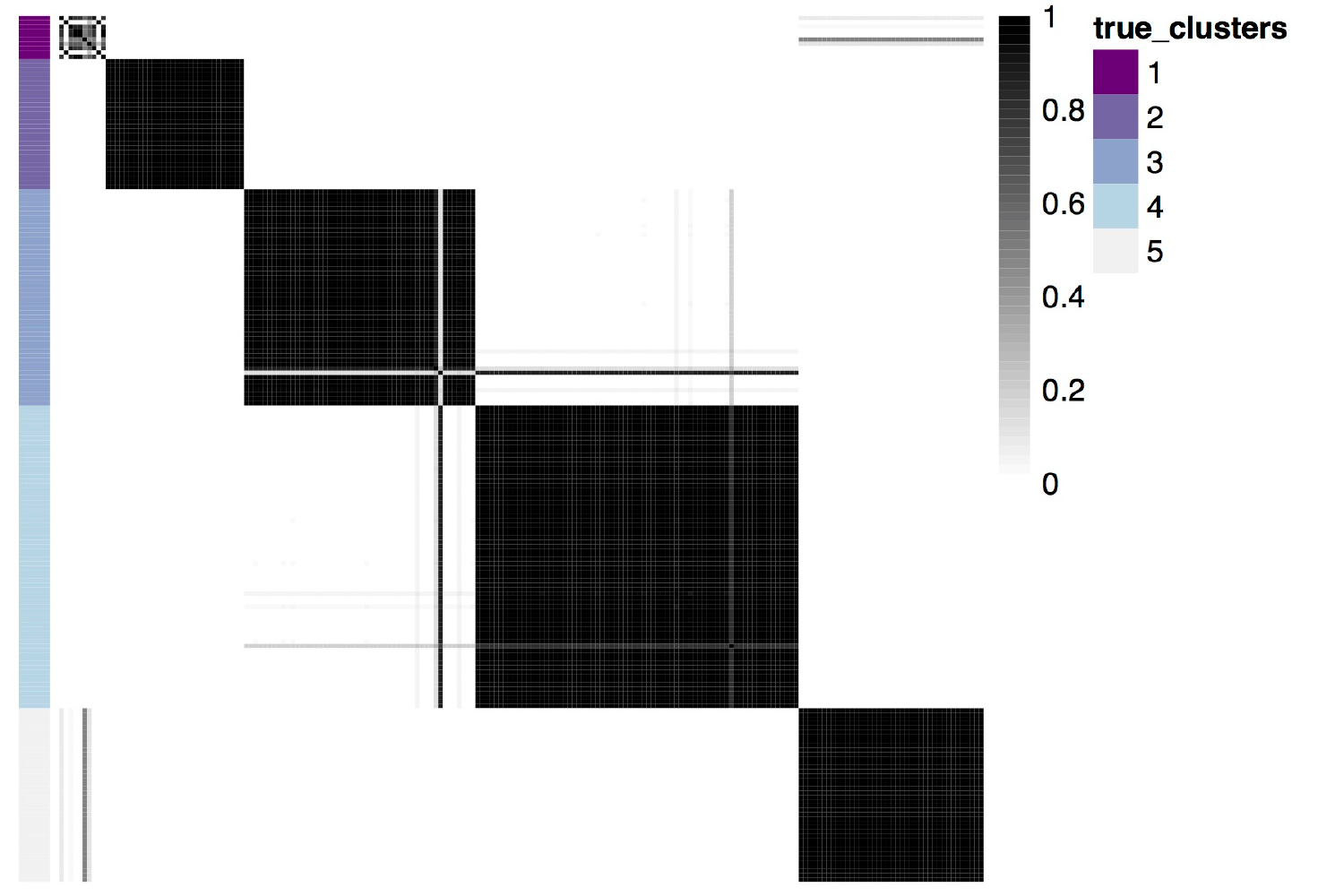}\\  \vspace{1cm}
 		\includegraphics[width=0.45\linewidth]{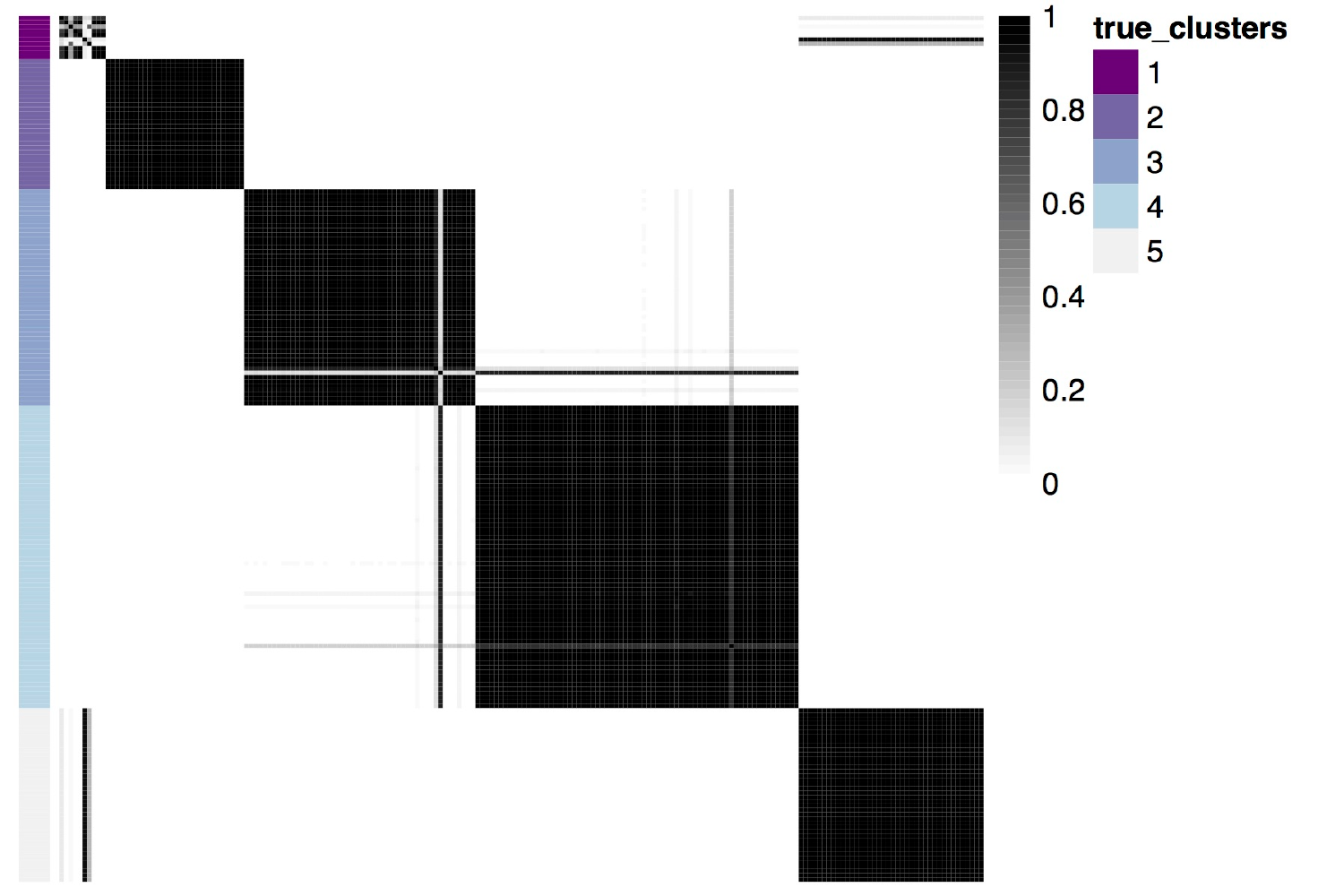}
 	\end{minipage}
 	\caption{Simulation study 4: Posterior similarity matrices of the 7 chains. The individuals are  in the same order to facilitate visual comparison. The true cluster allocations are represented in the left column, near each posterior similarity matrix.}
 	\label{PSM_simu}
 \end{figure}
 
 \clearpage
  \section{Supplementary simulation with 500 individuals}
  
  We simulated a dataset of 500 individuals to mimic the size of the dataset in the application.  The sizes of the generated clusters were 50,  75, 125, 150 and 100, respectively.   A set of 5 categorical covariates, with 3 categories each, were generated for each individual given the cluster allocations, with the cluster-specific parameters shown in Supplementary Table \ref{supphyper}. The outcome was generated from a cluster-specific Gaussian Processes with mean ${\bf m}_c$ and hyperparameters $\boldsymbol{\theta}_c$ for each cluster $c$ presented in Supplementary Table \ref{supphyper}. Regarding the model inference, we adopted the same priors as in Section 4.5.1.

  \begin{table}[h]
  	\begin{tabular}{c| c c c}
  		Cluster & 	${\bf m}_c$& $\boldsymbol{\theta}_c$ &$\boldsymbol{\phi}_c$ \\\hline
  		1&[9,8.5,8,6,5,4,3]&[0.5,0.1,-0.7]& 	[0.8,0.1,0.1] [0.8,0.1,0.1] [0.8,0.1,0.1] [0.8,0.1,0.1] [0.8,0.1,0.1]\\
  		2&[6,7,6,4,2,4,5]&[0.6,0.2,-0.3] & [0.1,0.8,0.1] [0.1,0.8,0.1] [0.1,0.8,0.1] [0.1,0.8,0.1] [0.1,0.1,0.8]\\
  		3&	[10,11,10,9,10,8,7]& [0.1,0.3,-0.7] &[0.4,0.5,0.1] [0.5,0.4,0.1] [0.1,0.4,0.5] [0.5,0.1,0.4] [0.5,0.3,0.2] \\
  		4&[8,5,8,9,10,11,9] &[0.3,0.4,-0.5] &[0.2,0.7,0.1] [0.7,0.2,0.1] [0.3,0.6,0.1] [0.2,0.1,0.7] [0.6,0.3,0.1] \\
  		5&[7,7.5,6,5,5,3,0] & [0.1,0.5,-0.7]&[0.6,0.3,0.1] [0.2,0.1,0.7] [0.3,0.6,0.1] [0.7,0.2,0.1] [0.2,0.7,0.1] \\
  	\end{tabular}
  	\caption{Hyperparameters for data generation (with $\boldsymbol{\theta}_c = \log((a_c, l_c, \sigma^2_c)^{\top})$)} 
  	\label{supphyper}
  \end{table}
  
  \begin{figure}[h!]
  	\centering
  	\includegraphics[width=0.8\linewidth]{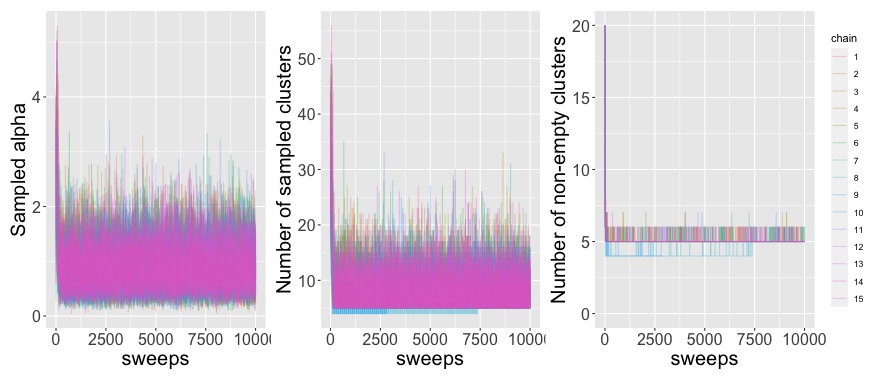}
  	\caption{From left to right: Traces of the $\alpha$ DP parameter, of the number of clusters and number of non-empty clusters for the 15 chains.}
  	\label{comparaison500}
  \end{figure}


  We ran 15 chains on the generated dataset, with 10000 iterations and compared the traces of $\alpha$, number of clusters (nClus), number of non empty clusters (Supplementary Figure \ref{comparaison500}).  The traces of these three parameters overlap across the 15 chains and the posterior similarity matrices in Supplementary Figure \ref{PSM500} show good recovery of the generated clustering structure. Supplementary Figure \ref{fig:chain1-trajectories-data} presents the generated individual longitudinal data, coloured by the clusters individuals were allocated to in the first chain. Supplementary Figure \ref{fig:chain1} presents the clusters obtained in the first chain, in terms of cluster sizes, empirical outcome means and covariate profiles. We observe that the later correspond to the generated profiles described in Supplementary Table \ref{supphyper}.
  
   \begin{figure}[h]
  	\begin{minipage}{\linewidth}
  		\vspace{-0.5cm}
  		\centering
  		\includegraphics[width=0.25\linewidth]{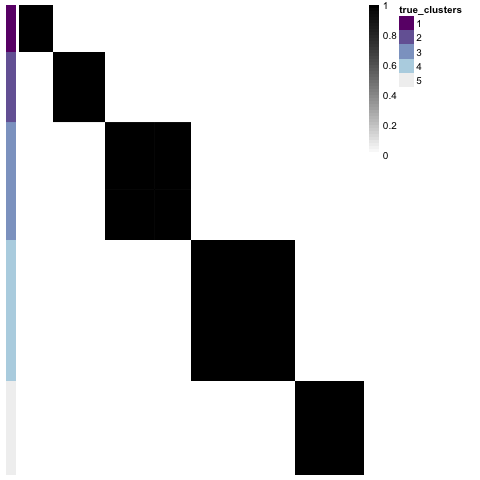}
  		\includegraphics[width=0.25\linewidth]{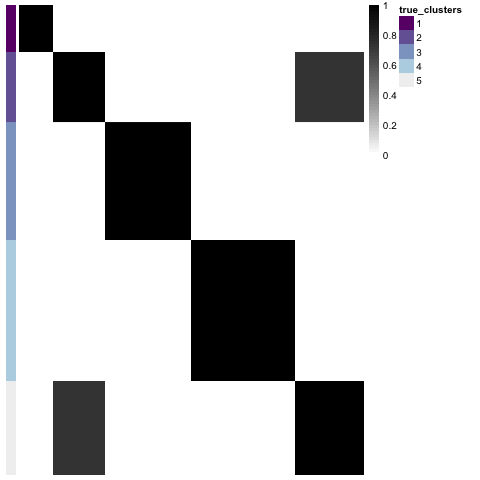}
  		  		\includegraphics[width=0.25\linewidth]{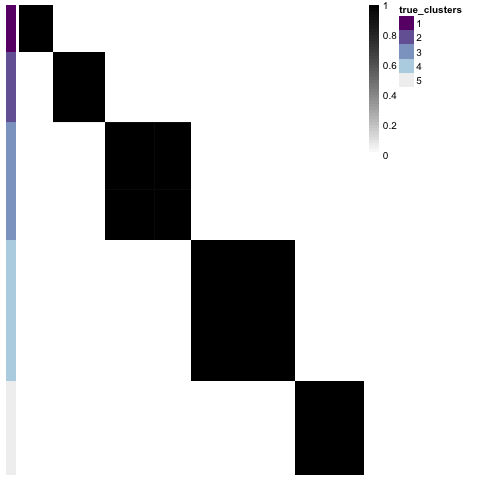}\\
  		  		  		\includegraphics[width=0.25\linewidth]{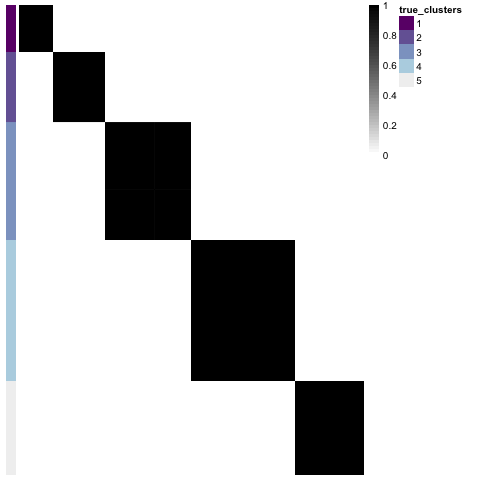}
  		  		\includegraphics[width=0.25\linewidth]{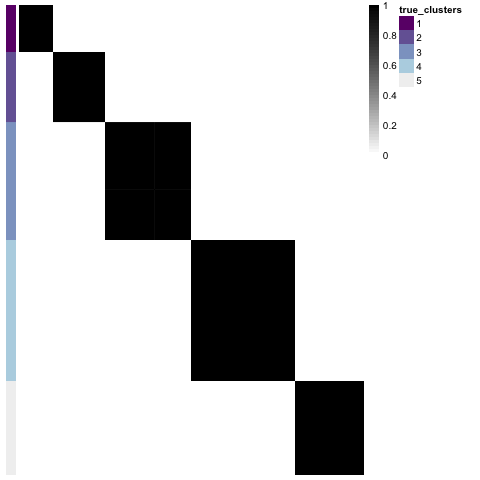}
  		  		\includegraphics[width=0.25\linewidth]{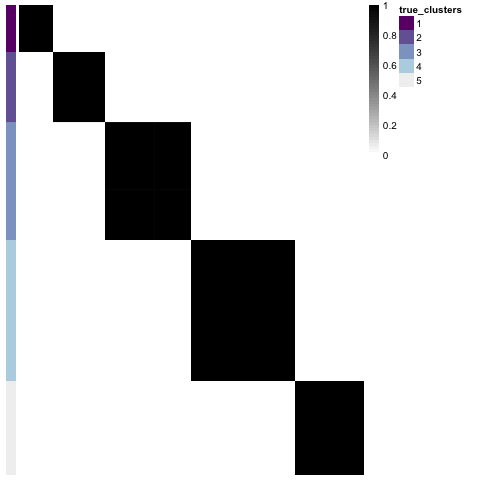}\\
  		  		  		\includegraphics[width=0.25\linewidth]{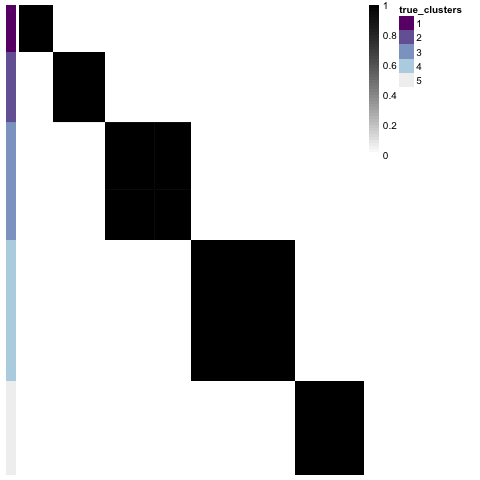}
  		  		\includegraphics[width=0.25\linewidth]{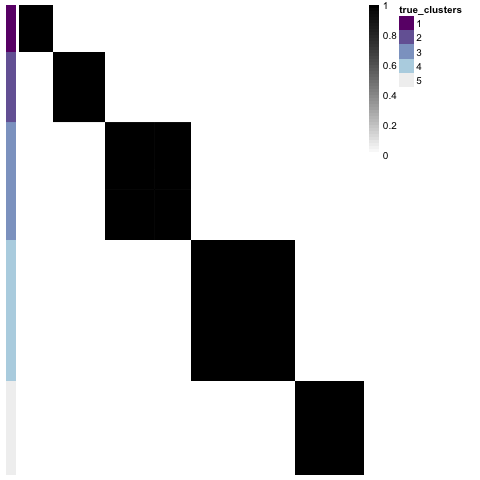}
  		  		\includegraphics[width=0.25\linewidth]{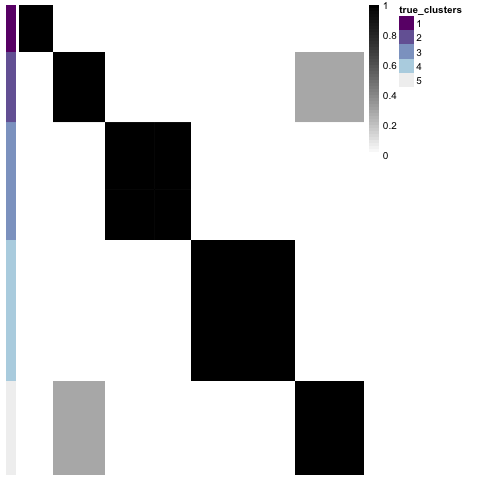}\\
  		  		  		\includegraphics[width=0.25\linewidth]{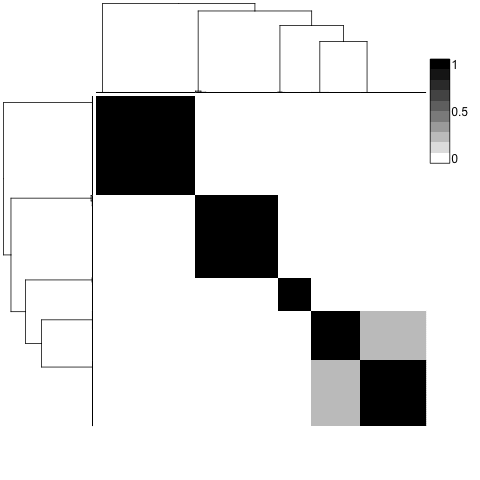}
  		  		\includegraphics[width=0.25\linewidth]{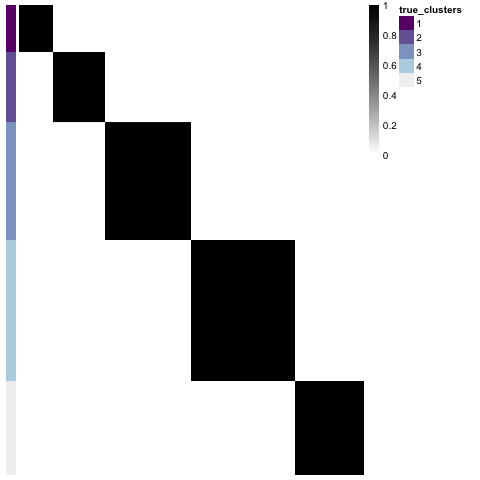}
  		  		\includegraphics[width=0.25\linewidth]{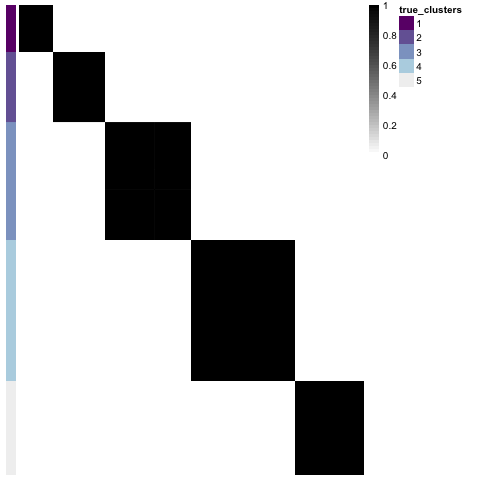}\\
  		  		  		\includegraphics[width=0.25\linewidth]{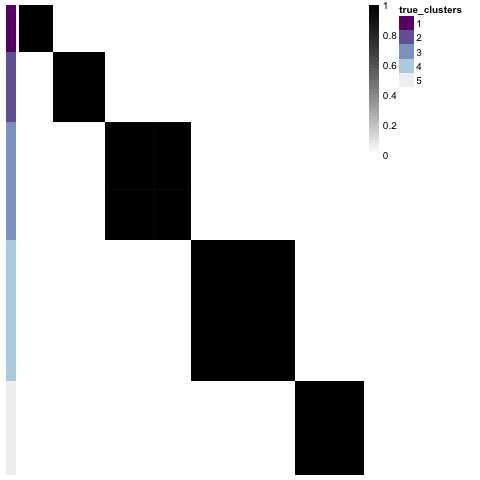}
  		  		\includegraphics[width=0.25\linewidth]{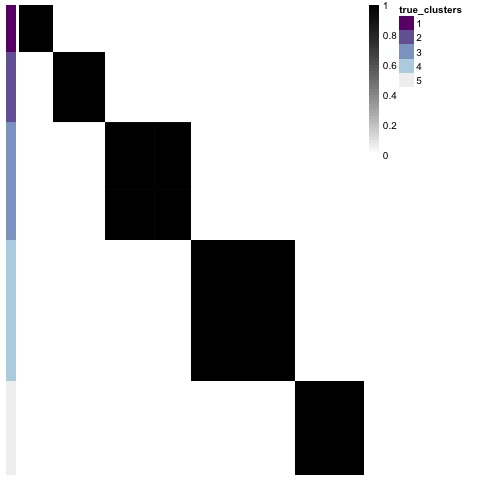}
  		  		\includegraphics[width=0.25\linewidth]{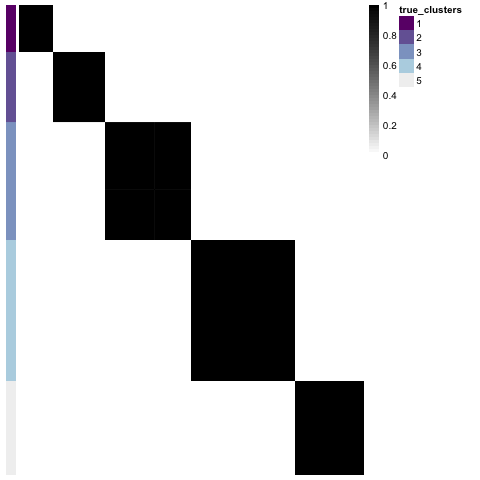}\\
  	\end{minipage}
  	\caption{Additional simulation study: Posterior similarity matrices of the 15 chains. The individuals are  in the same order to facilitate visual comparison. The true cluster allocations are represented in the left column, near each posterior similarity matrix.}
  	\label{PSM500}
  \end{figure}

  \begin{figure}[h!]
	\centering
	\includegraphics[width=0.7\linewidth]{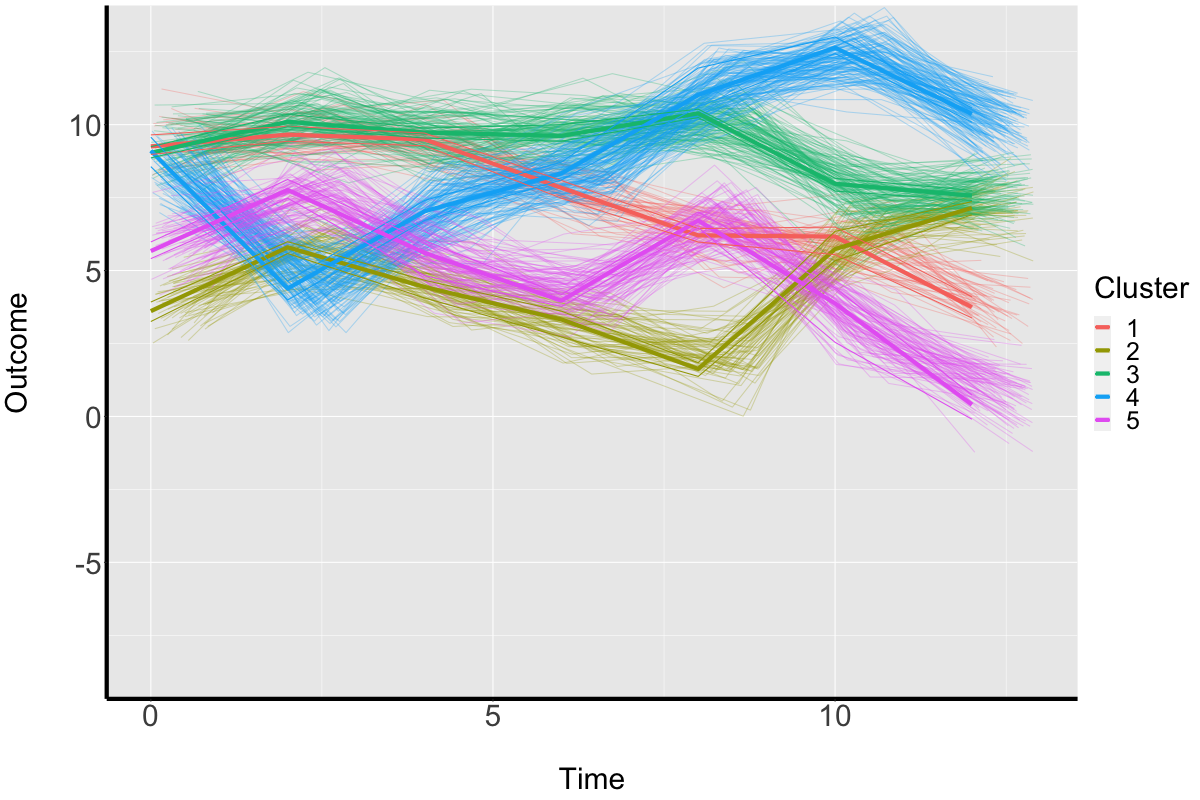}
	\caption{Generated longitudinal trajectories for 500 individuals (thin lines), coloured by clusters they are allocated to in the first chain, and cluster-specific estimated mean trajectories (thick lines).}
	\label{fig:chain1-trajectories-data}
\end{figure}

  \begin{figure}
  	\centering
  	\includegraphics[width=0.9\linewidth]{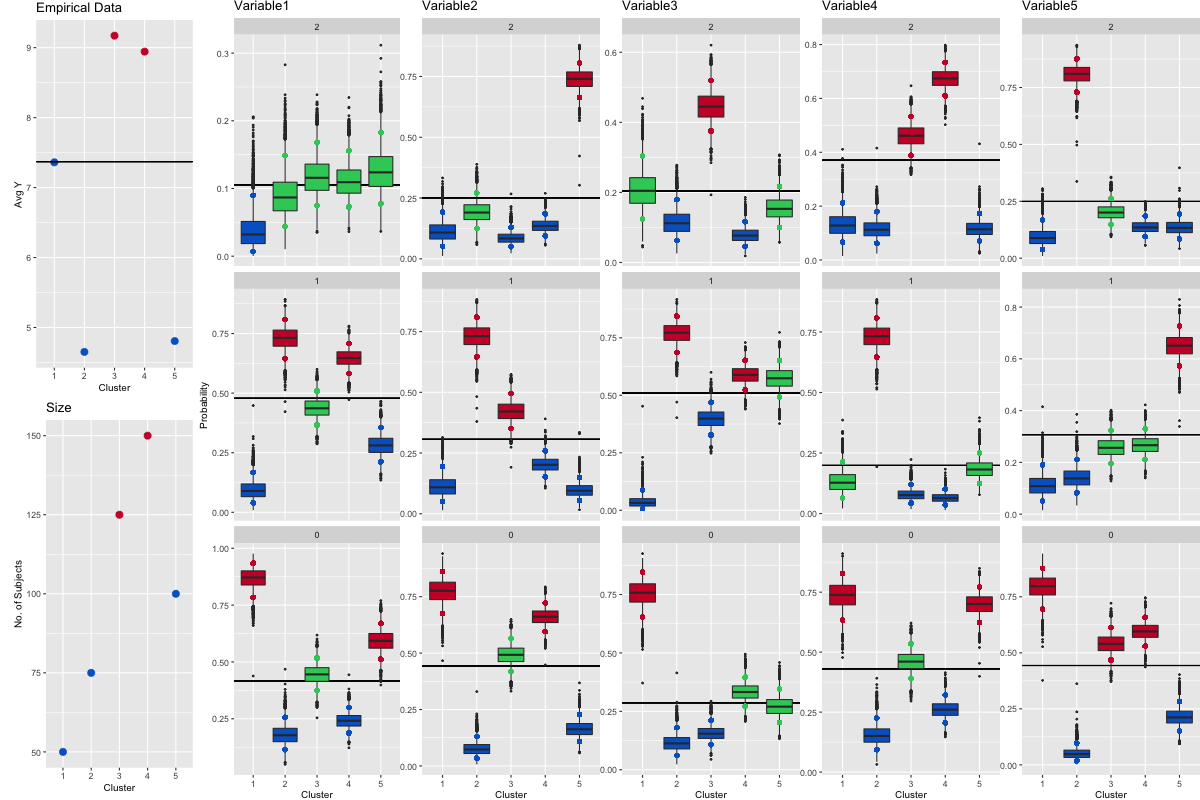}
  	\caption{Cluster profiles estimated by the first chain. The first column displays the cluster sizes (bottom), the cluster-specific empirical outcome means (dots in the top panel) and empirical outcome mean in the whole sample (solid line in top panel). The columns 2 to 6 represent the boxplots of the estimated probabilities for covariates 1 to 5 to be equal to 0 (bottom row), 1 (middle row) and 2 (top row) for each cluster. The solid lines represent the empirical proportions of individuals with covariates equal to 0, 1 or 2 in the whole sample. The boxplots are red if the 5th percentile is above the solid line, blue if the 95th is below, and green otherwise.}
  	\label{fig:chain1}
  \end{figure}
  
   \clearpage
 \section{Application}\label{Web_Application}
 \subsection{GP specification}
 \begin{figure}[ht] 
 	\begin{center}
 		\includegraphics[width=\textwidth]{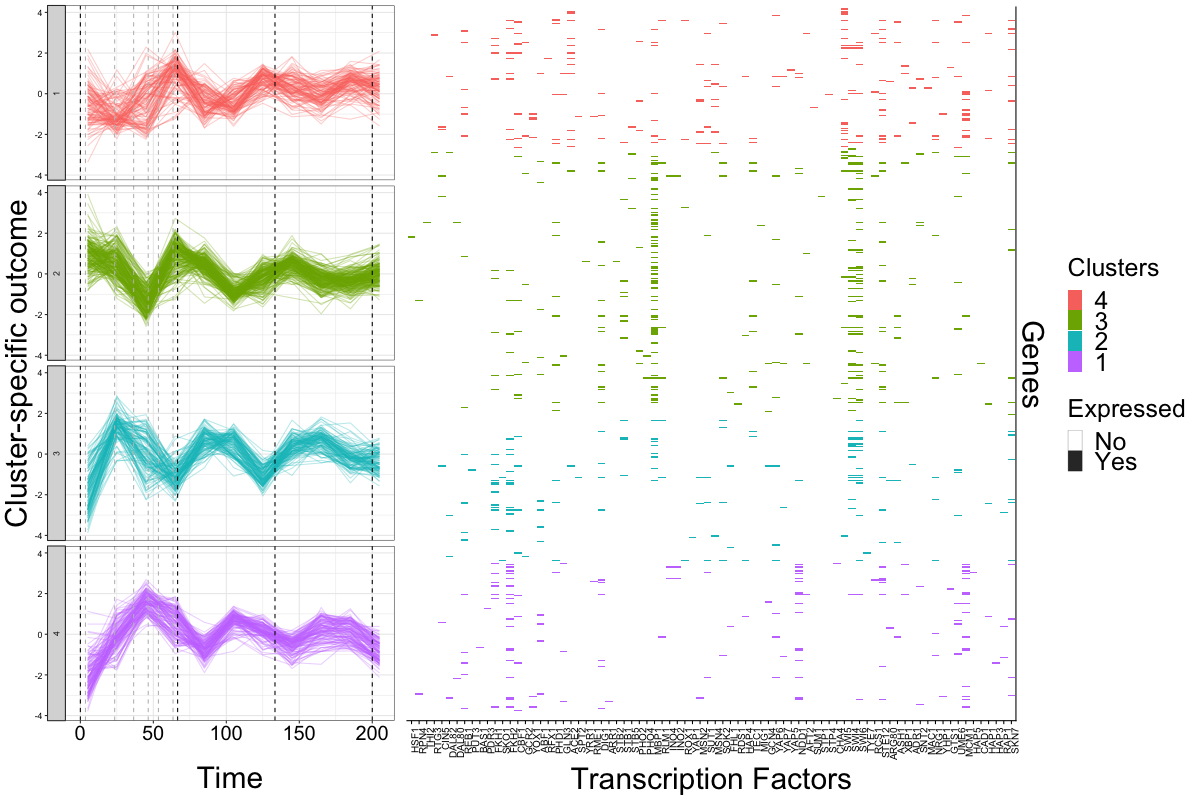}
 		\caption{Trajectories of gene expression and covariate profiles for the 4 clusters of the final partition. On the left panels, trajectories of the genes in the color associated with the cluster they are allocated to. Dashed black lines delimit the three cell cycles and gray ones delimit the different phases, in the first cycle: M/G1, G1, S, G2, G2/M, M and M/G1 again.  On the right panels: cluster-specific profiles with regard to the 80 transcription factors.}
 		\label{GP_results2}
 	\end{center}
 \end{figure}
 
 \begin{figure}[ht!]
 	\centering
 	\includegraphics[width=0.7\linewidth]{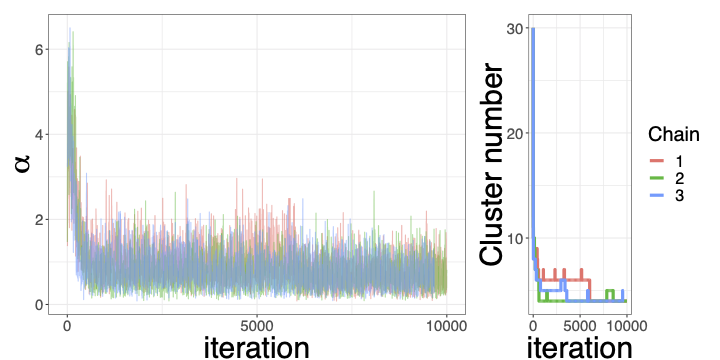}
 	\caption{Trace of the concentration parameter $\alpha$ (left panel) and number of non-empty clusters (right panel) across the 10000 iterations for the three chains}
 	\label{CV}
 \end{figure}
 
 \begin{figure}[ht!]
 	\centering
 	\includegraphics[width=0.4\linewidth]{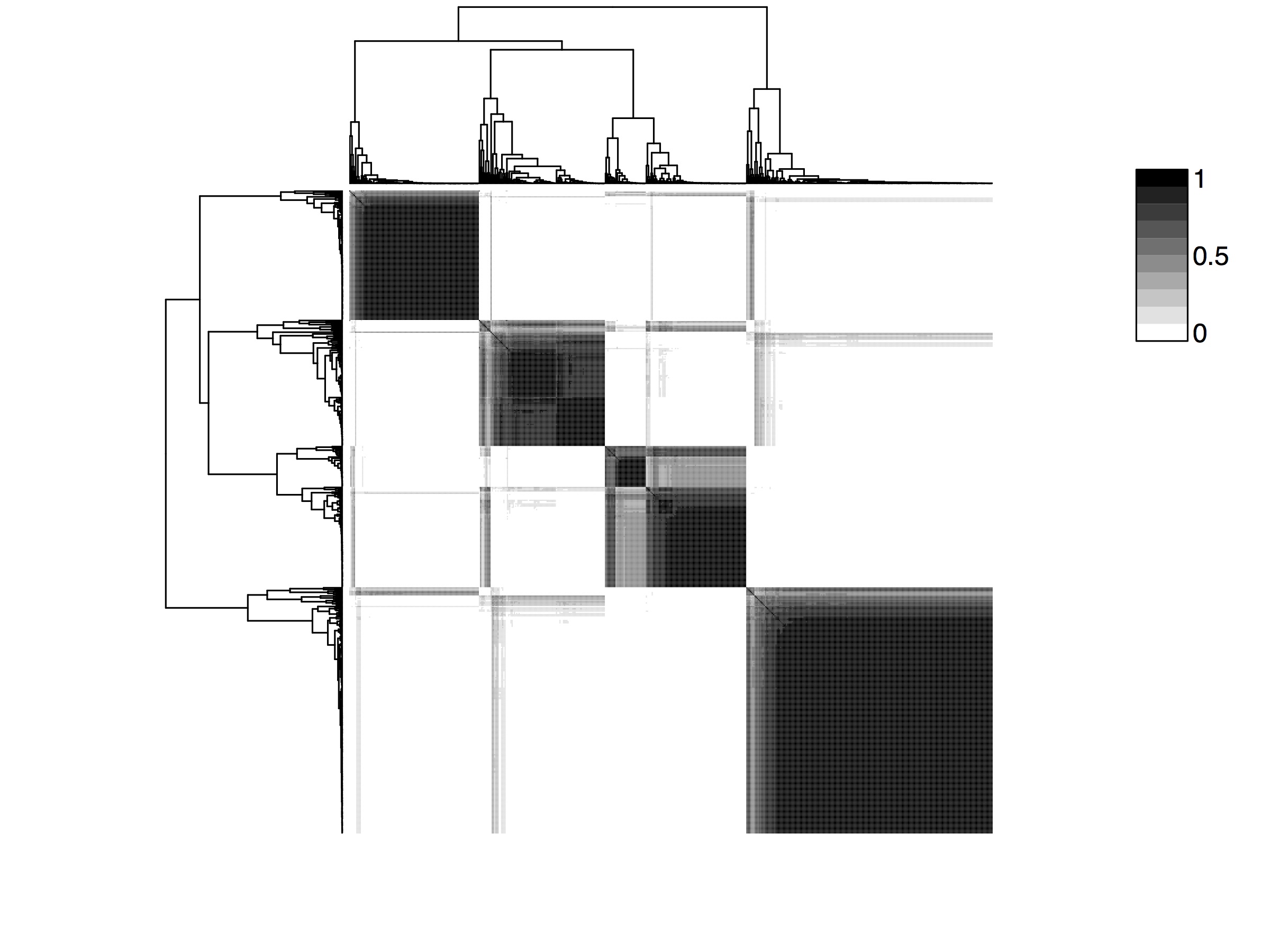}
 	\includegraphics[width=0.4\linewidth]{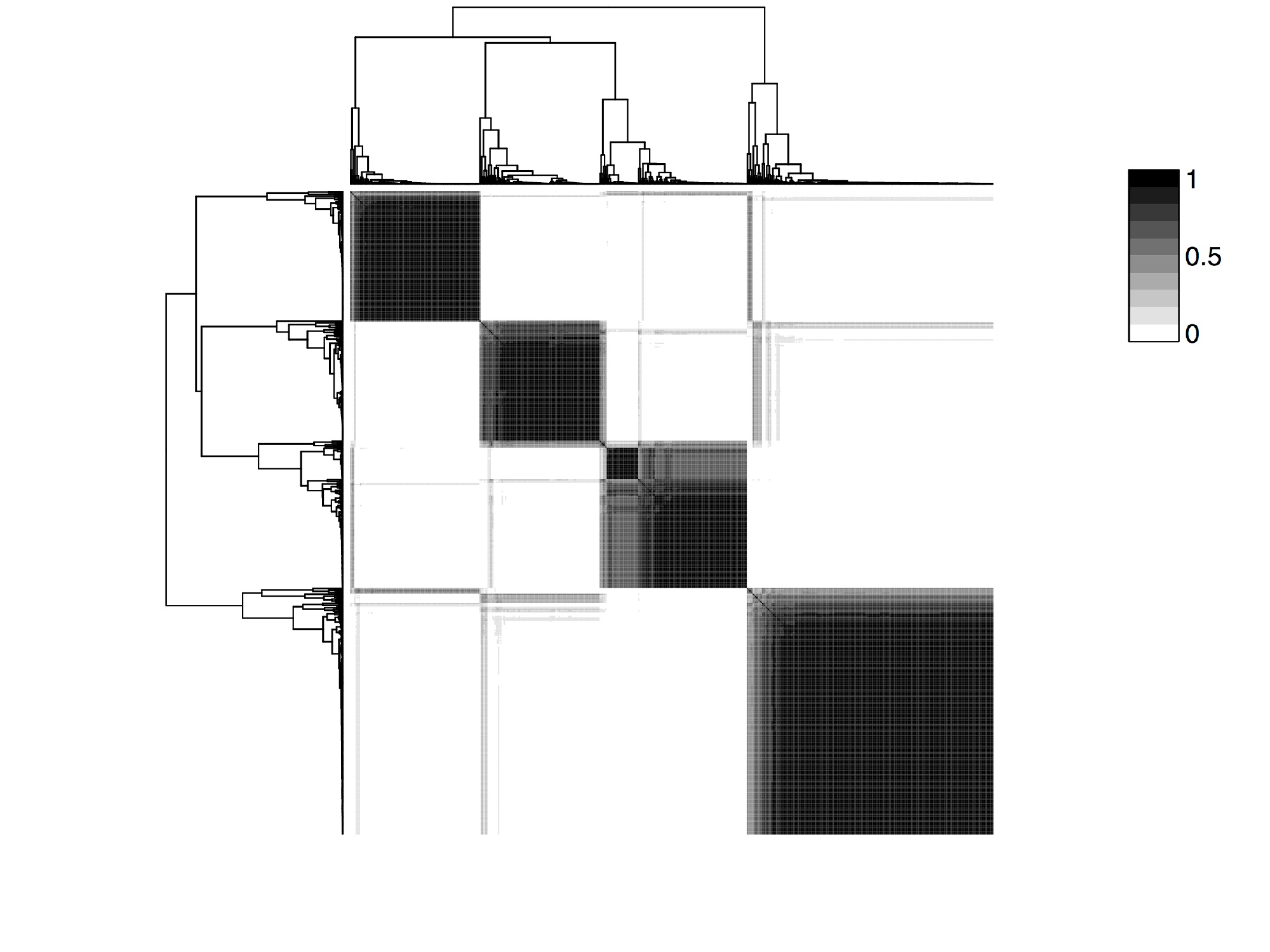}
 	\caption{Posterior similarity matrices of chains 2 and 3 obtained by PReMiuMlongi on the \textit{Saccharomyces cerevisiae} budding yeast data. Rows and columns are ordered in the same way for both matrices}
 	\label{PSMs}
 \end{figure}
 
 \begin{table}
 	\begin{center}
 		\begin{tabular}{c | c c c c c | c}
 			PEAR Index & 2 & 3 & 4 & 5 & 6 & mean \\\hline
 			2&-  &0.914& 0.980& 0.986 &0.862& 0.936\\
 			3& 0.914 &-& 0.895& 0.901 &0.897&0.902\\
 			4& 0.980 &0.895& -& 0.995 &0.855&0.931\\
 			5& 0.986 &0.901& 0.995& - &0.861&0.936\\
 			6& 0.862 &0.897& 0.855 &0.861 &-&  0.869\\
 		\end{tabular}
 	\end{center}
 	\caption{Comparison of partitions obtained with ratio values $r \in \{2, ...,  6 \}$, in terms of PEAR indices}
 	\label{RAND}
 \end{table}

\begin{figure}
	\centering
	\includegraphics[width=0.6\linewidth]{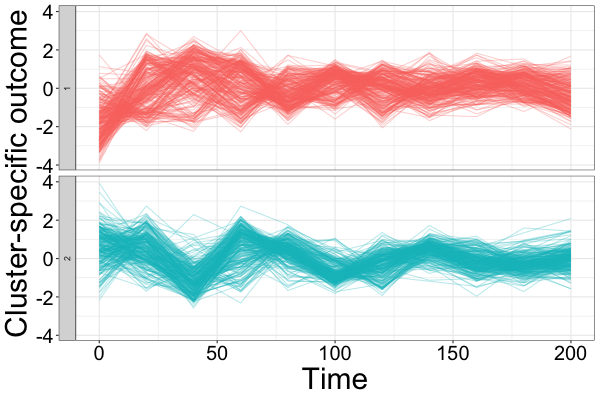}
	\caption{Individual trajectories of gene expression coloured by clusters obtained by the iCluster method in Kirk et al.(2012) \cite{Kirk2012}.}
	\label{fig:icluster-alltrajectories-data}
\end{figure}

\clearpage
 \subsection{MVN specification}
 \textcolor{black}{We analyzed the same dataset using the MVN specification. We adopted a gamma prior for  $\alpha\sim \text{Gamma}(2,1)$. For the covariate model, we adopted Dirichlet priors $\phi_{c,q,e} \sim  ~ \text{Dirichlet}(1)$ for each cluster $c$, covariate $q$ and covariate level $e$. For the outcome model, we set  $\kappa_0 = 0.01$ and $\nu_0=11$ as hyperparameters to the inverse Wishart prior.  The initial number of clusters was 20. We ran 3000 iterations for burn in and 30000 for sampling.}\\

 \begin{figure}[h]
 	\centering
 	\includegraphics[width=0.4\linewidth]{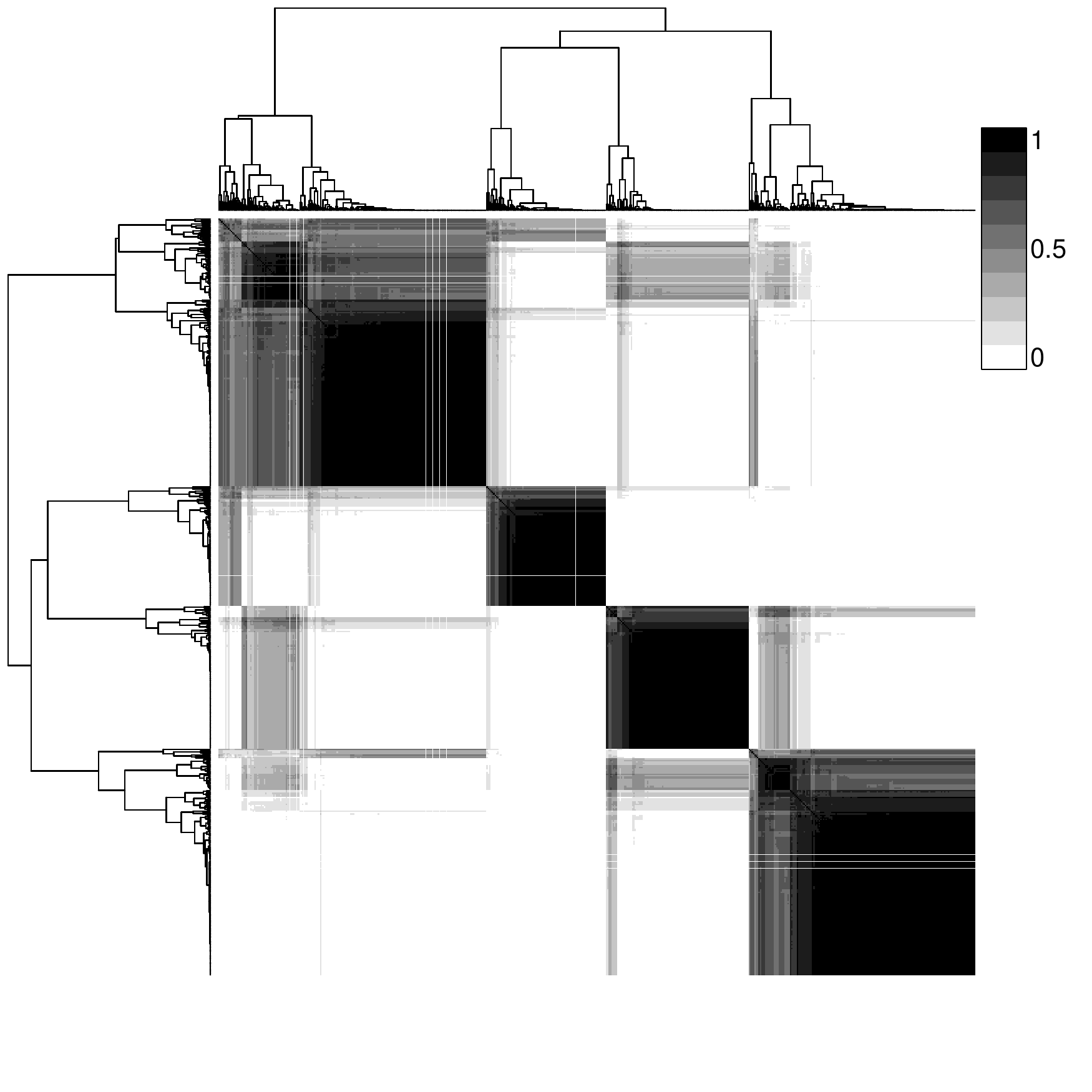}
 	\caption{Posterior similarity matrix from the model with MVN specification applied to the data presented in the Application section.}
 	\label{PSM_MVN}
 \end{figure}
 
\begin{figure}[h]
 	\centering
 	\includegraphics[width=0.9\linewidth]{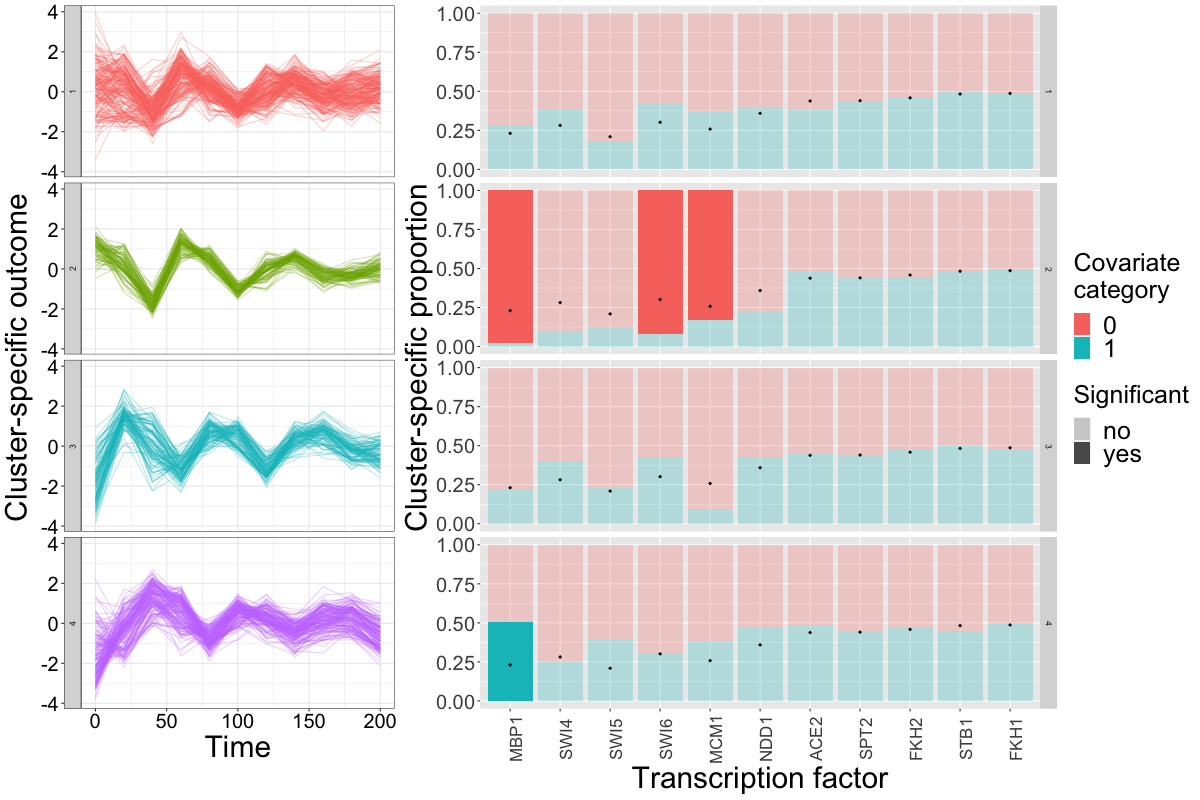}
 	\caption{Trajectories of gene expression and covariate profiles for the 4 clusters of the final  partition estimated using the MVN specification. On the left panels, trajectories of the genes in the color associated with the cluster they are allocated to. On the right panels: cluster-specific profiles with regard to the selected transcription factors. In each cell, associated with cluster $k=1,...,4$ (represented by rows) and transcription factor $q=1,...,10$ (represented by columns), the black pip represents the empirical proportion of genes with $x_q=1$ in the whole sample. The turquoise filled square and the red one represent the estimated cluster-specific proportions $\hat{P}(X_{c,q}=1)$ and $\hat{P}(x_{c,q}=0)$, respectively. Squares are dark filled if the 95\% credibility interval of $\hat{P}(x_{c,q}=1)$  does not contain the empirical proportion, meaning that the considered category is significantly different in the cluster compared to the whole sample. Red squares are dark filled if the 5th percentile is above the empirical mean, and blue ones are dark filled if the 95th is below.}
 	\label{plot_MVN}
 \end{figure}

  We obtained 4 clusters of sizes 194,  89, 105 and 163 genes, respectively (Supplementary Figure \ref{PSM_MVN}). Supplementary Table \ref{comp_MVN_GP} compares the final clustering to the one estimated using the GP specification. Clusters 3-4 in both partitions are similar, however MVN cluster 1 includes genes from both GP clusters 1 and 2. This translates into  more heterogeneous patterns in MVN cluster 1 (Supplementary Figure \ref{plot_MVN}), and two additional selected transcription factors (based on a 10\% lower threshold for the variable-specific relevance indicator $\rho_q$ (Equation 4)).
  
 \begin{table}[ht]
\centering
\begin{tabular}{rrrrr}
 & 1 & 2 & 3 & 4 \\ 
  \hline
1 &  73 &   0 &   0 &  36 \\ 
  2 & 114 &  89 &   3 &   0 \\ 
  3 &   6 &   0 & 102 &   5 \\ 
  4 &   1 &   0 &   0 & 122 \\ 
\end{tabular}
\caption{Cross-table of the clustering structures obtained with the GP specification (rows) and the MVN specification (columns).}
\label{comp_MVN_GP}
\end{table}

 When longitudinal data are highly variable such as gene expression, we recommend using the GP specification which is more parsimonious than the MVN outcome model, in terms of parameters to estimate,  while still offering high flexibility to recover the underlying clustering structure.
 
 \clearpage
 \section{Implementation in PReMiuMlongi}\label{implementation}
 
 Here, we outline the inclusion of the response models in the software package PReMiuMlongi. We treat each response model separately, considering the MVN response model in Section \ref{pmvn} and the GP response model in \ref{pgp}. The \includegraphics[height=10pt]{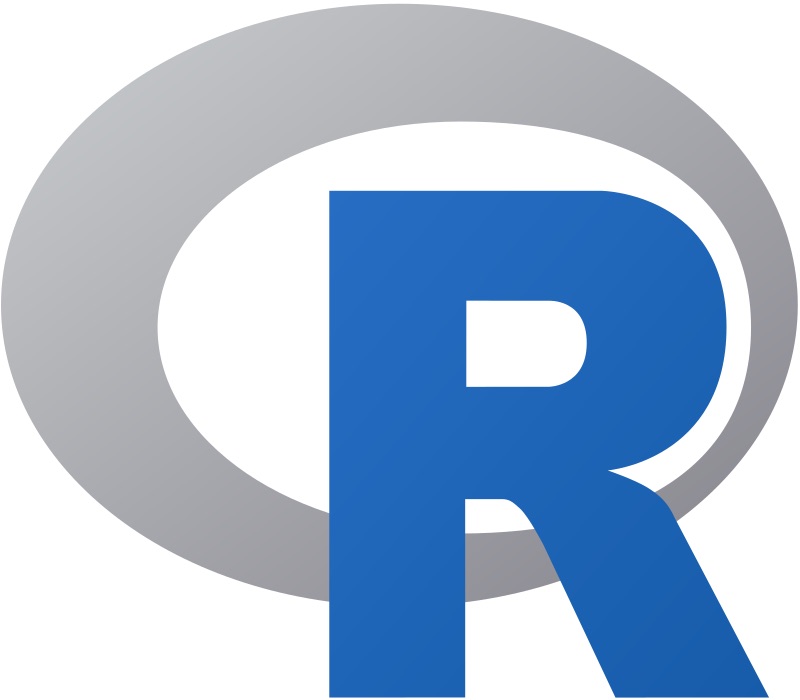} page for PReMiuMlongi, \url{https://github.com/premium-profile-regression/Longitudinal-PReMiuM}, contains both the package and detailed documentation.
 
 \subsection{Multivariate normal implementation}\label{pmvn}
 
 \subsubsection{Input data}
 
 \textcolor{black}{The utility function in PReMiuMlongi to generate data with an MVN outcome creates two clusters with 50 people each, with $Q=5$ discrete covariates each with $R=3$ categories. The covariate parameters are $\phi_1=\left\{\begin{array}{ccc} 0.8 & 0.1 & 0.1 \\ 0.8 & 0.1 & 0.1 \\ 0.8 & 0.1 & 0.1 \\ 0.8 & 0.1 & 0.1 \\ 0.8 & 0.1 & 0.1 \end{array}\right\}$ and $\phi_2=\left\{\begin{array}{ccc} 0.1 & 0.8 & 0.1 \\ 0.1 & 0.8 & 0.1 \\ 0.1 & 0.8 & 0.1 \\ 0.1 & 0.8 & 0.1 \\ 0.1 & 0.1 & 0.8 \end{array}\right\}$. The outcomes have mean values $[10, 10, 10]$ and $[10, 5, 0]$ respectively and both have covariance matrix $I_3$.}
 
 The data input to PReMiuMlongi from the \includegraphics[height=10pt]{Rlogo2} interface are structured as before, with the \texttt{outcome} column now a sequence of columns \texttt{outcome1}, \texttt{outcome2}, \texttt{outcome3}, etc., with the additional requirement that these column names are supplied in the function call.
 
  \textcolor{black}{The model is then run on these simulated data using hyperparameters $\nu_0=3$ and $\kappa_0=0.01$ for the MVN outcome model, $\text{Dirichlet}(1)$ priors for the covariate model and $\text{Gamma}(2,1)$ prior for $\alpha$.} Below we show default output plots from PReMiuMlongi using data generated with these functions.
 
 \subsubsection{C++ back end}
 
 Within the C++ code for PReMiuMlongi, functions have been added to evaluate likelihoods for the multivariate outcomes, and to sample new parameters following the distributions in Equation \ref{iwish}.

 \subsubsection{Outputs}
 
\begin{figure}[ht] 
 	\begin{center}
 		\includegraphics[width=0.8\textwidth]{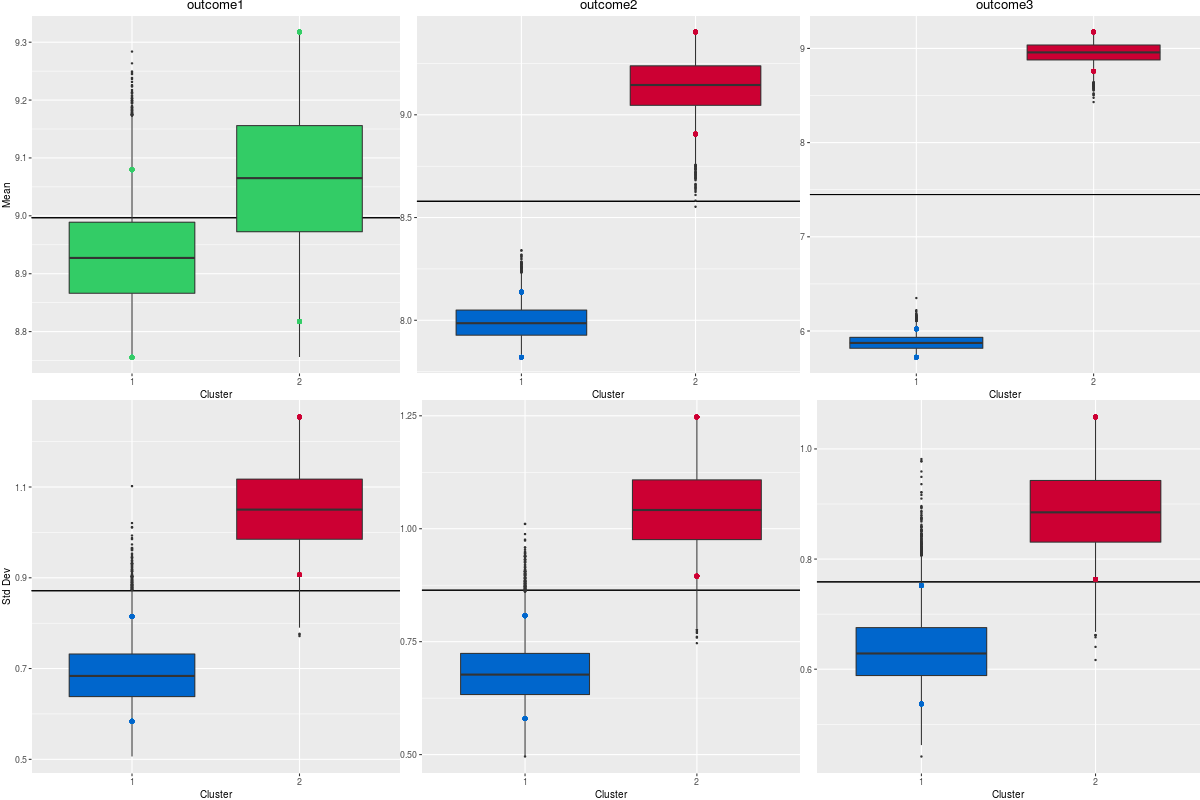}
 		\caption{Default PReMiuMlongi output summarising outcomes for the MVN response model. There is one column per outcome, with the mean shown on the top row and the standard deviation on the bottom. Within each panel, there is one box per cluster. Green colour indicates an overlap of the 90\% credible interval with mean value for all individuals, represented by the solid black horizontal line. Red indicates that the 90\% credible interval is above the mean value, and blue, below. NB this is the same format of output as is used for MVN covariates.}
 		\label{summarymvn}
 	\end{center}
 \end{figure}

 Additional output files are written for the parameters that are sampled at each iteration of the algorithm: $\mu$ and $\Sigma$. These files are used to construct the visualisation of the posterior parameter distributions, namely box plots of mean and standard deviation for each variable within each cluster (see Figure \ref{summarymvn} for an example).
 
\begin{figure}[ht] 
 	\begin{center}
 		\begin{tabular}{c}
 			\includegraphics[width=0.6\textwidth]{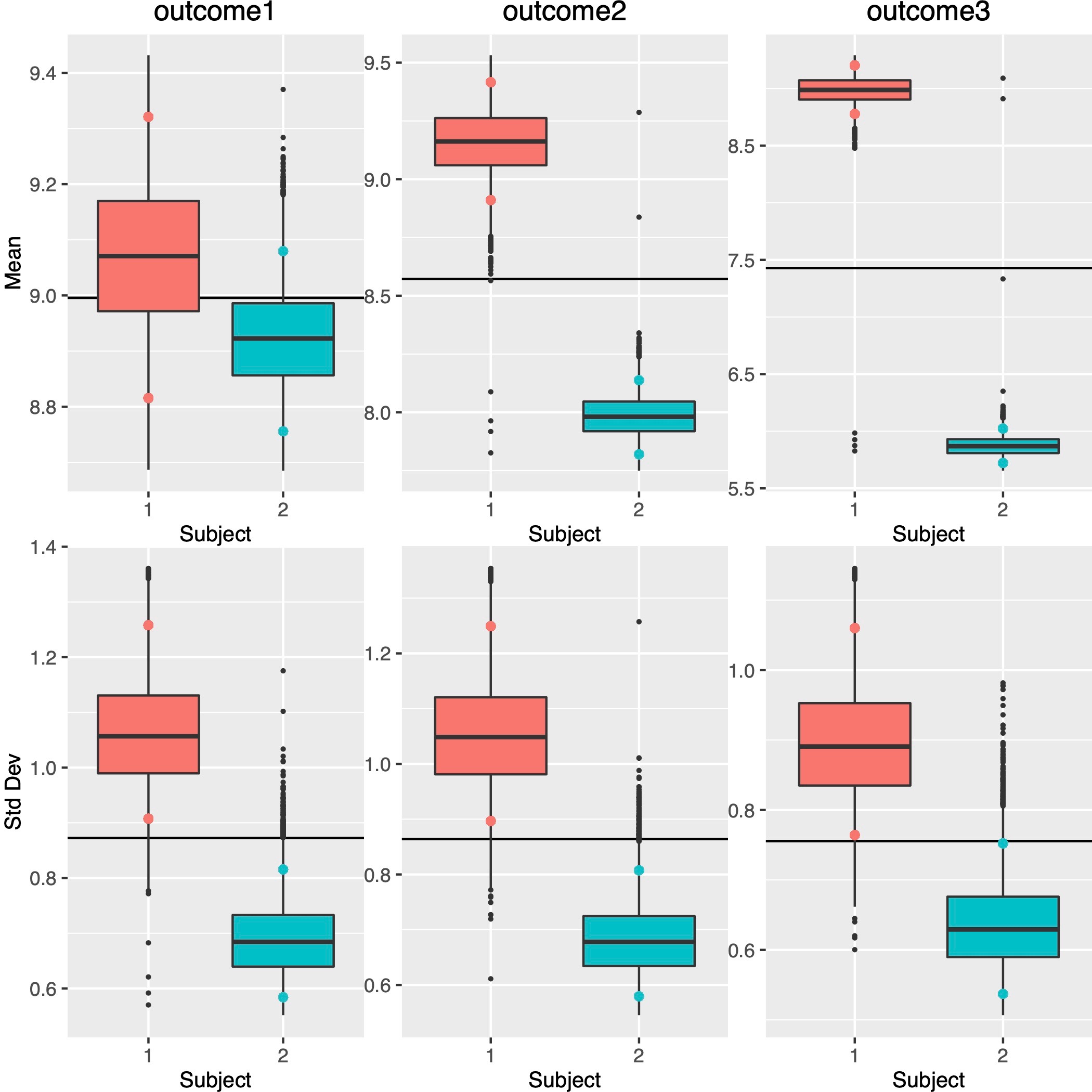}
 		\end{tabular}
 		\caption{Default PReMiuMlongi output for individual outcome predictions from the MVN response model. There is one column per outcome, with on the top row the box-plots of the posterior distributions for the means of individuals 1 and 2 and on the bottom the box-plots of the posterior distributions of the corresponding standard deviations. Within each panel, there is one box and one colour per individual for whom a prediction is made.
}
 		\label{predmvn}
 	\end{center}
 \end{figure}

 These files also form the basis for prediction of a new individual's outcome(s) based on their covariate data. These plots have a similar form; see Figure \ref{predmvn} for an example. The other graphical output, the covariate profile, is unchanged with the exception of the omission of the `risk' plot (see Figure \ref{summary}).

 \subsection{Gaussian process implementation}\label{pgp}
 
 \subsubsection{Input data}
 
\textcolor{black}{ The utility function to generate data with a longitudinal outcome creates two clusters with 30 people each, with $Q=5$ discrete covariates each with $R=3$ categories. The covariate parameters are $\phi_1=\left\{\begin{array}{ccc} 0.8 & 0.1 & 0.1 \\ 0.8 & 0.1 & 0.1 \\ 0.8 & 0.1 & 0.1 \\ 0.8 & 0.1 & 0.1 \\ 0.8 & 0.1 & 0.1 \end{array}\right\}$ and $\phi_2=\left\{\begin{array}{ccc} 0.1 & 0.8 & 0.1 \\ 0.1 & 0.8 & 0.1 \\ 0.1 & 0.8 & 0.1 \\ 0.1 & 0.8 & 0.1 \\ 0.1 & 0.1 & 0.8 \end{array}\right\}$. The outcomes have $M=7$ timepoints with observations at times $[0,2,4,6,8,10,12]$. The mean vectors are $[11,10,9,8,6,5,4]$ and $[9,7,6,4,2,1,0]$. Both have $\{a_c,l_c,\sigma_c\}=\{\exp(-0.5),\exp(-0.1),\exp(-0.5)\}$.}
 
 The R interface requires an additional data frame to be supplied, and a column to be added to the original `\texttt{data}' data frame. The additional data frame consists of a column of all the measurements, a column of the time points at which the measurements were taken, and a column of IDs, identifying from whom the measurements were taken. The ID column must correspond to the additional column in the original data frame so that covariate data can be correctly matched with outcome data.
 
 \textcolor{black}{The model is then run on these simulated data using standard lognormal priors for the hyperparameters $(a, l, \sigma^2)$ for the GP outcome model, $\text{Dirichlet}(1)$ priors for the covariate model and $\text{Gamma}(2,1)$ prior for $\alpha$.} Below we show default output plots from PReMiuMlongilongi using data generated with these functions.
 
 \subsubsection{C++ back end}
 
 Within the C++ code for PReMiuMlongi, functions have been added to evaluate likelihoods for the longitudinal outcomes, and a Metropolis-within-Gibbs step that proposes new parameter values, and evaluates them through comparison of their conditional distributions, which have the form $p(\boldsymbol{\theta}_c)\cdot p(\textbf{y}^{(c)}|t^{(c)},\textbf{w}^{(c)},\boldsymbol{\theta}_c,\boldsymbol{\beta})$.

 \subsubsection{Outputs}
 
\begin{figure}[ht] 
 	\begin{center}
 		\begin{tabular}{c}
 			\includegraphics[width=0.5\textwidth]{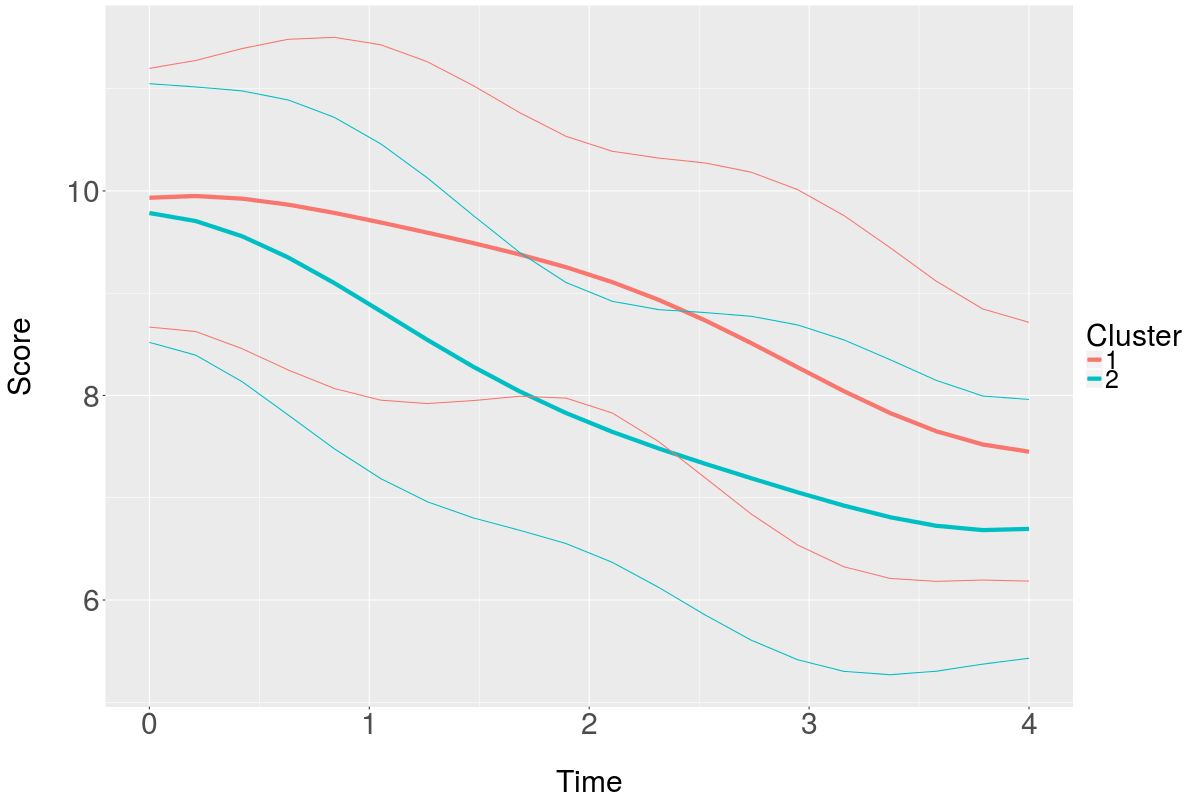}
 			\includegraphics[width=0.5\textwidth]{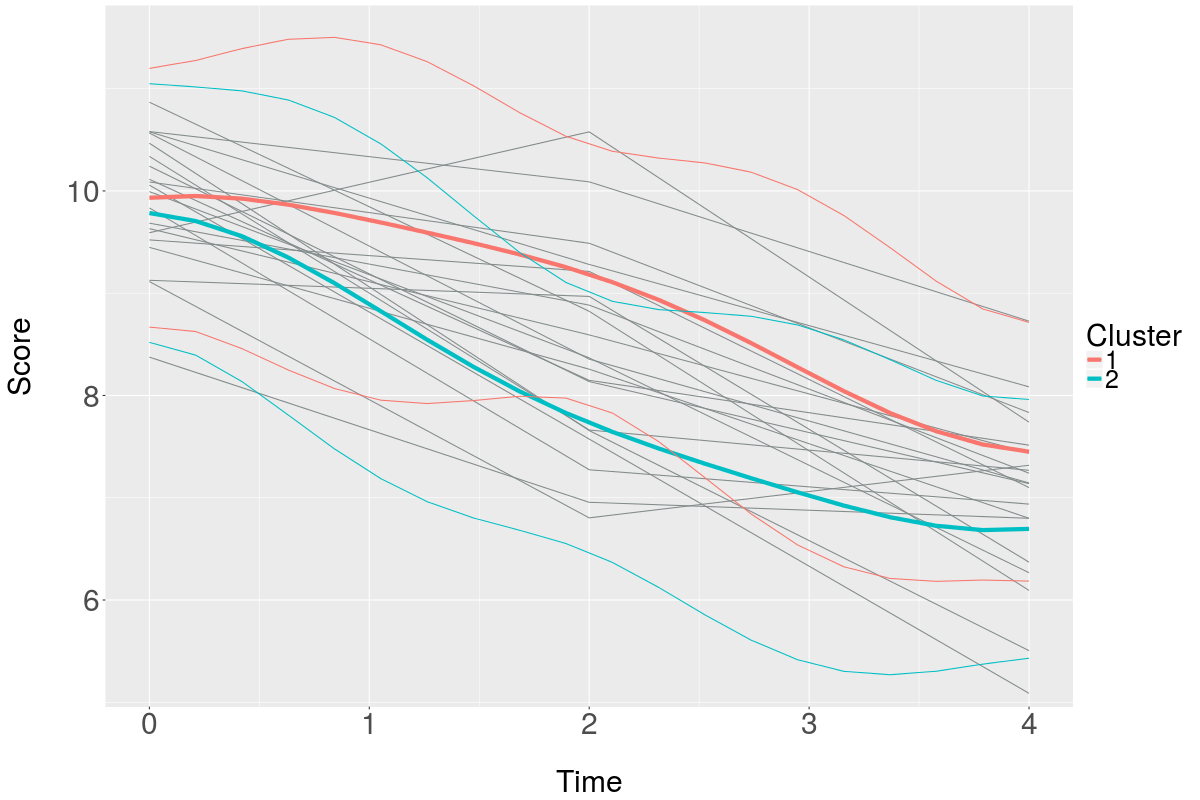}
 		\end{tabular}
 		\caption{Default PReMiuMlongi output summarising estimated cluster-specific outcome trajectories, with the outcome data on the y axis and time on the x axis. Colours identify the separate clusters. Bold lines show the posterior mean function and thin lines show the 90\% credible intervals. The right panel shows the individual trajectories (gray thin lines) together with the cluster patterns.}
 		\label{summarygp}
 	\end{center}
 \end{figure}

 An additional output file is written for the $\boldsymbol{\theta}$ parameters, which are sampled at each iteration of the algorithm. These values are used to reconstruct the posterior parameter distributions over $g^{(c)}$, which are visualised as a function of time with 90\% credible intervals shown on either side. The standard deviation is a sum of the standard deviation about the mean throughout all samples and the mean standard deviation of all samples. These plots are realised with and without the data in the background. See Figure \ref{summarygp} for examples.
 
\begin{figure}[ht] 
 	\begin{center}
 		\begin{tabular}{c}
 			\includegraphics[width=0.5\textwidth]{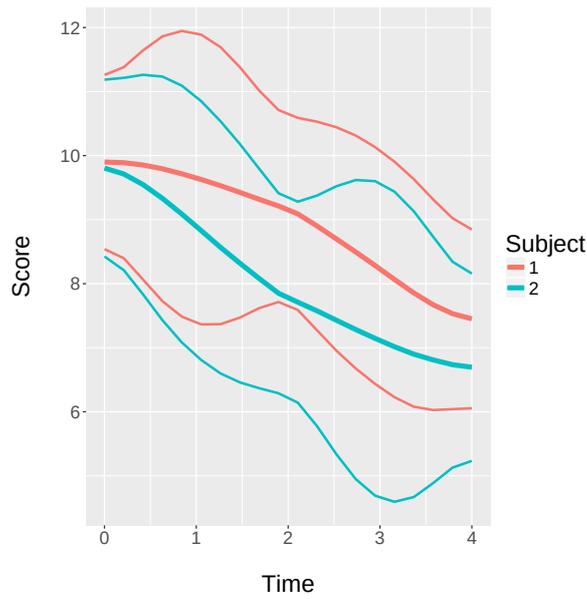}
 		\end{tabular}
 		\caption{Default PReMiuMlongi output showing predictions for the longitudinal response model of two individuals, with the outcome data on the y axis and time on the x axis. Colours identify the separate individuals for whom predictions are made. Bold lines show the posterior mean function and thin lines show the 90\% credible intervals.}
 		\label{predgp}
 	\end{center}
 \end{figure}

\begin{figure}[ht] 
 	\begin{center}
 		\includegraphics[width=0.8\textwidth]{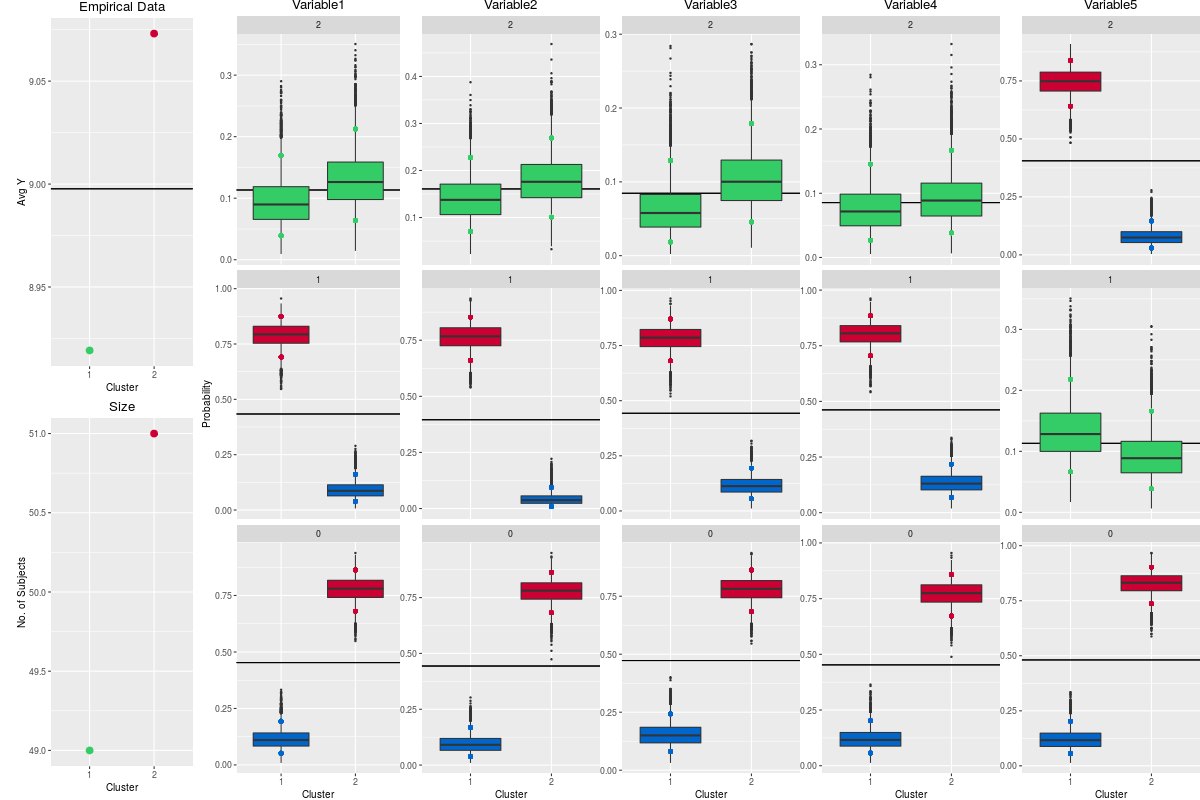}
 		\caption{Default PReMiuMlongi output summarising cluster-specific estimations. The first column presents the empirical outcome mean by cluster and the cluster sizes. The other columns present the box-plots of the posterior distributions of the probability for each variable to take values 2, 1 and 0 (from top to bottom). Green colour indicates an overlap of the 90\% credible interval with mean value for all individuals, represented by the solid black horizontal line. Red indicates that the 90\% credible interval is above the mean value, and blue, below.
 }
 		\label{summary}
 	\end{center}
 \end{figure}

 The same procedure also forms the basis for prediction of a new individual's outcome based on their covariate data. These plots have a similar form; see Figure \ref{predgp} for an example. The other graphical output, the covariate profile, is unchanged with the exception of the omission of the `risk' plot (see Figure \ref{summary}).
 
\end{document}